\documentclass[5p,sort,compress]{elsarticle}
\usepackage{fancyhdr}
\usepackage{amsfonts}
\usepackage{url}
\usepackage{amsmath}
\usepackage[labelfont=bf,justification=raggedright,singlelinecheck=false]{caption}
\usepackage{array}
\usepackage{tabu}
\usepackage{multirow}
\usepackage{graphicx}
\usepackage{amsmath}
\usepackage{algorithm}
\usepackage[noend]{algpseudocode}
\usepackage{pdflscape}
\usepackage{rotating}
\usepackage{wrapfig}
\usepackage[inline, shortlabels]{enumitem}
\usepackage{graphbox}
\usepackage{mathtools}
\usepackage{amsmath}
\newcommand*\rfrac[2]{{}^{#1}\!/_{#2}}
\newcolumntype{v}{>{\centering\arraybackslash}m{.18\linewidth} }
\usepackage{hyperref}
\hypersetup{
	colorlinks=true,
	linkcolor=blue,
	filecolor=magenta, 
	urlcolor=cyan,
}
\usepackage[labelfont=bf, justification=justified, format=plain]{caption} 
\begin{document}
	\journal{Arxiv}
	\title{WSMN: An optimized multipurpose blind watermarking in Shearlet domain \\ using MLP and NSGA-II}
	\author[fum,MVLAB]{Behrouz Bolourian Haghighi}
	\ead{b.bolourian@stu.um.ac.ir}
	\author[fum,MVLAB]{Amir Hossein Taherinia\corref{cor1}}
	\ead{taherinia@um.ac.ir}
	\author[fum,MVLAB]{Ahad Harati}
	\ead{a.harati@um.ac.ir}
	\author[fum]{Modjtaba Rouhani}
	\ead{rouhani@um.ac.ir}
	\cortext[cor1]{Corresponding author}
	\address[fum]{Computer Engineering Department, Ferdowsi University of Mashhad, Mashhad, Iran }
	\address[MVLAB]{Machine Vision Laboratory, Ferdowsi University of Mashhad, Mashhad, Iran}
	\begin{abstract}
		Digital watermarking is a remarkable issue in the field of information security to avoid the misuse of images in multimedia networks. Although access to unauthorized persons can be prevented through cryptography, it cannot be simultaneously used for copyright protection or content authentication with the preservation of image integrity. Hence, this paper presents an optimized multipurpose blind watermarking in Shearlet domain with the help of smart algorithms including MLP and NSGA-II. In this method, four copies of the robust copyright logo are embedded in the approximate coefficients of Shearlet by using an effective quantization technique. Furthermore, an embedded random sequence as a semi-fragile authentication mark is effectively extracted from details by the neural network. Due to performing an effective optimization algorithm for selecting optimum embedding thresholds, and also distinguishing the texture of blocks, the imperceptibility and robustness have been preserved. The experimental results reveal the superiority of the scheme with regard to the quality of watermarked images and robustness against hybrid attacks over other state-of-the-art schemes. The average PSNR and SSIM of the dual watermarked images are 38 dB and 0.95, respectively; Besides, it can effectively extract the copyright logo and locates forgery regions under severe attacks with satisfactory accuracy.
	\end{abstract}
	\begin{keyword}
		Digital watermarking\sep Copyright protection \sep Image authentication\sep Discrete Shearlet Transform\sep Multilayer Perceptron \sep Texture analysis \sep Multi-objective optimization
	\end{keyword}
	\maketitle
	\section{Introduction}
	Simultaneous with technological advancements, tampering and spreading of digital media such as images, voices, videos, and documents are expanding rapidly. Hence, exploitation of digital media without observance of copyright, document tampering, and forged documents have taken emerging and extending \cite{ref36}. Particularly, availability and easement access to image editing tools and rapid expansion of the Internet have made people able to modify and publish images with low costs and without any quality degradation. These serious issues have updated some problems including violation of copyright and integrity protection in the last decade. The best solution for protecting copyright and content integrity of images is digital watermarking \cite{ref36, ref3,ref5,ref4}.
	
	Digital watermarking is a category of data hiding in which confidential information called watermark is embedded in the host image through the embedding process. Otherwise speaking, a watermark is a pattern of bits that are embedded in multimedia data to identify the copyright owner and authenticate information. A simple digital watermark can be considered signatures or stamps on an image for property identification. The watermark should be imperceptible, and its bits must be scattered through the signal so that no one can identify or modify them \cite{ref1, ref2}. Generally speaking, the embedding watermarks should be done in a way that not only causes the least changes from the viewpoint of the human visual system and machines but also protects the watermark against attacks and tampering. The embedded watermark should be identified by the extraction unit with acceptable accuracy. To sum up, digital watermarking in data hiding science is an expanding technology for facilitating and ensuring data validation, security, and preservation of copyright and integrity for digital media \cite{ref1, ref2}.
	
	Evidently, the watermark extracting process can be divided into three categories \cite{ref1,ref2,ref6}. The method calls non-blind in the event that the host image is required \cite{ref9, ref23}. Subsequently, it entitles blind by extracting watermark without required any data \cite{ref8, ref10, ref12, ref14, ref19, ref22, ref24}. On the other hand, the extra information, including a histogram, statistical parameters, etc. plays a pivotal role in semi-blind works \cite{ref11, ref13, ref20, ref21}.
	
	Generally, robustness, imperceptibility, and capacity should be taken into consideration in a watermarking system \cite{ref1,ref2, ref3, ref4, ref5}.
	
	Robustness refers to the ability to extract watermarks from a media after basic operation of image processing, compression, and multiple attacks. Accordingly, it can be divided into three branches as robust, semi-fragile, and fragile. Robust watermarking is mostly used for proving copyright information with the aim of damage prevention of embedded watermark via frequent tampering, filtering, compression, etc. \cite{ref36, ref8, ref9, ref10, ref11, ref12, ref13, ref14}. During the tamper detection phase, the semi-fragile type can tolerate a certain degree of changes in a watermarked image such as compression noises and basic operation of image processing \cite{ref19, ref20, ref21}. Lastly, fragile methods are only used for authentications and document protections, which are highly sensitive toward signals due to their fragility and lack of robustness against attacks \cite{ref7, ref6, ref15, ref16, ref17, ref18}.
	
	On the other hand, the embedded marks should not be detectable by the eyes. It should only be recognizable and extractable by specific processes and jurisdictions. In other words, the mark hides such that does not attract the attention of viewers or reduce the quality of the media. Hence, the concept of imperceptibility is used for this purpose. Therefore, the development and expansion of imperceptible watermarking have great importance \cite{ref1,ref2,ref6}.
	
	The capacity of the system refers to the maximum amount of information that can be inserted into a host media. In other words, the number of hidden bits in the image represents the capacity of the system. Also, it indicates the possibility of embedding multiple watermarks in parallel forms. The capacity property has always encountered conflicts in comparison to robustness and imperceptibility \cite{ref1,ref2,ref6}.
	
	A remarkable point about the mentioned subjects is that all these characteristics cannot be provided simultaneously. Based on the desired functions and priorities, some of them are weak or strong. Nevertheless, watermarking methods should provide an intelligent trade-off. For this aim, the optimization algorithms are employed to search the optimum threshold steps for embedding watermark \cite{ref12, ref13, ref14, ref23}.
	\subsection{Literature Review}
	In this subsection, the state-of-the-art single and dual-purpose watermarking schemes presented for copyright and integrity protection of images are investigated. In this way, first, the single application inducing copyright and image integrity protection are discussed. Then, the details and key points of the novel multipurpose method are mentioned.
	
	Nowadays, researchers have developed different protection schemes to defend the rightful ownership of digital images. In \cite{ref8}, a watermarking scheme based on Discrete Wavelet Transform (DWT) and the q-deformed chaotic map was proposed. This work introduced an algorithm that tries to improve the problem of failure of encryption, including small keyspace, encryption speed, and level of security. The authors claimed that the excellent efficiency of the watermarking system is derived from the existence features of deformation. In another works \cite{ref9}, a secure watermarking scheme using logistic and RSA encryption was proposed. The aim of this method is to guarantee the security of the embedded data, improve robustness, and high computational efficiency. The watermark data embedded into the approximate sub-band of DWT. The experiment results demonstrated that the scheme had better quality and robustness, and also provided large embedding capacity compared to the previous method. Another common domain for hiding the watermark signal is a Discrete Cosine Transform (DCT). Accordingly, an image can be divided into high, low, and medium frequencies, and a watermark embeds in the best coefficients. One of the most popular compression formats based on this transform is JPEG, which is highly used due to its appropriate compression and preservation of quality, Simultaneously. Therefore, the robustness of the embedded watermark against compression is recently taken into consideration. An improved robust watermarking based on DCT and YCoCg-R color space was introduced in \cite{ref10}. The authors contend that the triplet planes of YCoCg-R color space have an excellent decorrelation, and modifying one component leads to the low effect on the rest. Hence, it can be employed to increase the robustness of watermarking against various attacks. Contrary to the previous method, a simple strategy based on the complexity and energy of each block was utilized to select candidate block and adaptive embedding strengths, respectively. Thanks to this mechanism, the robustness has increased against JPEG. 
	
	Lately, there has been a trend of employing machine learning and optimization algorithms to propose a more robust and intelligent watermarking scheme. In \cite{ref11}, a watermarking scheme based on Support Vector Machine (SVM) and Principal Component Analysis (PCA) was proposed. The copyright binary logo embedded in the approximate sub-band of the third level of Lifting Wavelet Transform (LWT). In the extraction phase, different statistical parameters were calculated for each block coefficients as features. Then, the features set were reduced and trained based on watermark as a label with the help of PCA and SVM, respectively. The energy compaction property of LWT improved the robustness compared to the traditional wavelet. To guarantee the robustness of the watermarking system, a new approach based on the optimization algorithm as an Artificial Bee Colony (ABC) was proposed in \cite{ref12}. The optimization objective was to provide the maximum possible correlation without raising a predetermined distortion limit. Therefore, the quality of the watermarked image guaranteed through that constraint optimization. It is evident from the results that the proposed approach achieves admissible robustness for image watermarking and satisfies the imposed quality constraint. In \cite{ref13}, a Bi-directional Extreme Learning Machine (B-ELM) for semi-blind watermarking of compressed images was presented. In this work, B-ELM was used to perform the watermarking of JPEG images by embedding a binary logo into it. In this way, DWT was selected to embed a binary watermark. Also, the watermark embedding strength had estimated based on Complex-Valued Neural Network (CVNN). With the help of this technique, the imperceptibility of the watermarked image was improved.
	Totally, the proposed scheme is suitable for image watermarking in the compressed domain. In another work, a new DCT based compressed image watermarking method using Teaching Learning Based Optimization (TLBO) was presented \cite{ref14}. The embedding phase is based on the relationships between the DCT coefficients. As well as, YCbCr color space was used due to its components have a good decorrelation. Moreover, it has been designed to gain more robustness against most of the attacks in the JPEG format. The embedding parameters and suitable position for embedding the watermark were determined based on TLBO, which has been applied rarely so far in watermarking works. TLBO is a novel method of optimization that has become an interesting issue in recent years. The algorithm is based on the principle of teaching and learning. In detail, teachers improve the knowledge of students, and the students learn from interaction among themselves. According to the experimental results, the quality of watermarked images and resistance are satisfactory.
	
	As discussed, nowadays, due to the common signal processing tools as well as the difference tampering, authentication of images has changed to one of the critical fields of digital image processing, which can be guaranteed via watermarking methods. Unfortunately, the majority of works that have been published in the last decade are fragile. Hence, the concept of the robust watermark, which tolerates against different attacks, is meaningless in these works \cite{ref7, ref6, ref15, ref16, ref17, ref18}. To cover drawbacks of fragile schemes which work in spatial domain, the semi-fragile method based on the frequency domain is always suggested. In \cite{ref21}, a quantization based semi-fragile watermarking scheme for image content authentication with tampering localization was presented. The method used a quantization technique to modify one chosen wavelet approximation coefficient of each non-overlapping block. Thanks to this strategy, its robustness against incidental attacks and fragileness against malicious attacks were guaranteed. Also, the security of the watermark was provided based on the hash function and Mersenne Twister Algorithm (MTA) by different keys. The experimental results demonstrated that the scheme identifies intentional and incidental modification and localizing tampered regions. In \cite{ref19}, the authors proposed DWT and Arnold Cat Map (ACM) based method for tamper detection. In this method, the approximate coefficient of each block is embedded in detail coefficients of another block. In the embedding stage, both blocks are shuffled using ACM. The results indicated the efficiency of the proposed method in tamper detection and recovery. In \cite{ref20}, the authors suggested a semi-fragile method for authentication and localization of tampering based on Singular Value Decomposition (SVD) transform. In this method, the watermark is obtained by applying exclusive-or via singular values. After that, the watermark is embedded in 4$\times$4 blocks of DWT domain to generate the watermarked image. The watermark is extracted and recovered in the authentication unit, and consequently, the tampering areas are marked. Totally, the main disadvantage of the presented methods is the low robustness and quality of the watermarked image.
	
	Finally, multiple or dual watermarking techniques are attracting more research attention in recent years for its wide variety of applications in the information security field. These schemes facilitate the integration of copyright protection and integrity verification into the same scheme Simultaneously. In this kind of works, ownership watermark is usually robust, while content integrity protection based watermarks are fragile watermarks. In \cite{ref22}, a dual watermarking method for copyright protection and integrity verification was introduced. In this work, the generated authentication and copyright bits are combined and embedded in the Least Significant Bits (LSBs) of the cover image. In the tamper detection phase, the calculated authentication bits and extracted ones were compared to discover tampered blocks. To enhance the efficacy of the scheme, a random chaotic mapping of blocks was employed. In detail, the Logistic Chaotic Map (LCM) was utilized to guarantee the security of the scheme. According to results, the scheme tolerated well against a different set of attacks like noises, filtering, histogram equalization, JPEG compression. In \cite{ref23}, a novel strategy for copyright protection and content validation in the domain of DWT based on SVD and ABC was proposed. For this aim, a robust watermark is embedded into DWT coefficients based on modified singular values in accordance with the principal components of the watermark. On the other hand, the last two LSBs of the cover image are reserved to carry the authentication information. In this work, the optimization algorithm is employed to obtain maximum robustness corresponding to the user-specific threshold of imperceptibility. The sequential insertion of robust and fragile watermarks made the method suitable for dual applications. Another work in \cite{ref24} proposed a multipurpose scheme based on DCT and LSBs. To do so, the authentication of the content has been ensured thanks to a fragile watermark embedded in the spatial domain. Also, the ownership protection is guaranteed with the help of the robustly correlated logo in DCT coefficients. Moreover, to thwart an adversary and ensure security, a novel encryption algorithm in conjunction with Arnold transform was employed. The required side information during the extraction phase, weak robustness, poor quality of watermarked images, the fragility of content validation watermark, etc. are the significant drawbacks of the discussed developed schemes.
	\subsection{Key Contributions of WSMN}
	Robust and semi-fragile watermarking schemes in the frequency domain have their own advantages; however, the main disadvantages are the low capacity of selected transform domain and quality of the watermarked image. Additionally, the most presented schemes are single-application and fragile against hybrid attacks in recent research. According to the reviewed methods as literature, disadvantages, and challenges are mentioned below:
	\begin{enumerate}[1),itemsep=0mm,leftmargin=5mm]
		\item One of the main challenges of this scope is content authentication protection or tamper detection at the time of multiple intentional and unintentional attacks. Most of the presented methods were worked on the spatial domain to focus on higher control of quality and capacities. In other words, they had not been concentrated on appropriate robustness against attacks, and the embedded watermark is too vulnerable to attacks. 
		\item According to a lack of sufficient embedding space, one of the important challenges of the previous works while tampering and applying hybrid attacks are low accuracy in the copyright extraction phase. The copyright mark is extracted from the forged region, which is extremely destroyed. So, the ownership claim cannot be proved clearly.
		\item Simultaneous quality protection of watermarked images and robustness against attacks is another challenge that played an important role in the current research. To the best of the authors’ knowledge, an efficient strategy to the trade-off between quality and robustness is not considered in the watermark embedding phase. In most works, the strength threshold of watermark bits is manually determined and is constant for whole blocks.
		\item One of the key characteristics of the watermarking method should be guaranteed security and randomness of the embedded marks. The attacker should not be able to detect the embedded regions of marks. On the other hand, most of the presented methods are not responsible to detect security tamperings such as copy-move, protocol, and vector-quantization attacks.
		\item Lastly, the proposed method should be generalizable to color images. Also, it should be intelligently preserved watermark bits in texture case, which has different rough and flat regions to be robust and imperceptible.
	\end{enumerate}
	In WSMN, to tackle the mentioned challenges, novel robust and semi-fragile optimized multipurpose blind watermarking in Shearlet domain along with texture analysis based on MLP and NSGA-II for copyright and content protection is presented. To the best of the authors' knowledge is the first time that robust and semi-fragile schemes are employed to prove the ownership and integrity verification of digital images simultaneously. The aim of WSMN is to enhance robustness against intentional and unintentional attacks and quality improvement of the watermarked image simultaneously. The key innovations and contributions of WSMN are listed as below:
	\begin{enumerate}[1),itemsep=0mm,leftmargin=5mm]
		\item Multi-resolution and multi-direction of DST can provide more capacity compared to rest transform Ex. DWT, especially for dual purpose applications.
		\item Applying texture descriptors to cluster the blocks makes the embedded watermarks more robust and imperceptible.
		\item Using difference strength thresholds for each direction of DST in the proposed correlation techniques dramatically improve the quality and robustness.
		\item An intelligent threshold estimation strategy for both content and copyright marks based on NSGA-II leads to an optimum trade-off between quality and robustness.
		\item Considering four chances for ownership protection to increase the accuracy in case of large destruction by finding out in the authentication phase.
		\item Acceptable leaning and generalizing ability of MLP able to withstand against signal distortion and aid to extract mark with the highest possible correlation that enhances the rate of tamper detection in normal and security cases.
		\item Simultaneously, robust and semi-fragile plans for multi-objective applications, including copyright and content protection, which neither cover nor watermark images are transferred to the extraction phase.
		
	\end{enumerate}
	\subsection{Road map}
	The remainder of this paper is organized as follows: Section \ref{sec:Preliminaries}, briefly explains some background material for WSMN. In Section \ref{sec:Proposed}, the design, and implementation of WSMN are described in detail. Next, the experimental evaluation scenario and details of comparison with the state-of-the-art methods are described in Section \ref{sec:Experimental}. Finally, the conclusion and future scope of WSMN is found in Section \ref{sec:Conclusion}.

	\section{Preliminaries}
	\label{sec:Preliminaries}
	This section presents the background material of WSMN. First, a brief introduction to NSGA-II is explained. Next, the details of Shearlet transform is demonstrated, and finally, a significant descriptor as Gabor is described.
	
	\subsection{Multi-objective Optimization (NSGA-II)}
	Nowadays, Multi-Objective Optimization (MOO) has attracted a wide interest over the past decades for engineering optimization problems. MOO is a scope of multiple criteria decision making, that is optimized mathematical problems involving more than one objective function, simultaneously. In other words, MOO refers to the minimizing or maximizing of two or three objectives functions. The formulation of MOO is presented as Eq. \ref{eq:MOO}:
	\begin{align}
	F(x)=\text{min} & \overrightarrow{f}(x) \quad x\in X, \overrightarrow{f}: X \to \mathbb{R}^m \nonumber \\
	\text{min} & \overrightarrow{f}(x) : \Big(f_1(x), f_2(x)\Big) \nonumber \\
	\text{S.t.} \quad x &= (x_1, x_2, ..., x_m) \in \Omega \subseteq \mathbb{R}
	\label{eq:MOO}
	\end{align}
	where $x \in \Omega$ and $\Omega$ are decision vector named a feasible solution and the feasible set (decision space), respectively. The vector $f(x) \in f(\Omega)$ is named an objective vector, and $F(x)$ represents the objective vector containing two objective functions. The decision vector (decision variables) defines the decision space where the feasible region is the set of solutions in the space which satisfy any constraints considered on the decision variables.
	Also, the bound constraints imposed in the decision variable represented by Eq. \ref{eq:x}:
	\begin{equation}
	x^l_i\leq x_i\leq x^u_i \quad i= 1, 2, .., m
	\label{eq:x}
	\end{equation}
	where $i$ is the index number of dimension, and $x^l_i$ and $x^u_i$ are the lower and upper limits of the variable, respectively. A decision vector $x$ that satisfies all the constraints is considered as feasible solution.
	
	Unlike the single objective technique that has one or several optimal solutions, MOO has a set of trade-off solutions (No single global). In other words, the solutions balance the objective and the best trade-offs among objective call Pareto Front (Pareto Optimal). All the Pareto optimal solutions form the Pareto set, and the objective values of the Pareto optimal solutions form the Pareto front. Generally, a limited number of solutions are gained when solving MOO called non-dominated solutions. Suppose, $a$ and $b$ are two decision vectors $(x_1, x_2) \in \Omega$ for maximizing each function. The decision vector $x_1$ is said to dominated $x_2$ if and only if:
	\begin{align}
	f_i(x_1) &\geq f_i(x_2) \nonumber\\
	f_j(x_1) &\geq f_j(x_2)\nonumber\\
	\forall i &\in {1, 2, ..., n}\nonumber
	\label{eq:dom}
	\end{align}
	In other words, a set of decision variables is non-dominated when no point in the set is dominated by another one.
	
	In the last decades, several significant MOO were proposed. In this scheme, the Non-dominated Sorting Genetic Algorithm (NSGA-II) \cite{ref26, ref27} is chosen to find Pareto front and optimum thresholds for embedding watermark. The reason for employing NSGA-II is that it has been demonstrated to be among the most efficient algorithms for multi-objective optimization in several benchmark functions. The basic structure of NSGA-II is similar to the previous version. The NSGA-II algorithm employs crossover and mutation to generate offspring population. Also, apply a fast non-dominated sorting technique to determine the non-dominated rank of individuals. In the selection phase, an elite preservation strategy is utilized to pick a new generation from the parent and offspring population. In NSGA-II, one individual is said to dominate another if its solution is:
	\begin{itemize}
		\item No worse than another in all objectives.
		\item Strictly better than another in at least one objective.
	\end{itemize}
	For more information about the NSGA-II, refer to \cite{ref26, ref27}.
	\subsection{Shearlet Transform}
	Most of the image processing functions use sparse displays. The most common transform in this kind of image processing application is discrete wavelet transform. Although wavelet is appropriate for the approximation of one-dimensional signals, it is not efficient for two or multi-dimensional data. In images with two-dimensional smooth signal pieces, the smooth regions are separated from each other by the edges. The edges are located along smooth curves; therefore, they are inherently geometrical. Two-dimensional wavelets are efficient for the separation of discontinuities of edge points; however, they do not consider continuities along the smooth curves. Since two-dimensional wavelets are obtained via tensor multiplication of one-dimensional wavelets, multiple expressions should be used for the representation of curves. As the scale gets smaller, the number of these expressions increases, and the weakness of the wavelet becomes clearer. However, the display of this boundary and curve is carried out by multi-directional wavelets in sparse and optimal modes.
	
	One of the multi-directional wavelet transforms is Shearlet \cite{ref31, ref32, ref34, ref37}. Shearlet is a multi-scale and multi-directional transform with a simple mathematical model. Its two main characteristics are the ability to use multi-scale transforms and preserve of data geometry. This transform is an affine system including a mother Shearlet function with three scale, shear, and translation parameters. The Shearlet transform on function $f\in L^2(R^2)$ is defined as Eq. \ref{eq:shear1}:
	\begin{equation}
	SH_\Psi(f)(a, s, t) = \langle f, \Psi_{a, s, t}\rangle
	\label{eq:shear1}
	\end{equation}
	Whenever $f$ is a two dimensional image, analytical elements of $\Psi_{a, s, t}$ are called Shearlet and an affine system is formed along with well localized functions in continuous scale of $a > 0$, and $t \in \rm I\!R^2$ along the direction with gradient of $s\in \rm I\!R$ in frequency domain. Shearlets are defined as Eq. \ref{eq:shear2}:
	\begin{equation}
	\Psi_{a, s, t}(x) = a^{-\frac{3}{4}}\Psi(A^{-1}_aS^{-1}_s(x-t))
	\label{eq:shear2}
	\end{equation}
	where $A_a$ represents scale matrix, and $S_s$ shows Shearlet matrix which is defined in Eq. \ref{eq:shear3}:
	\begin{equation}
	S_s = \Bigg(
	\begin{matrix}
	1 & s \\
	0 & 1
	\end{matrix}
	\Bigg),
	A_a = \Bigg(
	\begin{matrix}
	a & 0 \\
	0 & \sqrt{a}
	\end{matrix}
	\Bigg)
	\label{eq:shear3}
	\end{equation}
	
\begin{figure}[t!]
		\centering
		\includegraphics[width=0.6\columnwidth]{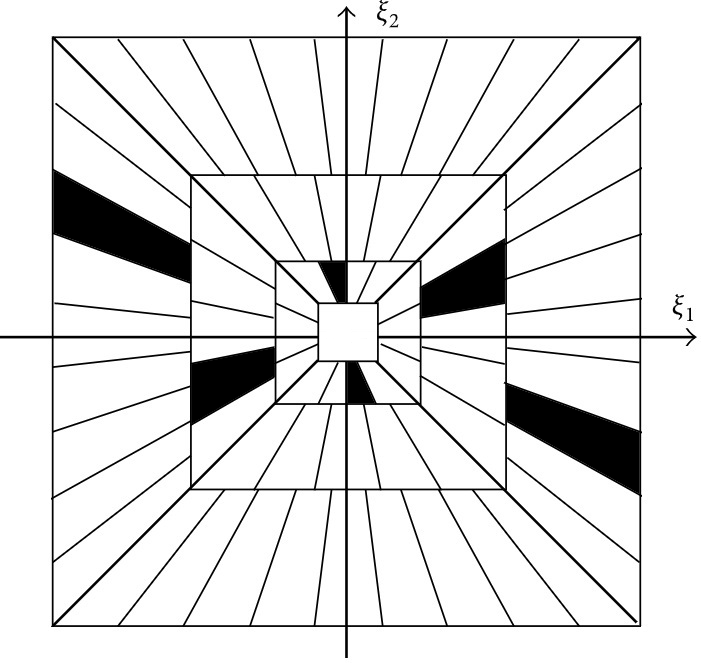}
		\caption{The tiling of the frequency planes based on Shearlet transform.}
		\label{fig:sheartrans1}
	\end{figure}

	$A_a$ matrix is an anisotropic delay, and $S_s$ Shearlet matrix parameterizes the directions via gradient related $s$ variable. Besides the advantages of other transforms, Shearlets can be divided step by step in frequency space, which increases their efficiency. Moreover, Shearlet transform can produce a collection of basic functions using the scale, transform, and rotation functions. This advantage is the superiority of Shearlet transform over wavelet transform. Each element of $\Psi_{a, s, t}$ has frequency support on a trapezoid-like pair \cite{ref31, ref32, ref34, ref37}. These trapezoids divide the frequency plate as Fig. \ref{fig:sheartrans1}.
	
\begin{figure}[t!]
		\center
		\setlength{\tabcolsep}{2pt}
		\begin{tabular*}{1\columnwidth}{ccc}
			\includegraphics[width=0.3\columnwidth]{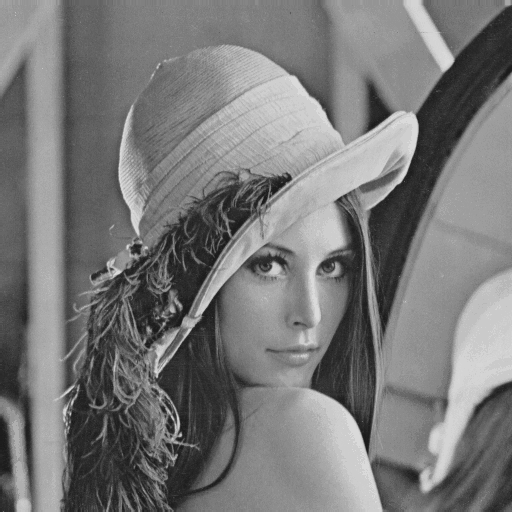} & \includegraphics[width=0.3\columnwidth]{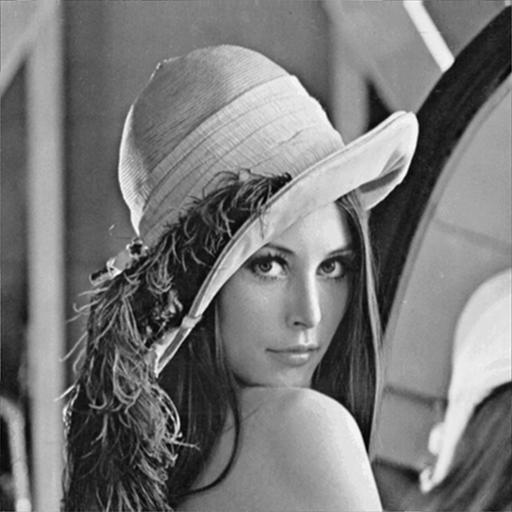}&
			\includegraphics[width=0.3\columnwidth]{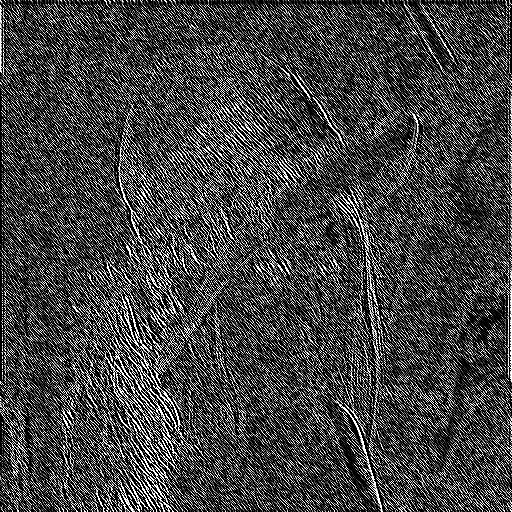} \\
			(a)&(b)&(c)\\
			\includegraphics[width=0.3\columnwidth]{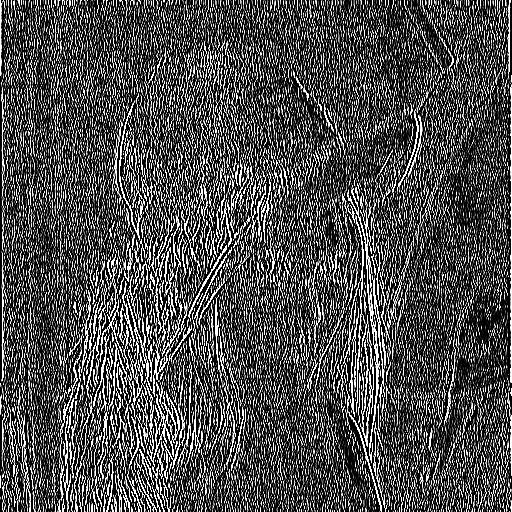} &\includegraphics[width=0.3\columnwidth]{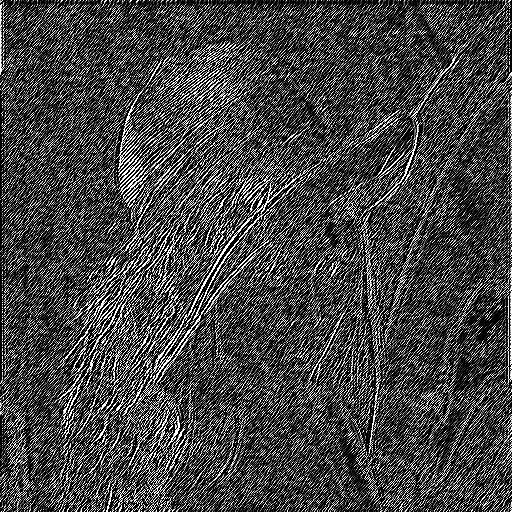} &
			\includegraphics[width=0.3\columnwidth]{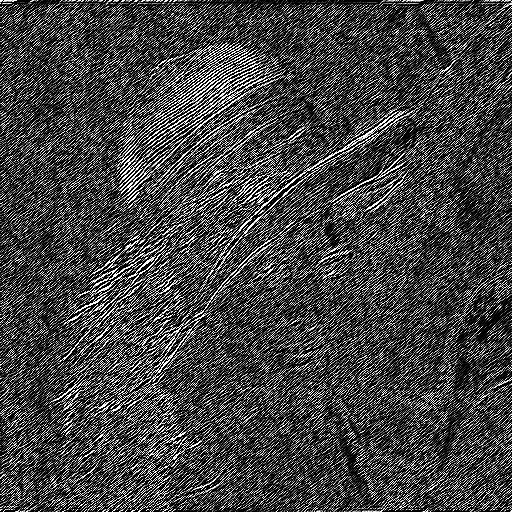}\\
			(d)&(e)&(f)\\
			\includegraphics[width=0.3\columnwidth]{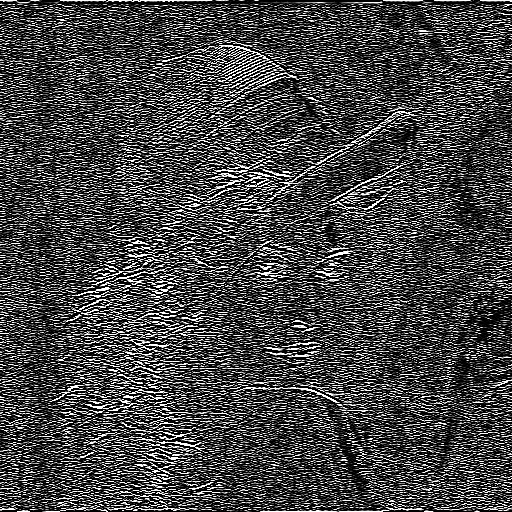} &\includegraphics[width=0.3\columnwidth]{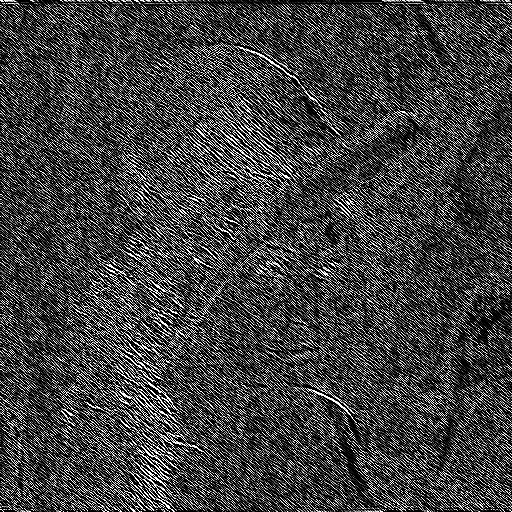} & \includegraphics[width=0.3\columnwidth]{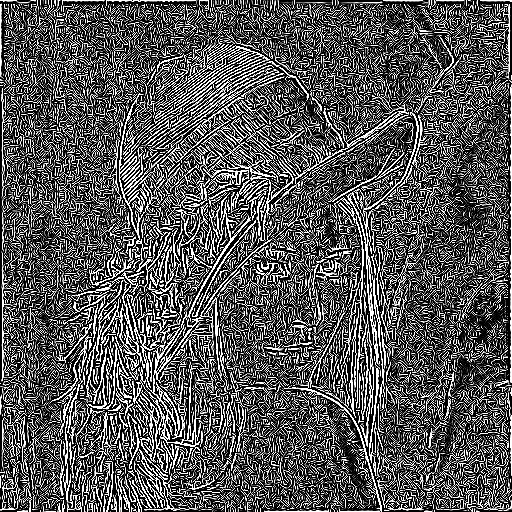} \\
			(g)&(h)&(i)\\
		\end{tabular*}
		\caption{(a) Original image, (b) Approximate, (c-h) Details, (i) Combined details based on STD.}
		\label{fig:sheartrans2}
	\end{figure}

	Fig. \ref{fig:sheartrans2} indicates the results of DST on Lena image. The number of levels in this transform is one and Shearlet vector is considered to be [0]. The results of this transform are six images in different directions as high frequency (details) and a image including low frequencies (approximate). Additionally, combining Shears with a maximum standard deviation in detail planes is indicated due to the application in WSMN.
	\subsection{Gabor Descriptor}
	Gabor descriptor is a potent tool for texture analysis due to its high and multiple resolutions in spatial and frequency domains. Gabor filter banks are used for extracting texture properties. The filter banks have two scale and direction components \cite{ref25}. Gabor filter is a high-pass filter multiplication of a Gaussian function in a sin complex function. By applying Gabor filters on the candidate image or one of its regions, several filtered images are obtained. For each filtered image, mean and standard deviation criteria are estimated as texture properties. Two dimensional Gabor function is defined as Eq. \ref{eq:Gabor1}:
	\begin{equation}
	g(x, y) = \Bigg(\frac{1}{2\pi\sigma_x\sigma_y}\Bigg)exp\Bigg[-\frac{1}{2}\Bigg(\frac{x_2}{\sigma^2_x}+\frac{y_2}{\sigma^2_y}\Bigg)+2\pi jfx\Bigg]
	\label{eq:Gabor1}
	\end{equation}
	where
	\begin{equation}
	x = xcos\theta+ysin\theta, y = -xsin\theta+ycos\theta
	\label{eq:Gabor2}
	\nonumber
	\end{equation}
	Fourier transform of Gabor filter is indicated in Eq. \ref{eq:Gabor3}:
	\begin{align}
	G(u, v) &= exp\Bigg[-\frac{1}{2}\Bigg(\frac{(u-f)^2}{\sigma^2_u}+\frac{v^2} {\sigma^2_v}\Bigg)\Bigg] \nonumber \\
	&\sigma_u = \frac{1}{2\pi\sigma_x}, \sigma_v = \frac{1}{2\pi\sigma_y}
	\label{eq:Gabor3}
	\end{align}
	where $\sigma_x$ and $\sigma_y$ are scale parameters in horizontal and vertical directions, and $f$ is the central frequency of the filter.
	\begin{figure*}[t]
		\centering
		\includegraphics[width=1\textwidth,trim= 32cm 13.5cm 28cm 0.5cm,clip]{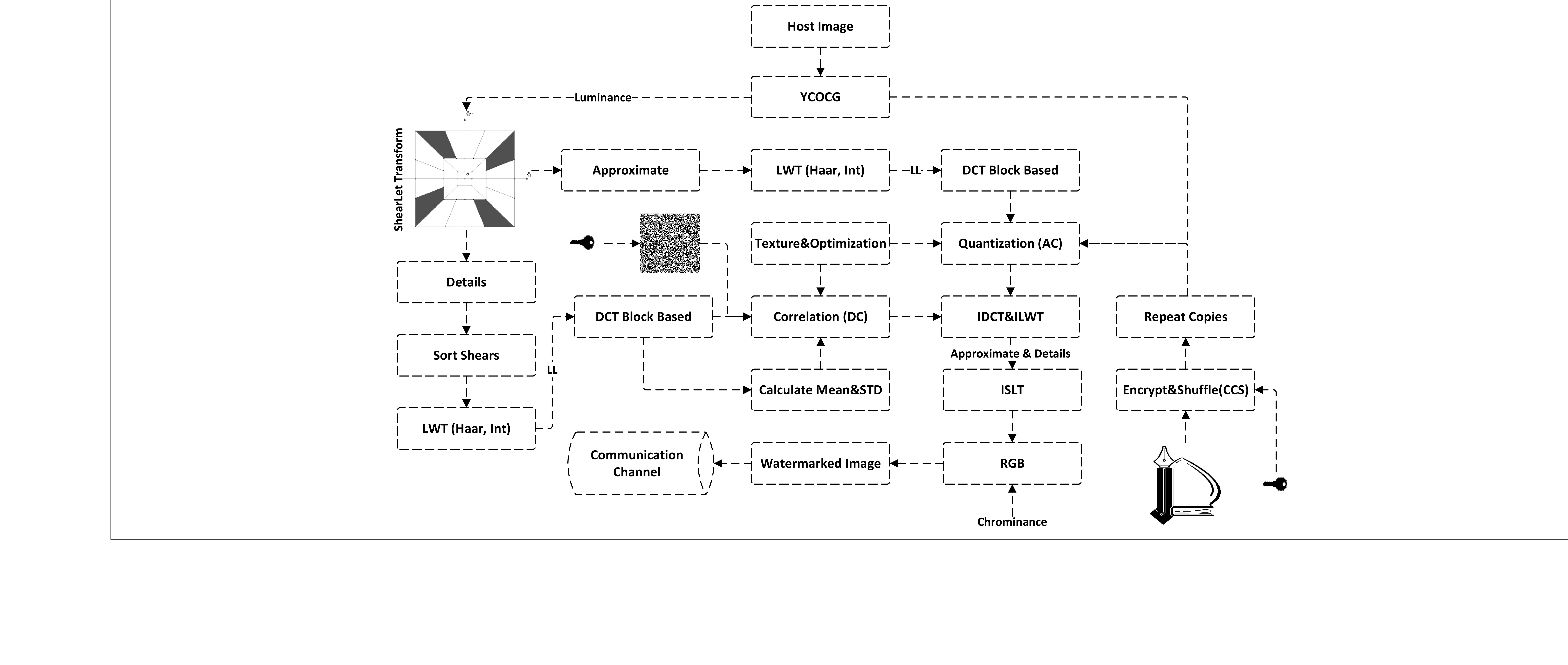}
		\caption{The block diagram of embedding phase of WSMN.}
		\label{fig:dembed}
	\end{figure*}
	\section{Proposed Method}
	\label{sec:Proposed}
	In this section, an optimized multipurpose blind watermarking in Shearlet domain using MLP and NSGA-II for image copyright protection and content authentication is proposed. The most proposed methods consider constant thresholds for embedding watermarks regardless of the texture of the block. However, some blocks require higher thresholds to gain more robustness against hybrid attacks; In contrast, some blocks require lower thresholds based on their characteristics to maintain the quality of blocks. In WSMN, various texture descriptors are employed to separate blocks as rough or flat. Afterward, the blocks are dynamically clustered based on K-Means algorithm. Besides, Shearlet transform which provides multi-resolution and multi-direction is used over the other transforms to conceal watermarks information. WSMN hides two watermarks including robust copyright logo and semi-fragile authentication mark in the approximate and details coefficients of Shearlet based on quantization and correlation techniques, respectively. Also, the NSGA-II algorithm is used for predicting the optimum embedding strengths to balance the robustness and imperceptibility of watermarked images. In this way, the meta-heuristic process finds the optimum values according to the fitness functions that evaluate each solution based on the optimization objectives including quality and robustness.
	
	On the receiver side, the four copyright marks are extracted by the inverse quantization process; On the other hand, the authentication watermark is evoked based on MLP due to its powerful and good generalization ability compared to different machine learning algorithms. Thereinafter, the authenticity of the received image is analyzed according to the generated random sequence and extracted watermarks. With the help of these properties, not only the quality of the watermarked image but also the robustness against several attacks are significantly improved. In other words, these properties enable a watermarking scheme to have high values of robustness and imperceptibility.
	
	In the following, the details of the proposed method are described into two subsections as \hyperref[Embedding_Watermarks]{Embedding Watermarks} and \hyperref[proof_ownership_and_tamper_detection]{Proof Ownership and Tamper Detection}.
	\subsection{Embedding Watermarks}
	\label{Embedding_Watermarks}
	In this subsection, the embedding phase is elaborated in detail. The process is done in six stages, including Watermarks Processing, Texture Analysis, Embedding Pre-processing, Embedding Processing, Embedding Post-processing, and Threshold Optimization. The block diagram is shown in Fig. \ref{fig:dembed}.
	
	First of all, Let's denote the candidate image as $X$ with a size of $M \times N$. Also, the size of the block is termed as $m \times n$. Thus, the total number of blocks is $\rfrac{M}{m}\times\rfrac{N}{n}$ which denoted as $\ell$. Plus, $[Y, Co, Cg]$ represents the color components of $X$ in the YCoCg color space. If $X$ is in gray-scale mode, the chrominance component is meaningless. The watermark embedding stages of WSMN are explained as follow:
	\subsubsection{Watermarks Processing}
	As mentioned before, WSMN considers two watermarks as a robust copyright logo and a semi-fragile random sequence for proofing ownership and integrity of the image, respectively. The details of this stage are described below in detail:
	\begin{enumerate}[1),itemsep=0mm]
		\item First, a binary random sequence is generated as $\chi \sim N(0,1)$ by $key_1$.
		
		\item The binary copyright logo as $\tilde{w}_c$ with size of $\rfrac{M}{2m}\times \rfrac{N}{2n}$ is encrypted by Eq. \ref{eq:encrypt}:
		\begin{align}
		\tilde{w}_{c}' = \tilde{w}_c \oplus \chi
		\label{eq:encrypt}
		\end{align}
		where $\oplus$ represents bit-wise exclusive-or operation.
		
		\item For aiming more security, the CCS algorithm is employed to shuffle the mark \cite{ref30, ref18}. In this way, a sequences $R$ is generated based on CCS by $key_2$. The permutation position $p$ is achieved by sorting $R$ in ascending order. Then, the shuffled mark are achieved by utilizing Eq. \ref{eq:shuffled}:
		\begin{align}
		\tilde{w}_{c}''(i) = \tilde{w}_{c}'(p(i)),\quad \forall i \in [1, \rfrac{\ell}{4}]
		\label{eq:shuffled}
		\end{align}
		It should be noted, in this process the input and output are converted into 1D and 2D matrix, respectively. For more information about CCS refer to \cite{ref30, ref18}.
		
		\item Next, an array containing four copies of $\tilde{w}_{c}''$ in the row and column dimensions is constructed to form $w_c$. Hence, four chances are obtained to increase the accuracy rate in the extraction phase.
		
		\item Lastly, a binary random sequence $w_a$ with size of $\rfrac{M}{m}\times \rfrac{N}{n}$ is generated by $key_3$ as authentication mark.
	\end{enumerate}
	\begin{figure}[t]
		\centering
		\includegraphics[width=1\columnwidth,trim= 20cm 3cm 40cm 3cm,clip]{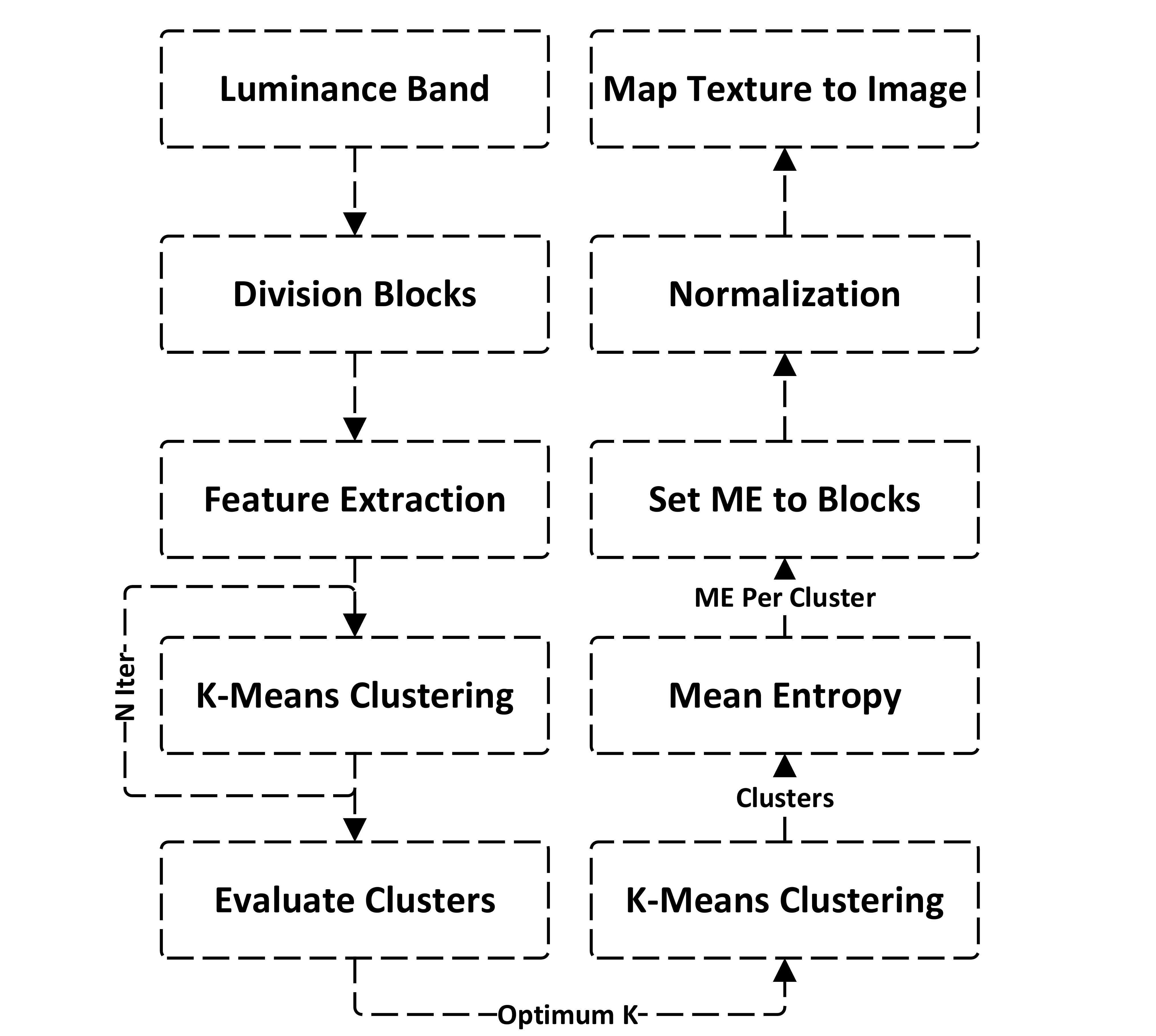}
		\caption{The block diagram of texture optimum analysis.}
		\label{fig:dtexture}
	\end{figure}
	\subsubsection{Texture Analysis}
	In the embedding phase, a constant scaling factor does not make sense because each image has a different type of region, including rough or flat. In other words, each coefficient has a different tolerance limit. Hence, to improve the quality of the watermarked image, and also increase the robustness against the severe hybrid attacks, the blocks of $X$ are separated into various types in terms of texture. The procedure of texture analysis is explained below in detail:
	\begin{enumerate}[1),itemsep=0mm]
		\item First, $X$ is divided into $m\times n$ non-overlap blocks.
		
		\item Texture descriptors include Local Binary Pattern (LBP), Entropy, Standard Deviation (STD), and Gabor filters are applied on each block to form features. Then, the mean elements of each feature is calculated by Eq. \ref{eq:feature}:
		\begin{align}
		\Phi^k_{i, j} &= \bar{F}^k_{i, j} \nonumber \\
		\quad \forall i \in [1, \rfrac{M}{m}]&, j \in [1, \rfrac{N}{n}], k \in \{1, 2, ..., 4\}
		\label{eq:feature}
		\end{align}
		where $\bar{F}$ represents the mean of corresponded feature.
		
		\item After normalizing features, k-means algorithm is employed by different $k$ for clustering features as Eq. \ref{eq:kmeans}:
		\begin{align}
		\Omega(k) = \text{k-Means}(\Phi, k), \quad \forall k \in \{1, 2, ..., \tau\}
		\label{eq:kmeans}
		\end{align}
		where $\tau$ shows list of number of clusters to evaluate.
		
		\item Next, an evaluating clustering solution is used, to find optimum $k$ as a number of clusters for $X$. For this aim, the Calinski-Harabasz is considered as criterion. The evaluating clusters is done according to Eq. \ref{eq:optkmeans}:
		\begin{align}
		k^* = \text{Evalclusters}(\Omega(k)), \quad \forall k \in \{1, 2, ..., \tau\}
		\label{eq:optkmeans}
		\end{align}
		where $k^*$ represents the optimum number of clusters. Then, $\Phi$ is clustered by $k^*$ and denoted results as $\gamma$.
		
		\item Lastly, the mean entropy of each block based on the corresponded cluster is computed by Eq. \ref{eq:entropyoptkmeans}:
		\begin{align}
		\bar{S_q} &= \frac{\sum_{p=1}^{\xi} S^q_p}{\xi} , \quad \forall q \in \{1, 2, ..., k^*\}
		\label{eq:entropyoptkmeans}
		\end{align}
		where $S$ and $\xi$ represent the entropy of each block and the number of elements per cluster. Then, assign $\bar{S}$ to the corresponded blocks by looking on the related cluster by Eq. \ref{eq:assigntexture}:
		\begin{equation}
		\xi_{i, j}= f(\bar{S}, \gamma), \forall i \in [1, \rfrac{M}{m}], j \in [1, \rfrac{N}{n}]
		\label{eq:assigntexture}
		\end{equation}
		where $f$ indicates function which set the values of $\bar{S}$ to correspond blocks. At the end, the range of elements of $\xi$ are scaled from $[0, log_2(256)]$ to $[0.6, 1]$ to update $\xi$ as texture coefficient of each block.
	\end{enumerate}
	The block diagram of texture analysis is illustrated in Fig. \ref{fig:dtexture}.
	\subsubsection{Embedding Pre-processing}
	In this stage, the frequency transforms are applied on the host image to form the coefficients for hiding watermarks. To do so, $X$ is decomposed according to the following steps:
	\begin{enumerate}[1),itemsep=0mm]
		\item First, the Shearlet transform is exerted on $X$ as Eq. \ref{eq:Shearlet}:
		\begin{equation}
		[\Psi_A, \Psi^s_D] = \text{DST}(X), s \in \{1, 2, ..., 6\}
		\label{eq:Shearlet}
		\end{equation}
		where $\Psi_A$ and $\Psi_D$ are approximate and details shear coefficients, respectively. For this aim, number of scales in DST is set to 1 and vector of Shearlet levels is determined to [0]. Therefore, 7 sub-bands are totally achieved as one low and six high frequency.
		
		\item Then, the coefficients of $\Psi_D$ are ordered based on standard deviation. For this aim, first, the block based STD for whole shears are separately computed by Eq. \ref{eq:stdShear}:
		\begin{align}
		\sigma^s(i, j) &= \text{STD}\Big(\Psi^s_D(i, j)\Big), \forall s \in \{1, 2, ..., 6\} \nonumber \\
		&\forall i \in [1, \rfrac{M}{m}], j \in [1, \rfrac{N}{n}]
		\label{eq:stdShear}
		\end{align}
		where $s$ represents the shears direction. Then, the coefficients of each direction are together exchanged based on STD by Eq. \ref{eq:orderedShear}:
		\begin{equation}
		[\Psi'_D, \kappa] = f(\Psi_D, \sigma)
		\label{eq:orderedShear}
		\end{equation}
		where $f$ and $\kappa$ express a function which order shears by considering the standard deviation values per blocks and the initial position of the coefficients in each direction, respectively.
		
		\item Next, one level of Lifting Wavelet Transform (LWT) is applied on the approximate and details by Eq. \ref{eq:LWT}:
		\begin{align}
		[LL_a, LH_a, HL_a, HH_a] &= \text{LWT}(\Psi_A, Haar) \nonumber \\
		[LL^s_d, LH^s_d, HL^s_d, HH^s_d] &= \text{LWT}(\Psi'^s_D, Haar)\nonumber \\
		\forall s \in \{1, 2, ..., 6\}&
		\label{eq:LWT}
		\end{align}
		
		\item In this step, the $LL$ sub-bands selected for data embedding is decomposed into $\rfrac{m}{2}\times\rfrac{n}{2}$ non-overlapping blocks. Next, DCT block based is performed on $LL$ coefficients by Eq. \ref{eq:DCT}:
		\begin{align}
		\varphi_a(i, j) &= \text{DCT}\Big(LL_a(i, j)\Big) \nonumber \\
		\varphi^s_d(i, j) &= \text{DCT}\Big(LL^s_d(i, j)\Big), \forall s \in \{1, 2, ..., 6\}\nonumber \\
		&\forall i \in [1, \rfrac{M}{m}], j \in [1, \rfrac{N}{n}]
		\label{eq:DCT}
		\end{align}
		
		\item Now, the mean and standard deviation of $DC$ coefficients of each shear block is computed by Eq. \ref{eq:MSDCT}:
		\begin{align}
		\mu^s &=\frac{1}{\ell}{\sum^{\rfrac{M}{m}}_{i=1}\sum^{\rfrac{N}{n}}_{j=1} \varphi^s_d (dc_{i, j})} \nonumber \\
		\sigma^s &=\Big(\frac{1}{\ell}{\sum^{\rfrac{M}{m}}_{i=1}\sum^{\rfrac{N}{n}}_{j=1} (\varphi^s_d (dc_{i, j})-\mu^s)\Big)^\frac{1}{2}} \nonumber \\
		&\forall s \in \{1, 2, ..., 6\}
		\label{eq:MSDCT}
		\end{align}
	\end{enumerate}
	\begin{figure*}[t]
		\centering
		\includegraphics[width=1\textwidth,trim= 49cm 13.5cm 48cm 9cm,clip]{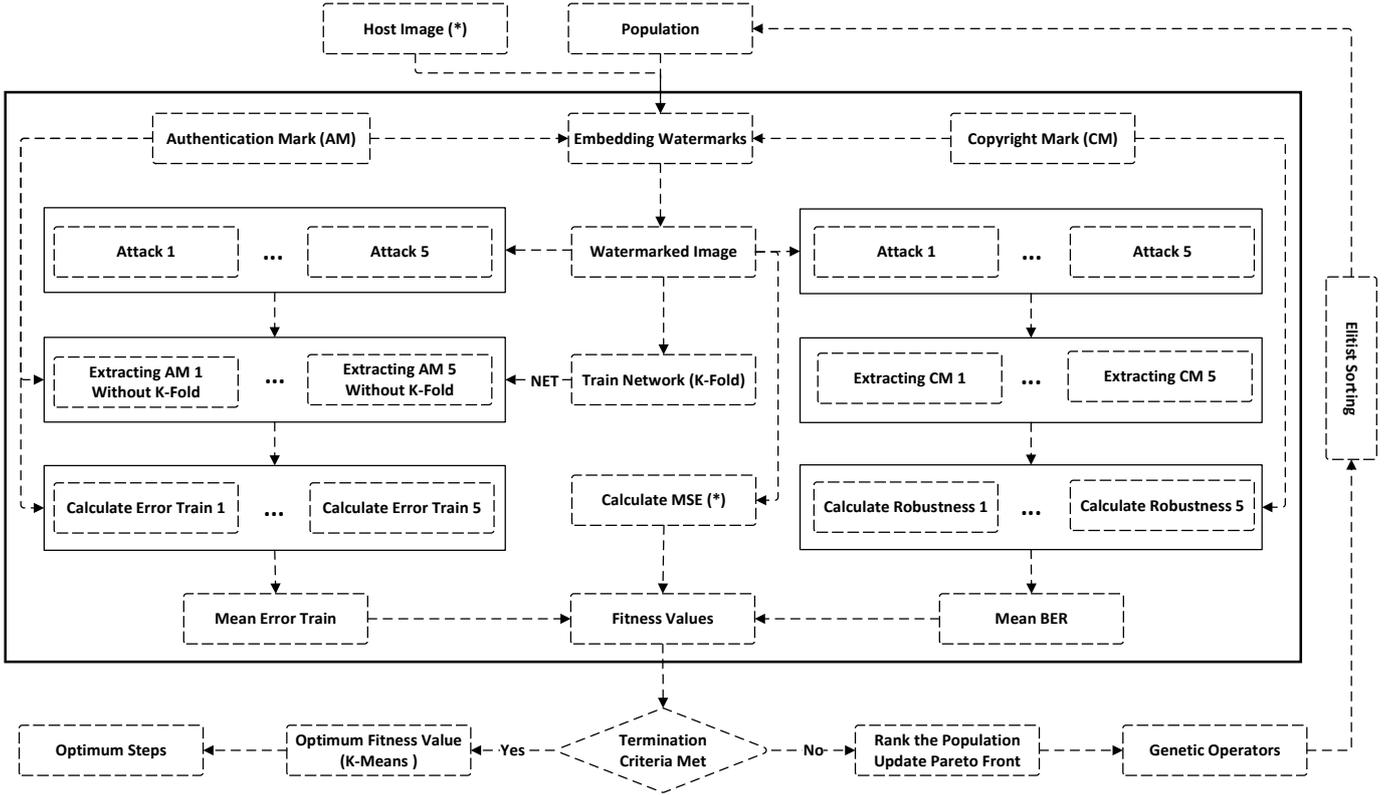}
		\caption{The block diagram of the optimization phase (Triple-Objective).}
		\label{fig:dopt}
	\end{figure*}
	
	\subsubsection{Embedding Processing}
	In this stage, the copyright and authentication marks are embedded in host signal by quantization and correlation techniques, respectively. The details are explained as below:
	
	\begin{enumerate}[1),itemsep=0mm]
		\item The copyright bits are embedded into each block using improved quantization technique \cite{ref33}. The quantized coefficients are gained by Eq. \ref{eq:quanitze}:
		\begin{align}
		\Gamma_{i, j} = 
		&\begin{dcases} 
		\delta'\Big\lfloor \frac{\varphi_a(ac_{i, j})}{\delta'} \Big\rceil & \text{if } w_c(i, j) = 0 \\
		\delta'\Big\lfloor \frac{\varphi_a(ac_{i, j})}{\delta'} -\frac{1}{2}\Big\rceil + \frac{\delta'}{2} & \text{if } w_c(i, j) = 1 \\
		\end{dcases} \nonumber \\
		&\forall i \in [1, \rfrac{M}{m}], j \in [1, \rfrac{N}{n}]
		\label{eq:quanitze}
		\end{align}
		where $\delta'$ and $ac$ represent the quantization step and second coefficients of AC components.
		\item Next, the copyright bit are embedded in host signal according to Eq. \ref{eq:embedquanitze}:
		\begin{align}
		\tilde{\varphi}_a(ac_{i, j}) &= \varphi_a(ac_{i, j}) + \xi_{i, j}\Big(\Gamma_{i, j}-\varphi_a(ac_{i, j})\Big)\nonumber \\
		&\forall i \in [1, \rfrac{M}{m}], j \in [1, \rfrac{N}{n}]
		\label{eq:embedquanitze}
		\end{align}
		where $\xi$ is the distortion compensation parameter which trade-off embedding distortion and robustness based on the texture coefficient of the corresponded block. In other words, the optimal values of $\xi$ are intelligently determined based on the texture of block.
		
		\item Finally, the authentication mark is embedded in the high frequency of Shearlet coefficients. For this aim, assuming $w_a^{i, j}$, $\xi_{i, j}$, and $\varphi^s_d (dc_{i, j})$ are represented by $w$, $\xi$, and $\varphi^s$, respectively. Now, $w$ is correlated with $\varphi^s$ according Eq. \ref{eq:embedcorr}:
		\begin{align}
		\rho^s &= \exp(\frac{1}{|\mu^s-\sigma^s|}) \nonumber \\
		\eta^s &= |(\mu^s+(\text{sgn}(w)\sigma^s\rho^s))-\varphi^s| \nonumber \\
		\vartheta^s &= \frac{2^{\delta''} exp(\xi)}{|\mu^s-\sigma^s|}\nonumber \\
		\tilde{\varphi}^s &= \begin{dcases} 
		\varphi^s + (\eta^s + \vartheta^s), & \text{if } \varphi^s < \mu^s+(\sigma^s\rho^s) \& w \nonumber \\
		\varphi^s - (\eta^s + \vartheta^s), & \text{if } \varphi^s > \mu^s-(\sigma^s\rho^s) \& \neg w\\
		\end{dcases} \nonumber \\ 
		& \forall s \in \{1, 2, ..., 6\}
		\label{eq:embedcorr}
		\end{align}
		where $\delta''$ represents correlation step. This process are done for whole blocks.
	\end{enumerate}
	
	As mentioned, the threshold step is playing a pivotal role in the watermarking algorithm. In other words, the key challenge is how to embed bits to achieve a maximum quality and rbustness. Therefore, to guarantee the quality and robustness of generated watermark, NSGA-II as a well known modern multi-objective optimization algorithm is employed. Notice that, the details of NSGA-II optimization for both quantization $\delta'$ and correlation $\delta''$ steps will further be explained in the Thresholds Optimization subsection \ref{sec:ThresholdsOptimization}. 
	\subsubsection{Embedding Post-processing}
	In the last stage of embedding phase, the inverse transforms are applied on coefficients to construct the watermarked image. The detail of this stage is mentioned below:
	\begin{enumerate}[1),itemsep=0mm]
		\item The inverse DCT block based are performed to reconstruct $LL$ coefficients by Eq. \ref{eq:invdct}:
		\begin{align}
		\tilde{LL}_a(i, j) &= \text{DCT}\Big(\tilde{\varphi}_a(i, j)\Big)^{-1} \nonumber \\
		\tilde{LL}^s_d(i, j) &= \text{DCT}\Big(\tilde{\varphi}^s_d(i, j)\Big)^{-1}, \forall s \in \{1, 2, ..., 6\}\nonumber \\ 
		&\forall i \in [1, \rfrac{M}{m}], j \in [1, \rfrac{N}{n}]
		\label{eq:invdct}
		\end{align}
		\item One level of inverse LWT is applied to reconstruct approximate and details coefficients of DST by Eq. \ref{eq:invlwt}:
		\begin{align}
		\tilde{\Psi}_A &= \text{LWT}([\tilde{LL}_a, LH_a, HL_a, HH_a], Haar)^{-1} \nonumber \\
		\tilde{\Psi}'^s_D &= \text{LWT}([\tilde{LL}^s_d, LH^s_d, HL^s_d, HH^s_d], Haar)^{-1}\nonumber \\ 
		&\forall s \in \{1, 2, ..., 6\}
		\label{eq:invlwt}
		\end{align}
		\item In this step, the coefficients of details are reordered to initial position according to step 2 of Pre-processing sub-section by Eq. \ref{eq:invorderedShear}:
		\begin{equation}
		\tilde{\Psi}_D = f(\tilde{\Psi}'_D, \kappa)^{-1}
		\label{eq:invorderedShear}
		\end{equation}
		
		\item Finally, the watermarked image is generated by the inverse Shearlet transform by Eq. \ref{eq:invShearlet}:
		\begin{equation}
		\tilde{X} = \text{DST}([\tilde{\Psi}_A, \tilde{\Psi}^s_D])^{-1}, s \in \{1, 2, ..., 6\}
		\label{eq:invShearlet}
		\end{equation}
		\item If case of color, it converts to the RGB space.
	\end{enumerate}
	Something which should be mentioned here is that WSMN can separately be employed for authentication or copyright protection. In other words, the method can be used as a single application by only embed corresponding watermark.
	\begin{table}[t]
		\footnotesize
		\caption{GA parameters used in the experiments.}
		\label{TABLE:GA}
		\renewcommand{\arraystretch}{1.5}
		\scalebox{1} {
			\begin{tabular*}{\columnwidth}{@{\extracolsep{\fill}}l@{}c@{}c}
				\cline{1-3}
				GA parameters&Value&Description\\
				\cline{1-3}
				Population size&50&Uniform\\
				Generation&100&-\\
				Number of variable&2&Vector\\
				Initialization range&\{(30, 50), (0, 2)\}&Double\\
				Fitness Function&\{(0, 1), (0, Inf)\}&BER/MSE\\
				Crossover rate&0.7&Arithmetic\\
				Mutation rate&0.2&Uniform \\
				Selection&5&Tournament\\
				Stopping criteria&100&\#Generation\\
				Pareto distance-population&0.35&Crowdy\\
				\cline{1-3}
		\end{tabular*}}
	\end{table}
	\begin{figure*}[t]
		\centering
		\includegraphics[width=1\textwidth,trim= 34cm 12cm 21cm 10cm,clip]{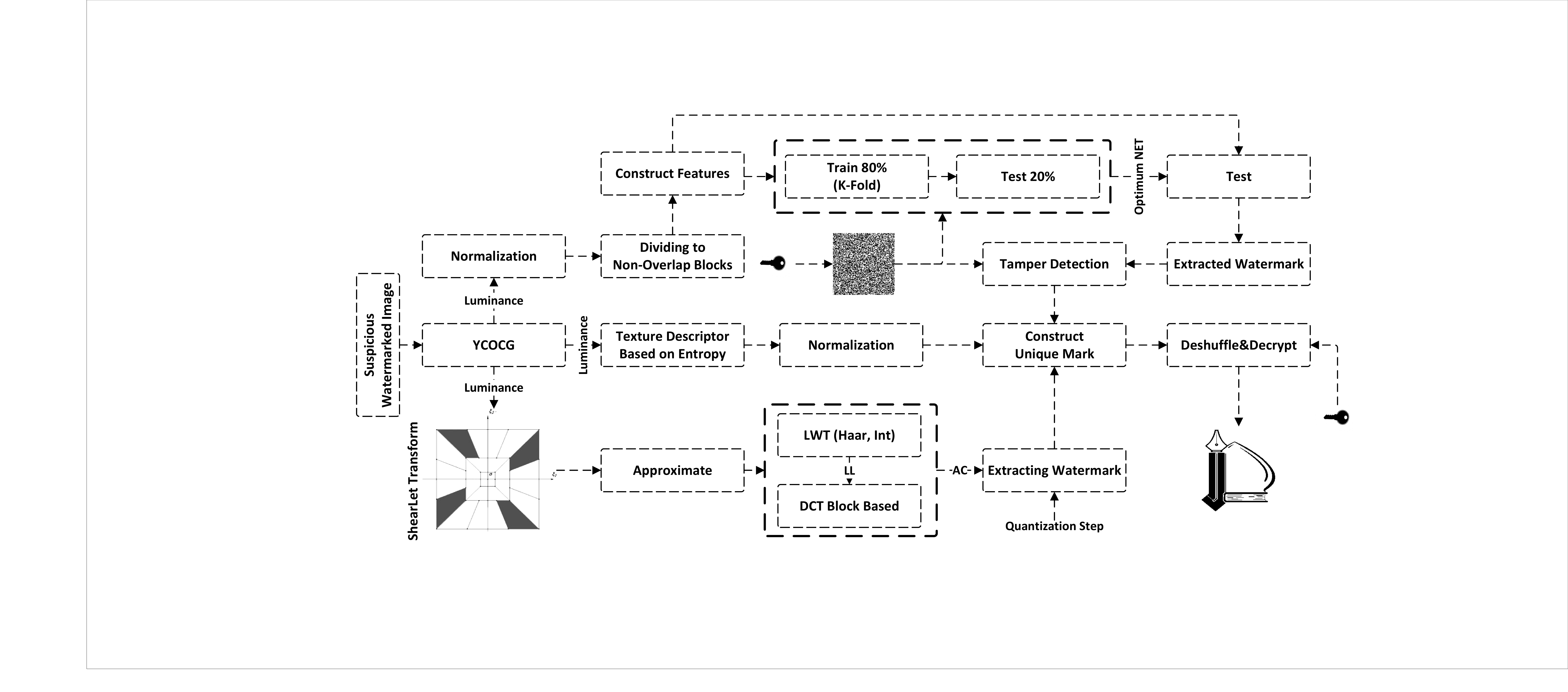}
		\caption{The block diagram of proof ownership and integrity verification phases of WSMN.}
		\label{fig:dextract}
	\end{figure*}
	\subsubsection{Thresholds Optimization}
	\label{sec:ThresholdsOptimization}
	In any robust (or semi-fragile) image watermarking scheme, a strength key plays a pivotal role and can be considered as an optimization problem. The embedding strength of the watermark determines the imperceptibility and robustness of the scheme. A high value of strength ensures better robustness but poor imperceptibility. In contrast, a low value ensures a better imperceptibility but weak robustness. Generally, the watermarking scheme should be provided a trade-off between various attacks. So, to tackle this problem, choose efficient strength is necessary for the scheme to provide a satisfactory value of both imperceptibility and robustness. In other words, a systematic mechanism is required for this purpose. Hence, the NSGA-II as a powerful and efficient optimizer is employed to intelligently select the optimum threshold steps (for both $\delta'$ and $\delta''$) that could improve the robustness with minimum degradation in the quality of the cover image, simultaneously.
	
	In this way, a common measure as Bit Error Ratio (BER) is used to calculate the number of extracted bits that have been altered for both $w_a$ and $w_c$ compared to original ones. BER value close to zero shows that both watermark are totally similar. Also, the Mean-Squared Error (MSE) is another measure which is used to compute the similarity between cover and watermarked image. Totally, the optimization process in WSMN can be formulated as Eq. \ref{eq:FF}:
	\begin{align}
	\text{min} F(x) &= \text{min} \Big(f_1(x), f_2(x), f_3(x)\Big) \nonumber \\
	f_1(x) &= \text{BER}(w_c, \tilde{w}_c) \nonumber \\
	f_2(x) &= \text{BER}(w_a, \tilde{w}_a) \nonumber \\
	f_3(x) &= \text{MSE}(X, \tilde{X}) 
	\label{eq:FF}
	\end{align}
	where $X$ and $\tilde{X}$ represent the host signal and the watermarked signal, respectively. Also, $w_a$ and $w_c$ show the original authentication and copyright marks, and $\tilde{w}_a$ and $\tilde{w}_c$ denote extracted ones. By assumption that $N$ types of attacks are applied on $\tilde{X}$, the objective function that calculated the robustness of watermarks are customized as Eq. \ref{eq:CFF}:
	\begin{equation}
	f(x) = \frac{1}{N}\sum_{i=1}^{N} \text{BER}(w, \tilde{w}_i)
	\label{eq:CFF}
	\end{equation}
	where in the experiments $N$ set by $5$ to trade-off between performance and efficiency. In other words, the watermarked image is analyzed under several attacks, and the watermarks are extracted from the destroyed watermarked images. Finally, the candidate solutions in $F(x)$ are obtained with the help of NSGA-II, and the individual with the minimum fitness of the final generation is used for watermark embedding. To do so, the optimum solution (candidate pair points) in Pareto front which satisfy priority of WSMN is selected by Eq. \ref{eq:PFF}:
	\begin{align}
	\delta^* &= S(\overline{f_1 + f_2}),\quad \text{s.t.} \quad \overline{f_1 + f_2} \leq T
	\label{eq:PFF}
	\end{align}
	where $S$ illustrates sort function which return the last solution as $\delta^*$ with highest fitness value in comparison to $T$. Otherwise speaking, the watermarking optimization should be done in such a way that the error ratio of robustness does not rise over a predefined threshold ($T = 0.1$). With the help of this strategy the strength parameters are chosen to guide toward the maximum robustness under admissible quality. The block diagram of the optimization process is shown in Fig. \ref{fig:dopt} and the parameters used are listed in Table. \ref{TABLE:GA}. 
	
	The overall optimization phase of WSMN is summarized below in details:
	\begin{enumerate}[1),itemsep=0mm]
		\item \textbf{Initialization:} In the first step, the random double initial solution population $X$ of size $N$ is generated, where each solution in $X_i$ has 2 dimensions as:
		\begin{align}
		X_i = [x'_i, x''_i]\in \rm I\!R^n, \text{where } i = 1, 2, ..., N \nonumber 
		\end{align}
		As mentioned, in WSMN, two embedding strength parameters as correlation and quantization steps should be optimized which modeled as a vector. All solutions are bounded between $X_{min}$ and $X_{max}$.
		\item \textbf{Embedding:} Now, the watermarked image is generated using the solutions in the population. For this aim, the authentication and copyright watermarks are embedded based on $x'_i$ and $x''_i$, respectively.
		\item \textbf{Attacking:} After the embedding phase, the watermarked image is subjected to planned attacks. In this way, several common attacks are applied using MATLAB such as Wiener and Sharpening filters, Histogram Equalization, Resizing, and Darkening. Thanks to the flexibility of WSMN, the other attacking scheme can easily be added or integrated with those used in the NSGA-II optimization process.
		\item \textbf{Extraction:} In this step, both copyright and authentication watermarks are extracted from the attacked images using the extraction procedure. Firstly, to improve the performance of extraction phase for extracting the authentication mark, the procedure is employed in the watermarked image without any attack. Then, the obtained model with trained weights is fine-tuned per each cases. In other words, a pre-trained network for the current watermarked image is fine-tuned to learn new specific features. Hence, the learning phase is much faster and easier than training a network from scratch.
		\item \textbf{Evaluation:} Now, the imperceptibility and robustness of both watermarks are computed using MSE and BER regarding original ones, respectively. A lower MSE and BER show that the watermarked image and the extracted marks resemble the original more closely.
		\item \textbf{GA Operations:} Next, the GA operators such as selection, crossover, and mutation are employed based on fitness values. For this aim, the tournament selection is adopted for the selection process of GA. In the end, a new generation of solutions is produced. 
		\item \textbf{NSGA-II Operators:} In this step, all individuals dominated by any other individuals are devoted to front number one. Then, individuals dominated only by the individuals in a front number one are devoted to front number two, etc. Now, the individual with the lowest front number is chosen if two individuals are from different fronts. Also, individuals with the largest crowding distance are selected if they are from the same front. In this step, the population of the next generation is produced form individuals (parent and children) based on the elitism mechanism.
		\item \textbf{Stopping Criteria:} In the following, predefined conditions including a number of generations and the objective values over the threshold are checked to decide for continuing or stopping optimization.
		\item \textbf{Optimum Solution:} Finally, after obtaining the Pareto-optimal front which consists of multiple solutions, a solution is chosen according to practical requirements for embedding watermarks. Notice that, each point in the Pareto front is a pair of optimal watermarking parameters which can effectively perform a trade-off between imperceptibly and robustness.
	\end{enumerate}
	\begin{table}[t]
		\footnotesize
		\caption{MLP hyperparameters used in the experiments.}
		\label{TABLE:MLP}
		\renewcommand{\arraystretch}{1.5}
		\scalebox{1} {
			\begin{tabular*}{\columnwidth}{@{\extracolsep{\fill}}l@{}c@{}c}
				\cline{1-3}
				Hyperparameters&Value&Description\\
				\cline{1-3}
				Layers&2&Hidden\\
				Neuron&[64, 32]&-\\
				Transfer Fcn&[tansig, poslin]&Hidden\\
				Transfer Fcn&softmax&Output\\
				Performance Fcn&MSE&-\\
				Epoch&1000&-\\
				Learning rate&0.1&-\\
				Max fail&20&Validation\\
				\cline{1-3}
		\end{tabular*}}
	\end{table}
	\subsection{Proof ownership and tamper detection} 
	\label{proof_ownership_and_tamper_detection}
	Through open communication channel, the watermarked image may accidental or malicious modified. To analyze the integrity and copyright of received image, this section describes the tamper detection and proof ownership phases of WSMN as three stages including Extracting Authentication Mark, Tamper Detection, and Extracting Copyright Mark. The block diagram of this phase is illustrated in Fig. \ref{fig:dextract}. It should be noted, as embedding phase, if $\tilde{X}$ is in grayscale mode, the chrominance component is meaningless.
	\subsubsection{Extracting Authentication Mark}
	\label{sec:Extracting_Authentication_Mark}
	In this stage, the watermark sequence is intelligently extracted from the received watermarked image based on shallow neural network. The trained MLP is used in the extraction procedure because this network is capable of memorizing the relation between the shear coefficients of the watermarked image and a corresponding pixel in the watermark image. The steps of this stage are listed below:
	\begin{enumerate}[1),itemsep=0mm]
		\item First, a binary random sequence $w_a$ with size of $\rfrac{M}{m}\times \rfrac{N}{n}$ is generated by $key_3$ as authentication mark. Also, $w_a$ plays role of labels in training and testing phase.
		\item Now, $\tilde{X}$ are divided into non-overlapping $m \times n$ blocks as feature vector $\bar{F}_{i, j}$ where $i, j$ determines the location of the block. Assuming, $\zeta$ and $\varrho$ are termed as train data and its labels. Mutually, the test data and desired output are represented by $\zeta'$ and $\varrho'$, respectively. Particularly, $\bar{F}$ and $w_a$ are partitioned to form train and test data as Eq. \ref{eq:partitioned}:
		\begin{equation}
		\bar{F} = \langle \zeta, \zeta' \rangle, w_a = \langle \varrho, \varrho' \rangle
		\label{eq:partitioned}
		\end{equation}

		\item The k-fold validation is used to gain optimum model. To do so, 80\% of data are used in training phase and the rest are selected for testing. The data are introduced to the network, and then feed-froward update the connection weights using the desired output $\varrho$. Notice that, 15\% of train data is chosen for validation. Finally, the model with minimum error is selected by Eq. \ref{eq:model}:
		\begin{align}
		\Delta_k &= \bigcup^k_{i=1} f(\zeta_k, \varrho_k), k = 5 \nonumber \\
		\Delta^* &= \min\Bigg(E\Big(\bigcup^k_{i=1}\Delta_k(\zeta'_k), \varrho'_k\Big)\Bigg)
		\label{eq:model}
		\end{align}
		where $\Delta^*$ shows the optimum model. Also, $f$ and $E$ are training and error estimation function, respectively. The hyperparameters of network are listed in Table \ref{TABLE:MLP}.
		\item Lastly, the authentication mark $\tilde{w_a}$ is extracted with the help of optimum model by Eq. \ref{eq:exauth}:
		\begin{equation}
		\tilde{w_a} = \Delta^*(\bar{F})
		\label{eq:exauth}
		\end{equation}
	\end{enumerate}
	
	\begin{figure}[t]
		\center
		\begin{tabular}{cc}
			\includegraphics[width=0.45\columnwidth]{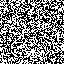} &
			\includegraphics[width=0.45\columnwidth]{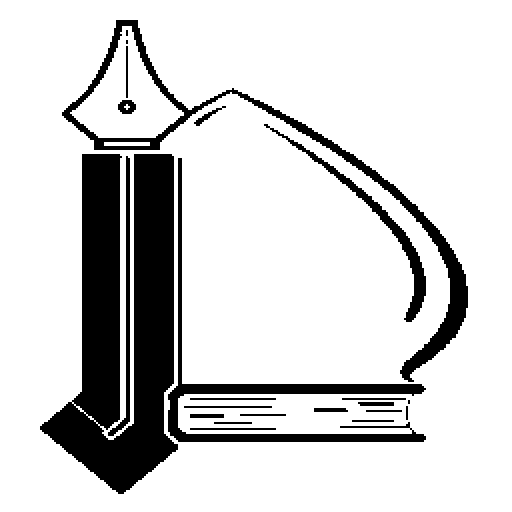} \\
			(a) & (b)
		\end{tabular}
		\caption{(a) Authentication (64$\times$64) and (b) Copyright (32$\times$32)  marks.}
		\label{fig:marks}
	\end{figure}
	\subsubsection{Tamper Detection}
	In verification and authentication stage, the tampered regions are marked based on extracted authentication mark. The details of final stage are elucidated as below:
	\begin{enumerate}[1),itemsep=0mm]
		\item Firstly, the bit-wise exclusive-or operation is applied between original and extracted mark by Eq. \ref{eq:tamper}:
		\begin{equation}
		\nu = \tilde{w}_a \oplus w_a 
		\label{eq:tamper}
		\end{equation}
		If $\nu_{i,j } = 1$, it means that the block at location $(i, j)$ is tampered; otherwise, it represents accurate block.
		\item The isolated pixels in $\nu$ which their length of connections is less than three and do not have candidate forged neighbors are eliminated. 
		\item Next, the closing morphology operator is employed on $\nu$ as post-processing to fill the gaps between tampered blocks that has been mistakenly marked as valid. To do so, a $5\times5$ square is used as a structure element.
	\end{enumerate}
	\begin{figure*}[t]
		\center
		\renewcommand{\arraystretch}{0}
		\begin{tabular}{@{\extracolsep{\fill}}c@{}c@{}c@{}c@{}c}
			\includegraphics[width=0.17\textwidth]{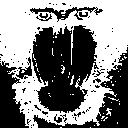} &
			\includegraphics[width=0.17\textwidth]{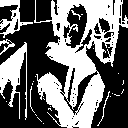} &
			\includegraphics[width=0.17\textwidth]{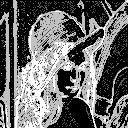} &
			\includegraphics[width=0.17\textwidth]{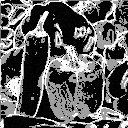} &
			\includegraphics[width=0.17\textwidth]{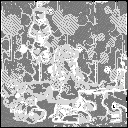} \\		
			\includegraphics[width=0.17\textwidth]{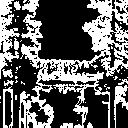} &
			\includegraphics[width=0.17\textwidth]{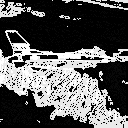} &
			\includegraphics[width=0.17\textwidth]{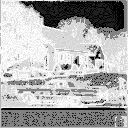} &
			\includegraphics[width=0.17\textwidth]{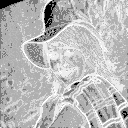} &
			\includegraphics[width=0.17\textwidth]{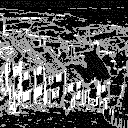} \\					
			\includegraphics[width=0.17\textwidth]{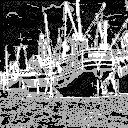} &
			\includegraphics[width=0.17\textwidth]{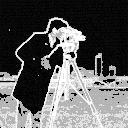} &
			\includegraphics[width=0.17\textwidth]{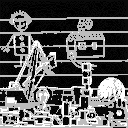} &
			\includegraphics[width=0.17\textwidth]{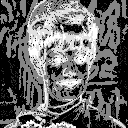} &
			\includegraphics[width=0.17\textwidth]{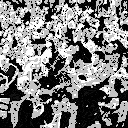} \\				
		\end{tabular}
		\caption{The results of texture analysis phase for fifteen standard color and gray images such as Baboon, Barbara, Lena, Pepper, Gril, Lake, F16, House, Elaine, Goldhill, Boat, Camera, Toys, Zelda, and Crowd.}	
		\label{fig:textureres}
	\end{figure*}
	\subsubsection{Extracting Copyright Mark}
	The extracting copyright mark is the inverse of the watermark embedding procedure which descried in previous section. In other words, the whole binary watermark is extracted by applying the previous steps to each block that holds a watermark bit. The details of this stage are explained below:
	\begin{enumerate}[1),itemsep=0mm]
		\item First, $\tilde{X}$ is decomposed based on Shearlet and LWT transforms by Eq. \ref{eq:ShearLwT}:
		\begin{align}
		\Psi_A &= \text{DST}(X) \nonumber \\
		LL_a &= \text{LWT}(\Psi_A, Haar) 
		\label{eq:ShearLwT}
		\end{align}
		where $\Psi_A$ and $LL_a$ represent approximate of Shearlet and LWT transform, respectively.
		\item In this step, $LL_a$ sub-band is divided into $\rfrac{m}{2}\times\rfrac{n}{2}$ non-overlapping blocks. Next, DCT block based is employed on coefficients by Eq. \ref{eq:DCT2}:
		\begin{align}
		\varphi_a(i, j) &= \text{DCT}\Big(LL_a(i, j)\Big) \nonumber \\
		&\forall i \in [1, \rfrac{M}{m}], j \in [1, \rfrac{N}{n}]
		\label{eq:DCT2}
		\end{align}
		\item Now, the copyright bits are fetched using dequantization techniques by Eq. \ref{eq:quanitze2}:
		\begin{align}
		w_c(i,j)&=\Big\lfloor \frac{2\varphi_a(ac_{i, j})}{\delta'} \Big\rceil \mod 2 \nonumber \\
		&\forall i \in [1, \rfrac{M}{m}], j \in [1, \rfrac{N}{n}]
		\label{eq:quanitze2}
		\end{align}
		\item In the following, due to the fake regions, the unique copyright logo is generated based on the four embedded chances in the signal. For this aim, first, the probability of block destruction is calculated by Eq. \ref{eq:prob}:
		\begin{equation}
		P(i, j) = 1 - S(i, j), \forall i \in [1, \rfrac{M}{m}], j \in [1, \rfrac{N}{n}]
		\label{eq:prob}
		\end{equation}
		where $S$ represents the entropy of each block which normalized from $[0, log_2(256)]$ to $[0, 1]$. Next, the valid mark are gained by Eq. \ref{eq:probcopy}:
		\begin{align}
		\tilde{w}_c''(i', j') &= \sum_{{\substack{
					i, j\\
					w_c=1}}}P(i, j) \geq \sum_{{{\substack{
						i, j\\
						w_c=0}}}}P(i, j)\nonumber \\
		&\forall i' \in [1, \rfrac{M}{2m}], j' \in [1, \rfrac{N}{2n}]
		\label{eq:probcopy}
		\end{align}
		where $P(i, j)$ represents the corresponding elements which belong to unique position in the original mark. In other words, each bit in $\tilde{w}_{c}''$ is constructed by looking at four corresponding bits in $w_c$.
		\item According to embedding phase, CCS algorithm is employed to place the copyright bits into initial position. Similarly, a sequences $R$ is generated based on CCS by $key_2$, and the permutation position $p$ is achieved by sorting $R$ in ascending order. Then, the unshuffle mark are obtained by utilizing Eq. \ref{eq:shuffled2}:
		\begin{align}
		\tilde{w}_{c}'(i) = \tilde{w}_{c}''(p(i))^{-1},\quad \forall i \in [1, \rfrac{\ell}{4}]
		\label{eq:shuffled2}
		\end{align}
		Notice that, in the unshuffling process input and output are converted into 1D and 2D matrix, respectively.
		\item Finally, a binary copyright logo as $\tilde{w}_c$ is obtained by decrypting $\tilde{w}_{c}$ based on $\chi \sim N(0,1)$ which is binary random sequence generated by $key_1$ according Eq. \ref{eq:decrypt}:
		\begin{align}
		\tilde{w}_{c} = \tilde{w}'_c \oplus \chi
		\label{eq:decrypt}
		\end{align}
		where $\oplus$ represents bit-wise exclusive-or operation.
	\end{enumerate}
	\begin{figure}[t]
		\center
		\begin{tabular}{c}
			\includegraphics[width=0.48\textwidth]{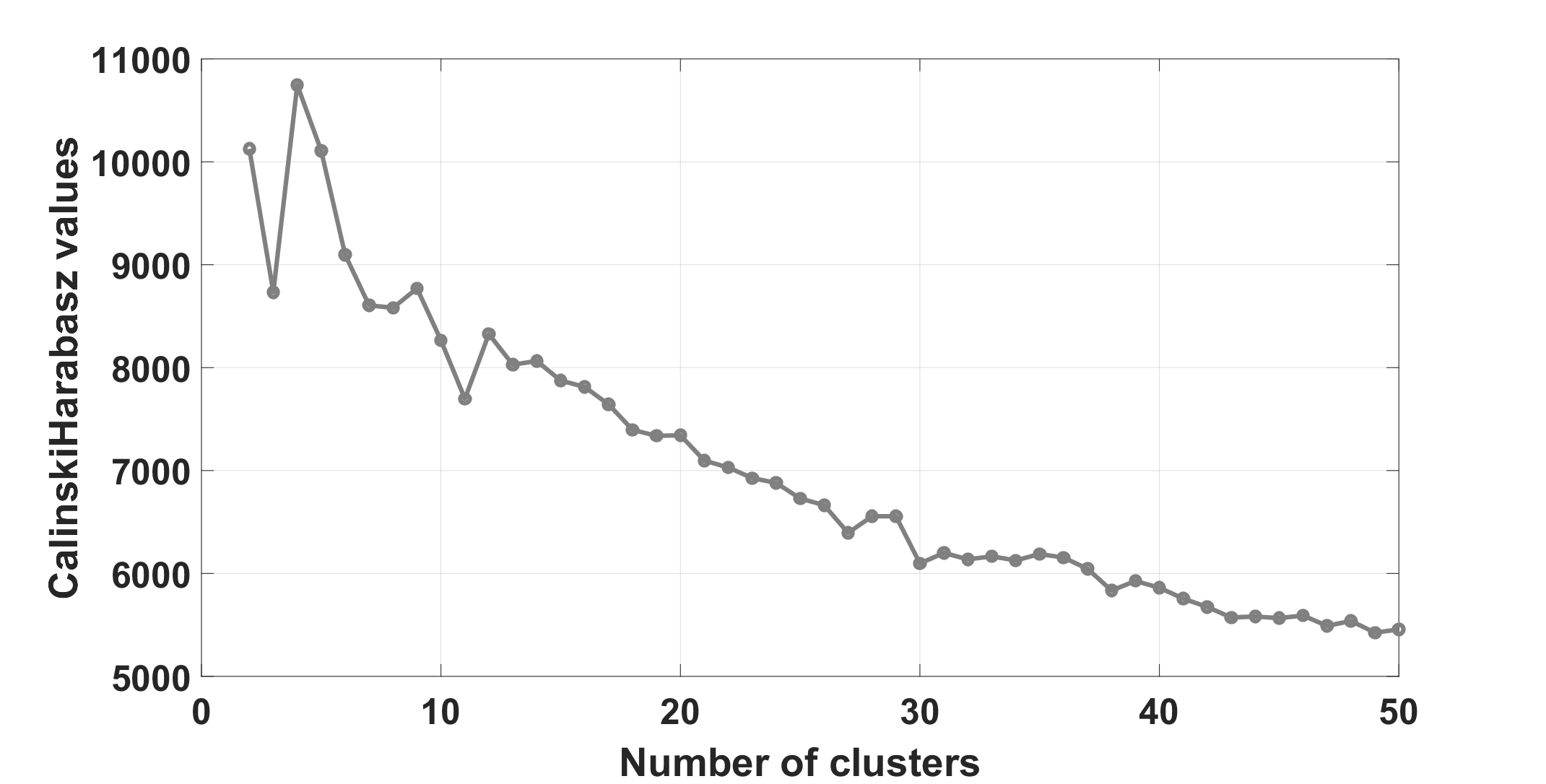} \\(a)\\
			\includegraphics[width=0.48\textwidth]{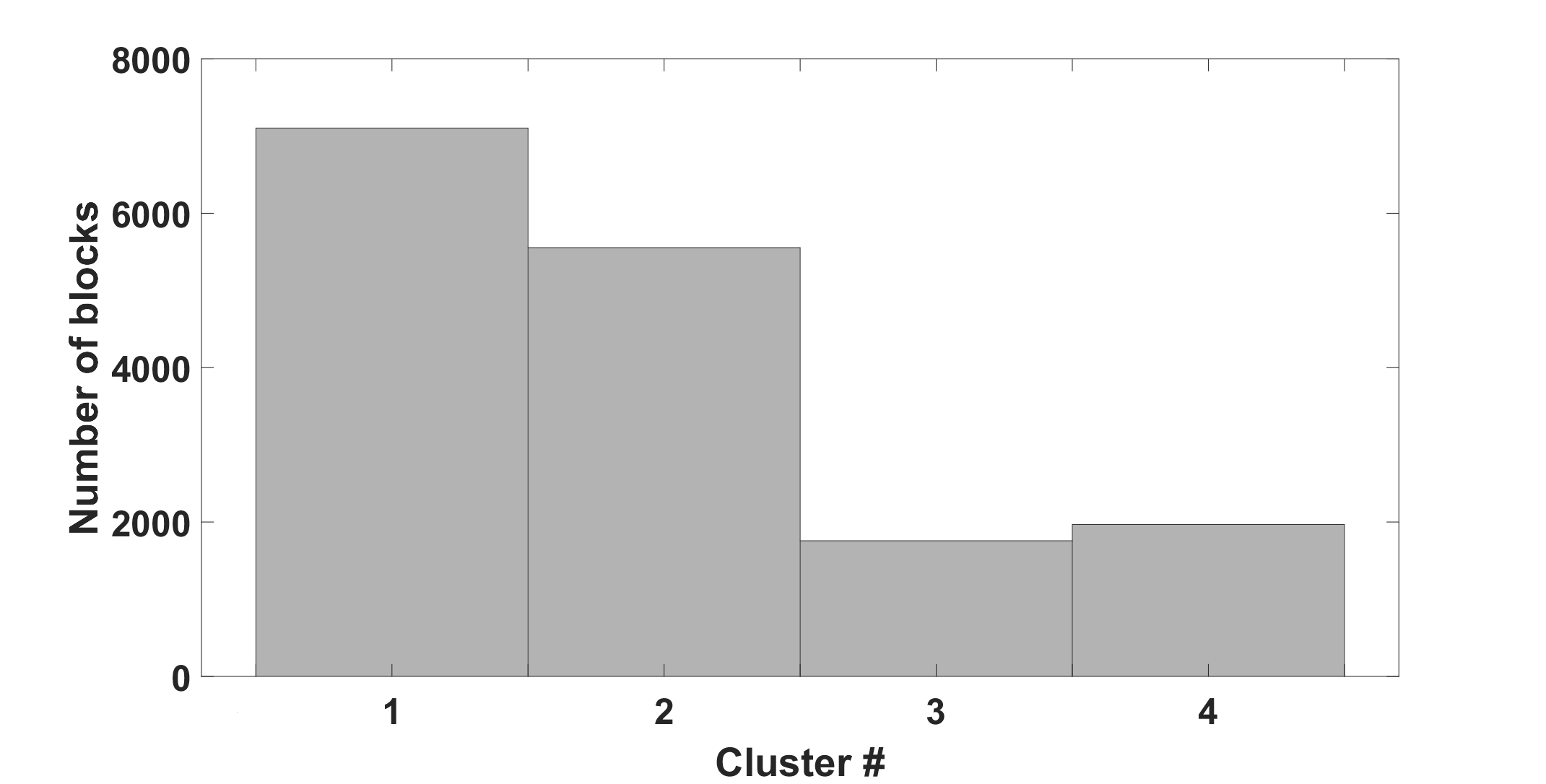}\\(b)
		\end{tabular}
		\caption{(a) Calinski-Harabasz value for each cluster, (b) Number of blocks in each clusters (Test image Lena).}
		\label{fig:texturelena}
	\end{figure}
	\section{Experimental Results}
	\label{sec:Experimental}
	In this section, first, the employed dataset and experimental settings are introduced in detail. After that, a series of experiments are reported to prove the superiority and efficiency of WSMN compared to prior state-of-the-art schemes.
	
	\subsection{Dataset and implementation details}
	In this work, fifteen non-compressed standard images, which consist of eight color and seven grayscale images of size 512$\times$512 are used to test the performance of the proposed scheme. The different types of textures, such as edge, smooth, and rough regions in these images lead to challenging the watermarking system in terms of imperceptibly and robustness. Also, the utilized authentication and copyright marks are illustrated in Fig. \ref{fig:marks}. The size of the authentication watermark is 64$\times$64; This means, WSMN is able to detect forged region with the accuracy of 8$\times$8 blocks. Moreover, WSMN inserts four copies of the copyright watermark with the size of $32\times32$ bits inside a $512\times512$ candidate host. On the other hand, all experiments were implemented on a computer with a 3.20 GHz Intel i7 processor, 24.00 GB memory with Windows 10 operating system. The programming environment was MATLAB R2019b; Also, the whole attacks are simulated with the help of its Image Processing toolbox. To make a forged image, the watermarked image is modified by Adobe Photoshop CC 2018. 
	\subsection{Texture analysis performance}
	Generally, there are various types of region in natural images, including rough, flat, smooth, etc. Due to this fact, considering a constant threshold is wasteful in the embedding phase of a watermarking system. In particular, the characteristics of a block play a crucial role in determining the strength threshold. Evidently, considering high value for the smooth region is unnecessary, and content can be watermarked by partial modifications for maintaining the quality of blocks. On the other hand, to guaranty the robustness of the rough blocks, the high threshold is required.
	
	In this study, to cope with such problems, a k-Means algorithm and texture descriptors were employed to distinguish the types of image blocks. The results of this strategy on the dataset are demonstrated in Fig. \ref{fig:textureres}. As can be seen, the variety of blocks in each image leads its content categorized into different parts. The number of optimum clusters is effectively determined by the Calinski-Harabasz criterion. Fig. \ref{fig:texturelena} shows the results of this criterion for choosing the number of clusters for the Lena image. Accordingly, the optimum $k$ is chosen by four in this case, which represents the number of texture variety. Totally, with the help of this analysis, which trade-off between imperceptibility and robustness, the performance of the system dramatically improved.
	
	\begin{figure}[b!]
		\center
		\begin{tabular}{c}
			\includegraphics[width=0.40\textwidth]{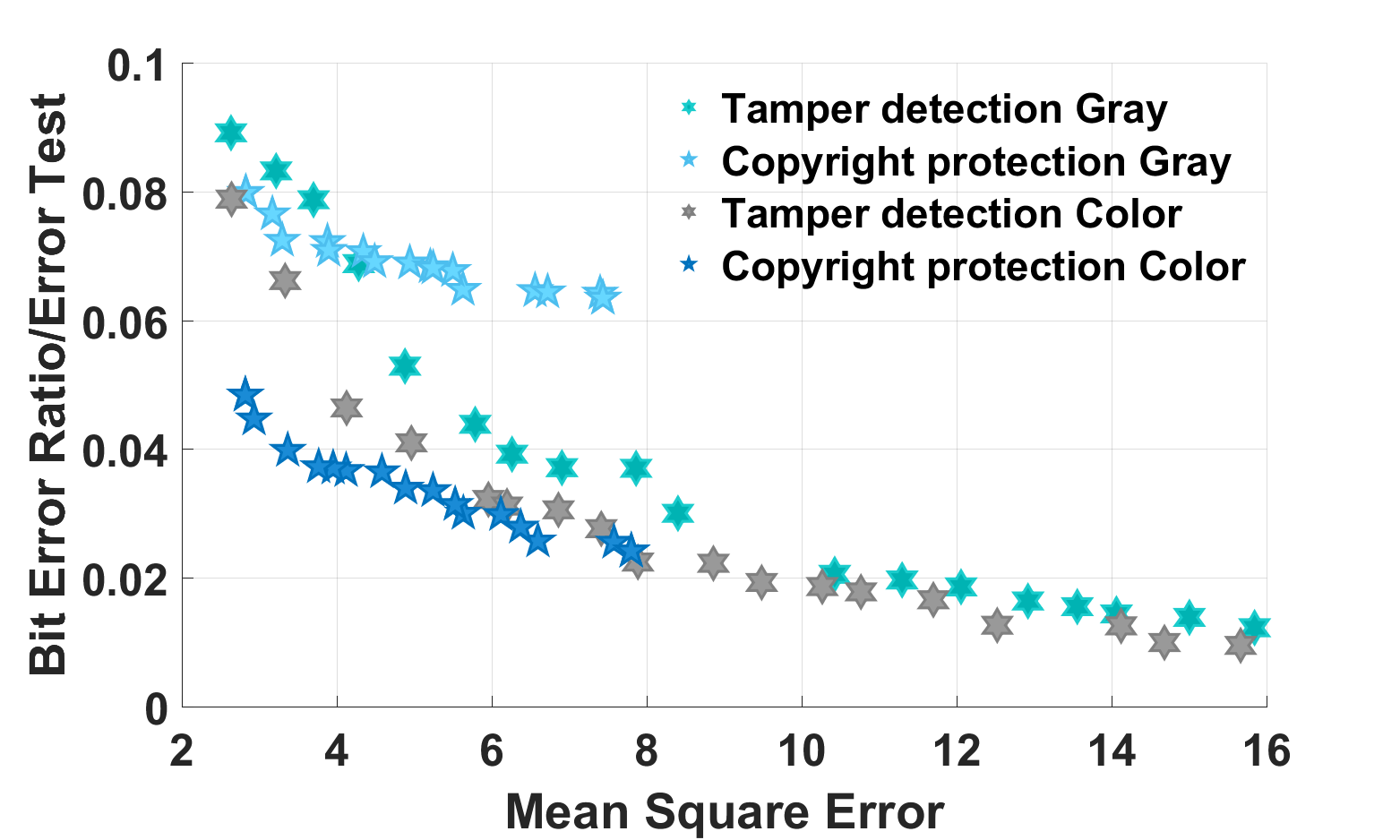}\\(a)\\
			\includegraphics[width=0.40\textwidth]{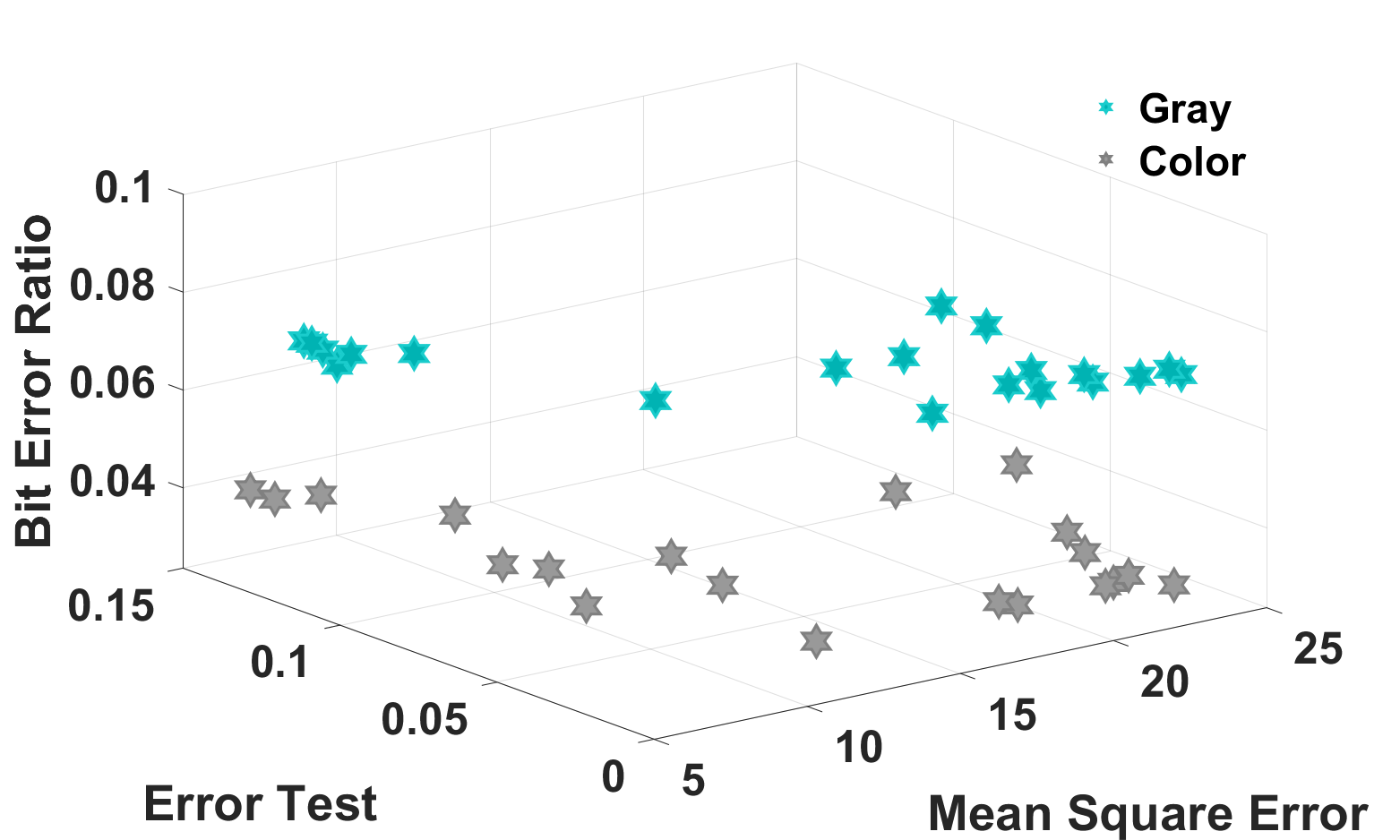} \\(b)
		\end{tabular}
		\caption{The Pareto fronts for (a) single (two objective) and (b) dual (three objective) applications.}
		\label{fig:optlena}
	\end{figure}
	\begin{figure*}[t!]
		\center
		\begin{tabular}{cc}
			\includegraphics[width=0.46\textwidth]{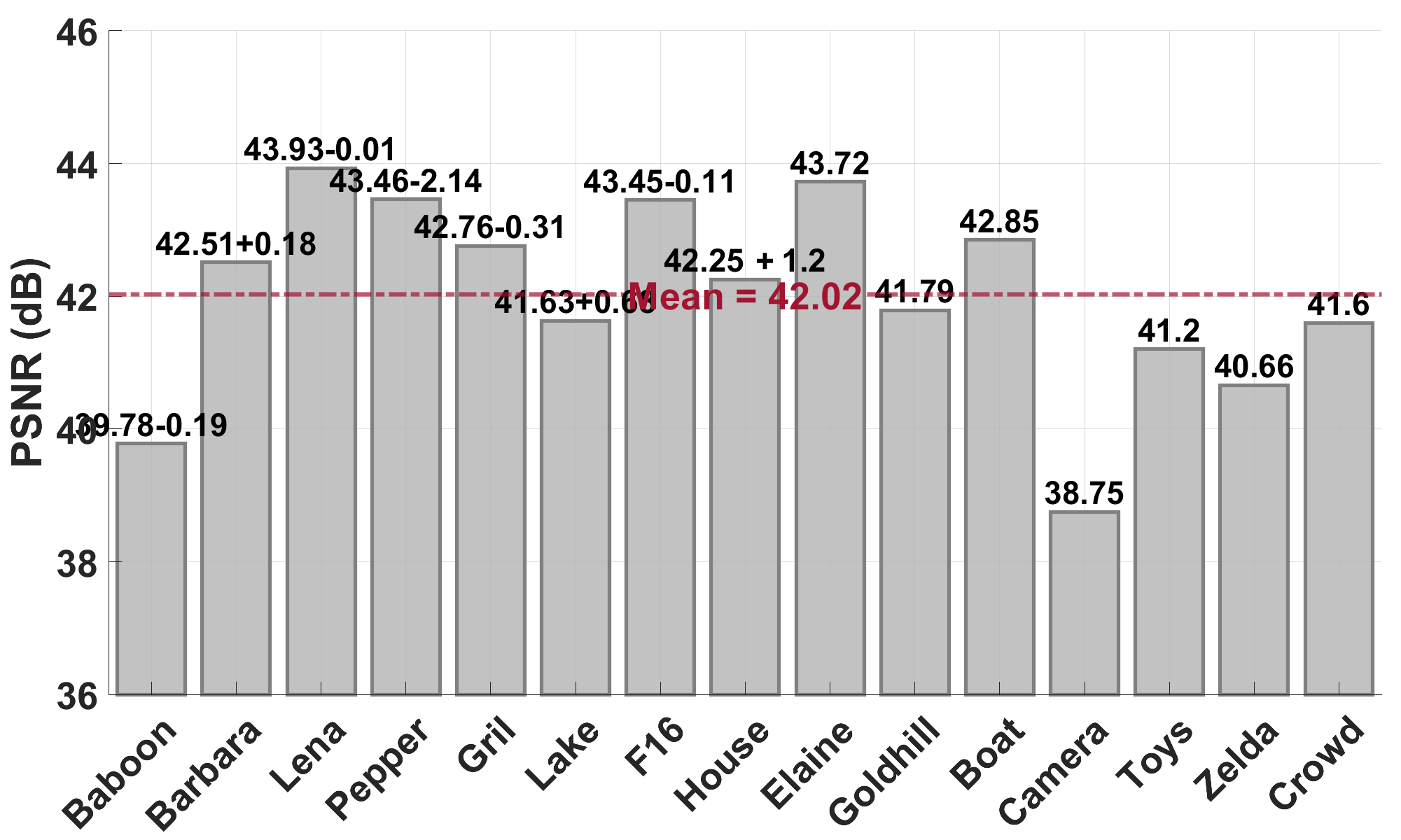} &
			\includegraphics[width=0.46\textwidth]{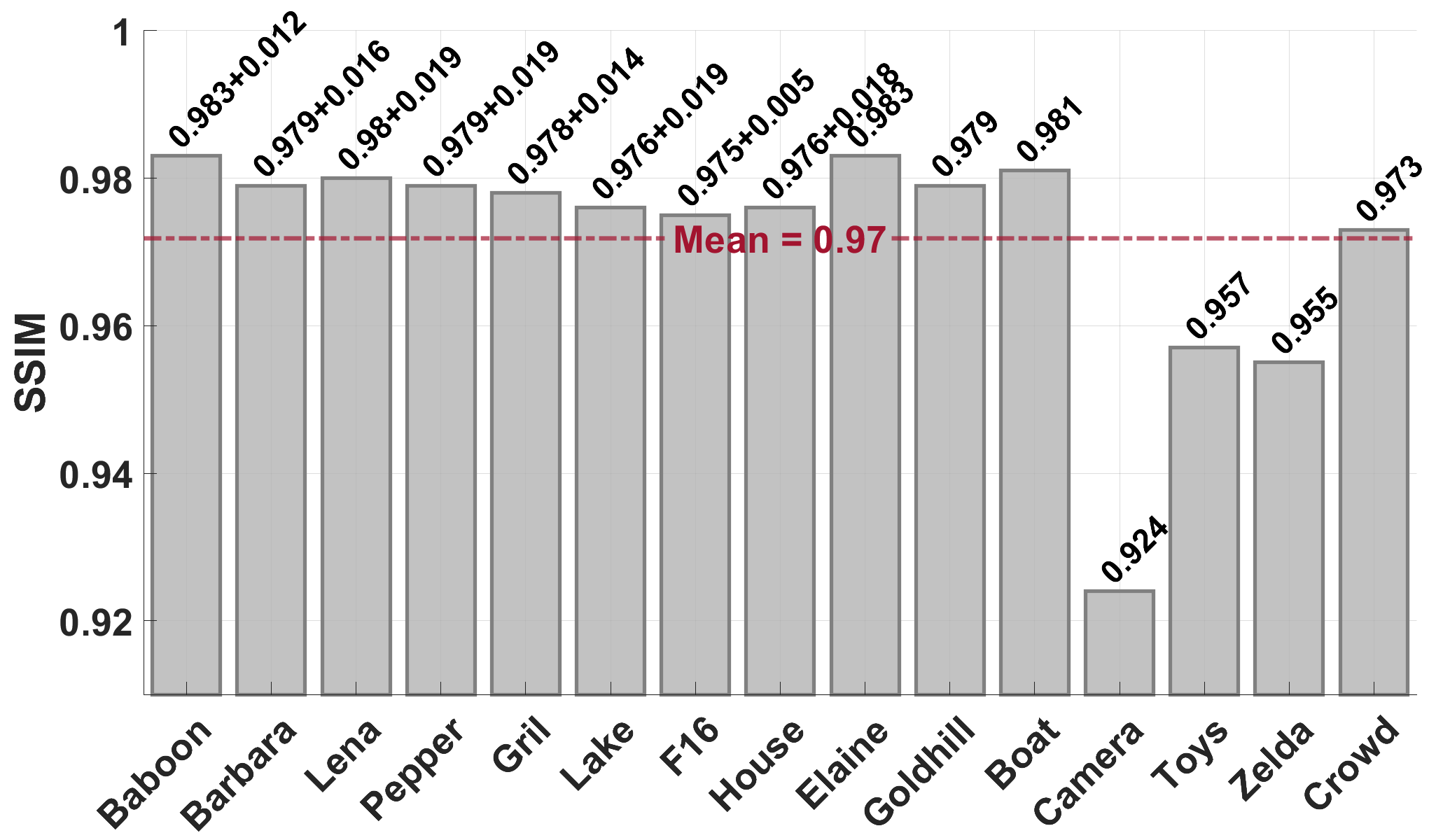} \\
			\multicolumn{2}{c}{(a)}\\
			\includegraphics[width=0.46\textwidth]{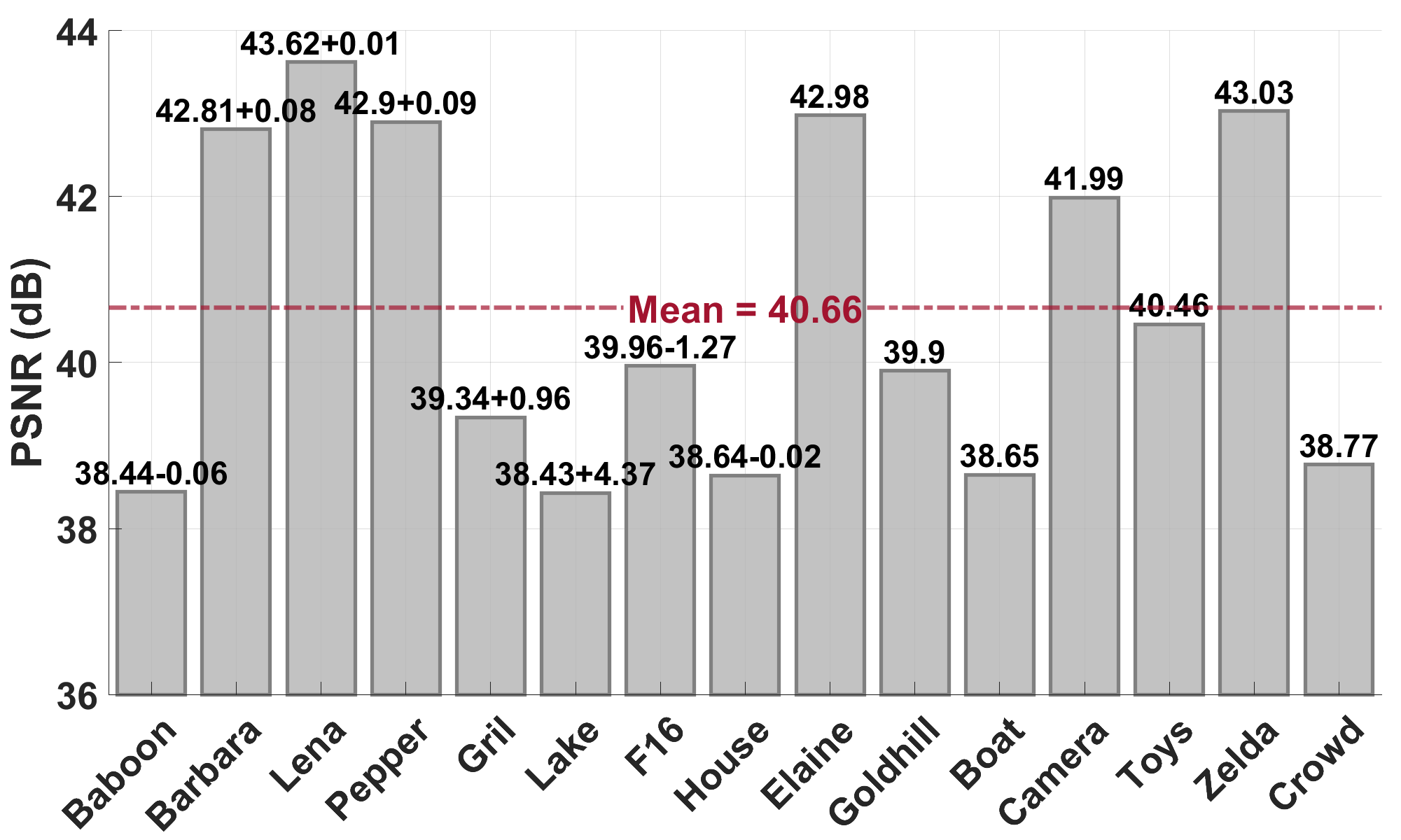} &
			\includegraphics[width=0.46\textwidth]{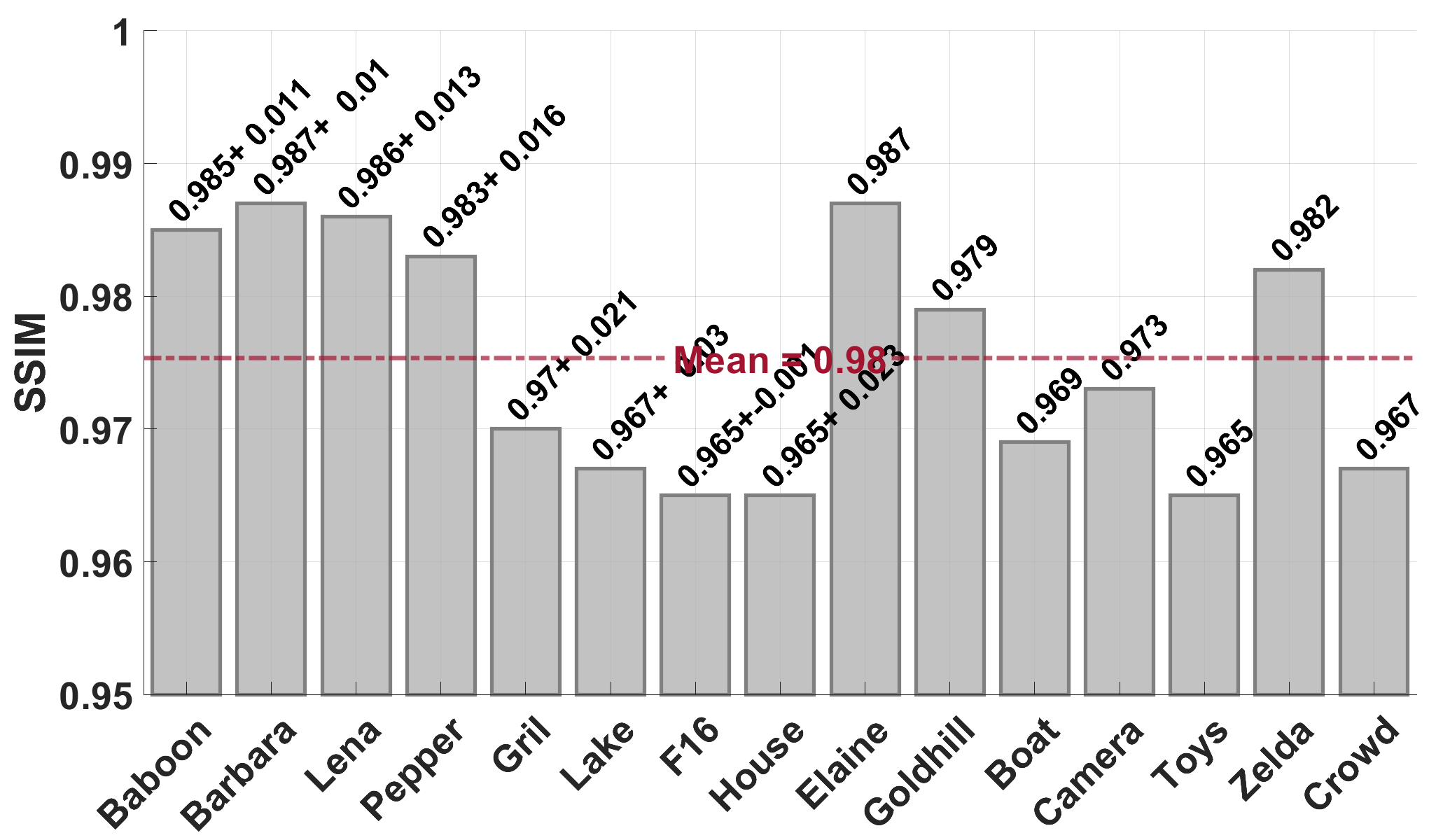} \\
			\multicolumn{2}{c}{(b)}\\
			\includegraphics[width=0.46\textwidth]{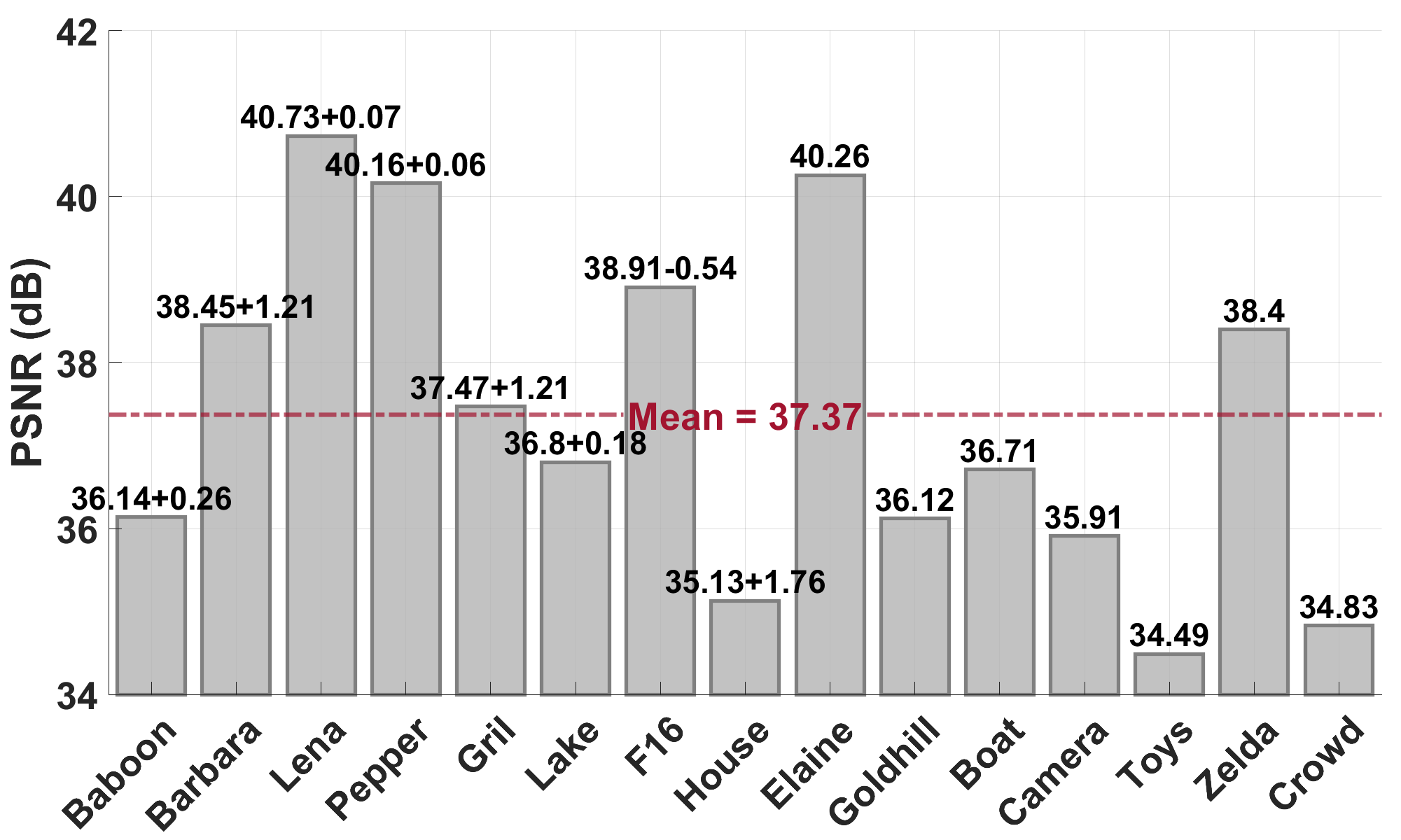} &
			\includegraphics[width=0.46\textwidth]{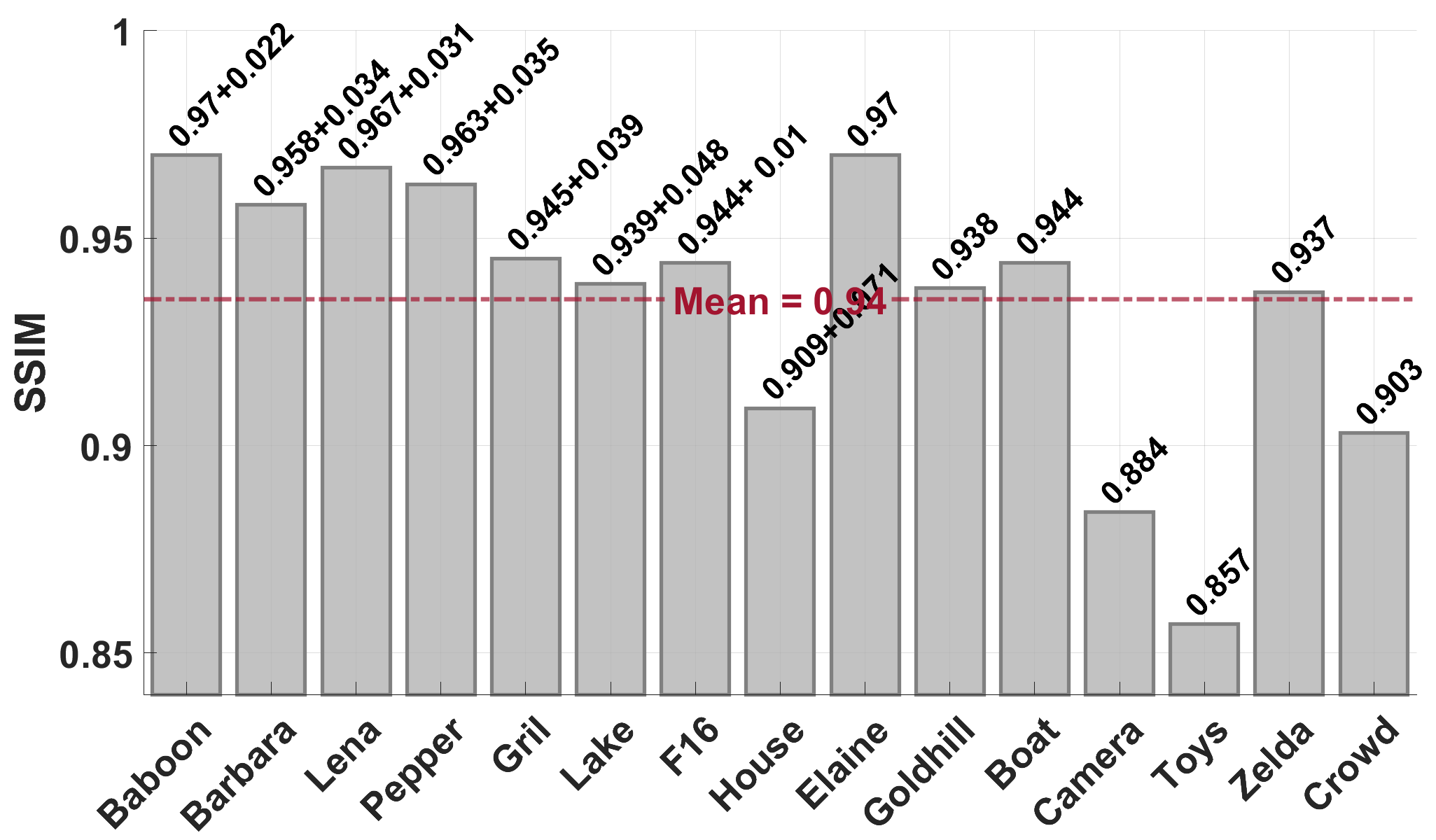} \\
			\multicolumn{2}{c}{(c)}\\
		\end{tabular}
		\caption{PSNR and SSIM values of watermarked images over fifteen standard color and grayscale images. (a) Content authentication protection (Integrity protection), (b) Proof ownership(Copyright protection), and (c) Dual application. Note, the second term in the top of the first eight bars shows the difference between grayscale and color mode.}		
		\label{fig:CURVPSNRSSIM}
	\end{figure*}
	\begin{figure*}[t!]
		\center
		\renewcommand{\arraystretch}{0}
		\begin{tabular}{@{\extracolsep{\fill}}c@{}c@{}c@{}c}
			\includegraphics[width=0.21\textwidth]{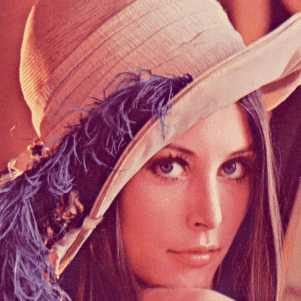}&
			\includegraphics[width=0.21\textwidth]{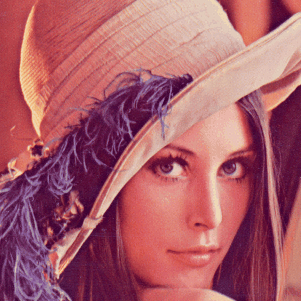}&
			\includegraphics[width=0.21\textwidth]{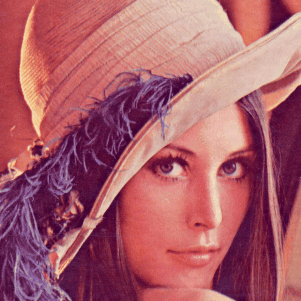}&
			\includegraphics[width=0.21\textwidth]{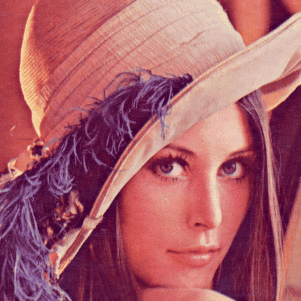} \\
			\includegraphics[width=0.21\textwidth]{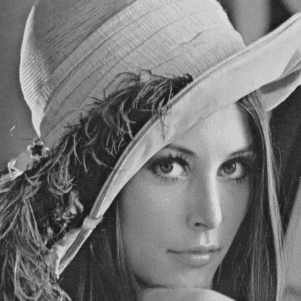}&
			\includegraphics[width=0.21\textwidth]{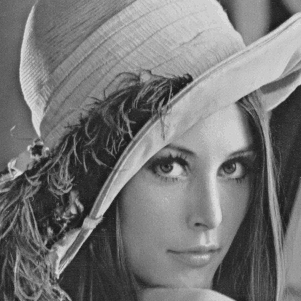}&
			\includegraphics[width=0.21\textwidth]{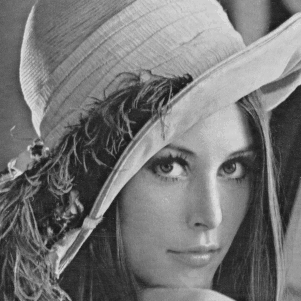}&
			\includegraphics[width=0.21\textwidth]{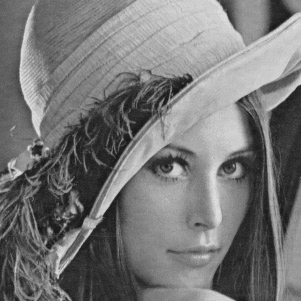} \\
		\end{tabular}
		\caption{Zoomed watermarked image. Left to right, Cover image and Content Authentication, Copyright Protection, and Dual applications cases.}
		\label{fig:lenazoom}
	\end{figure*}
	\subsection{Evolutionary algorithm performance}
	As mentioned earlier, the strength parameter plays a pivotal role in the watermarking design. In other words, it becomes a challenging problem to balance the imperceptibly and robustness of the various image due to the conflict behavior with each other. Nevertheless, an intelligent strategy should be utilized to explore the optimum values. To do so, a meta-heuristic optimization algorithm as NSGA-II was employed in WSMN to search for embedding parameters inside a search space that contains all feasible solutions. Thanks to the multi-objective property of this algorithm, the optimum strengths can be reached for both single and dual applications of WSMN. In this way, BER of extracted copyright mark, the error test of the learning algorithm in the extracting authentication phase, and MSE of the watermarked image were performed as fitness functions. In optimization stage, to simulate the robustness of WSMN against attacks, five image processing operations including Sharpen('Radius',1,'Amount',4), Wiener Filter(4$\times$4), Resize(-50\%), Darken(-50), and HistEq were employed. 
	
	Fig. \ref{fig:optlena} illustrates the Pareto fronts of both single and dual modes for color and grayscale Lena images. As discussed, the candidate solution (or Pair solutions) as an optimal point was selected under a predefined constraint to provide requirements. Hence, the performance of WSMN is effectively improved with such strategies that suggest the best compromise between imperceptibility and robustness.
	\subsection{Quality and imperceptibly analysis}
	The imperceptibility property is one of the primary purposes of any watermarking algorithm. It refers to the sufficiency of the system by which the cover and watermarked images are perceptually indistinguishable. In particular, the Human Visual System (HVS) should not discriminate between the original and modified version of the image. Generally, the quality of the signal will certainly reduce by any type of modification. Hence, the embedding mark should be carried out in such a way that the perceptual quality of the image does not remarkably degrade, which be perceptible to HVS.
	
	In this subsection, the efficiency of WSMN in terms of perceptual quality is proved by evaluating objective assessments as Signal to Noise Ratio(PSNR) and Structural Similarity Index(SSIM) criterion \cite{ref28, ref29}. These measures show the similarity between a host and watermarked images. In this way, Figs. \ref{fig:CURVPSNRSSIM} (a) and (b) indicate the quality of the watermarked image for both authentication and copyright protection goals, respectively. As can be seen, the results essentially express a good degree of imperceptibility of WSMN. By looking at details, due to the working domain, including low and high frequencies coefficients (details and approximate bands), the average copyright application is slightly lower than the authentication case in terms of PSNR. Subsequently, Fig. \ref{fig:CURVPSNRSSIM} (c) reports the quality of watermarked images in the dual case which carry both marks. As known, dual watermarking, especially in the frequency domain, reduces the imperceptibility of the watermarked image. To cope with such problems, WSMN can considerably handle this issue with the help of performed processing before the embedding phase. Although PSNR insignificant fall compared to single cases, SSIM values illustrate the superiority of WSMN. Totally, PSNR measure is not consistent enough with HVS and does not consider the local content and structure of the signal. On contrast, SSIM is sensitive to evaluate three features, such as brightness, contrast, and structure, which is more suitable for defining the distortion limit. Overall, from the histograms in Fig. \ref{fig:CURVPSNRSSIM}, the average PSNR and SSIM of authentication and copyright protection goals over the fifteen candidate images reach to (42.02 dB, 0.97) and (40.66 dB, 0.98), respectively. Likewise, the measures gain to the acceptable values equal to (37.37 dB, 0.94) for dual application. Notice that these results are derived with respect to the appropriate robustness under various types of attacks based on the best optimal value of strength in the optimization phase, simultaneously. Accordingly, WSMN achieves qualified imperceptibility and permissive robustness against attacks, which will be discussed in the next subsection. Moreover, to further explore the imperceptibility, the zoom of Lena's face is demonstrated for all three applications in Fig. \ref{fig:lenazoom}. As can be seen, the candidate part contains rough and flat texture regions. It is quite evident, in the whole case, the watermarked image is nearly identical to the cover image.
	
	\begin{figure*}[t!]
		\begin{tabular}{cc}
			\includegraphics[width=0.45\textwidth]{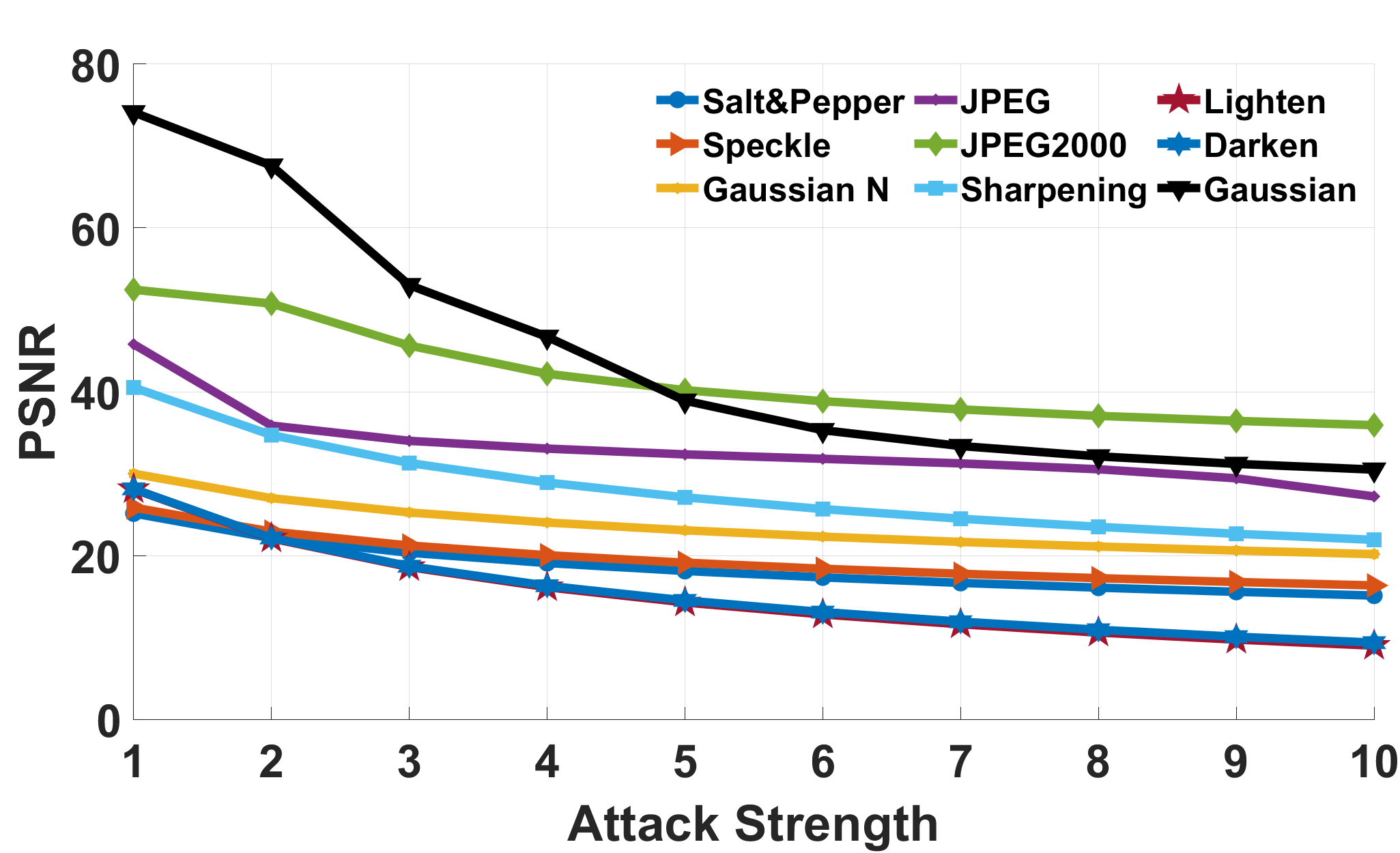} &
			\includegraphics[width=0.45\textwidth]{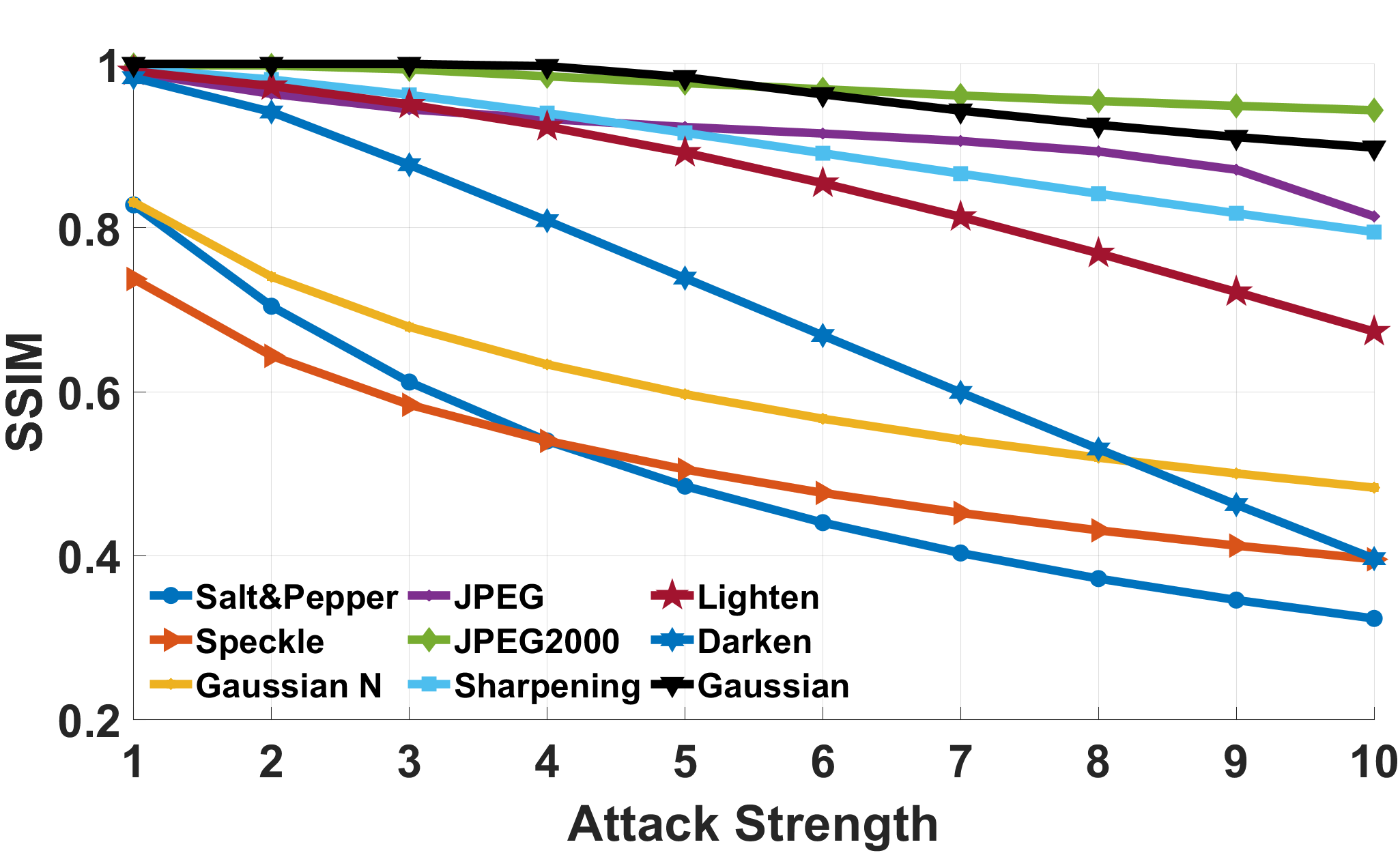} \\
			(a)&(b)
		\end{tabular}
		\begin{tabular}{ccc}
			\includegraphics[width=0.30\textwidth]{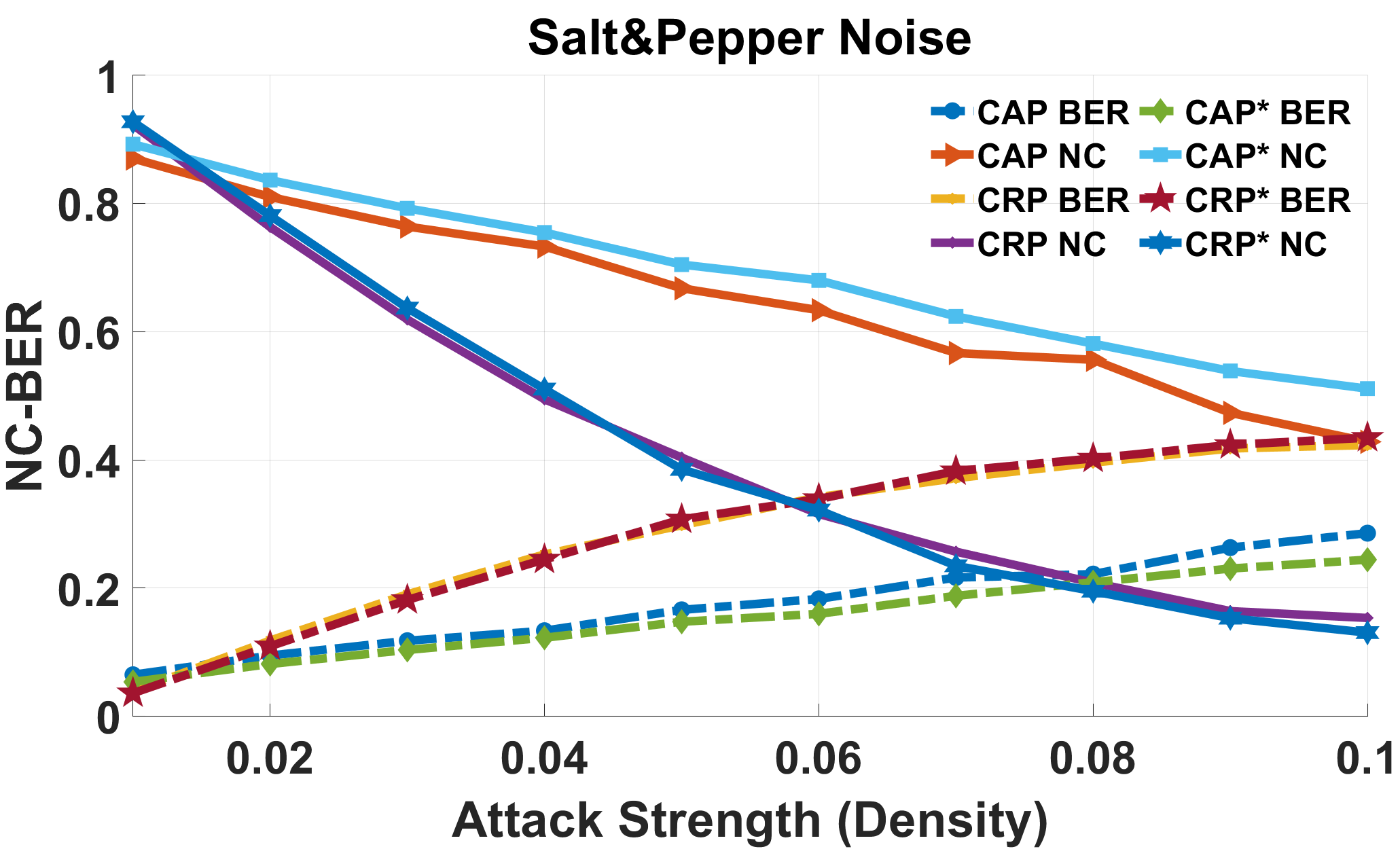} &
			\includegraphics[width=0.30\textwidth]{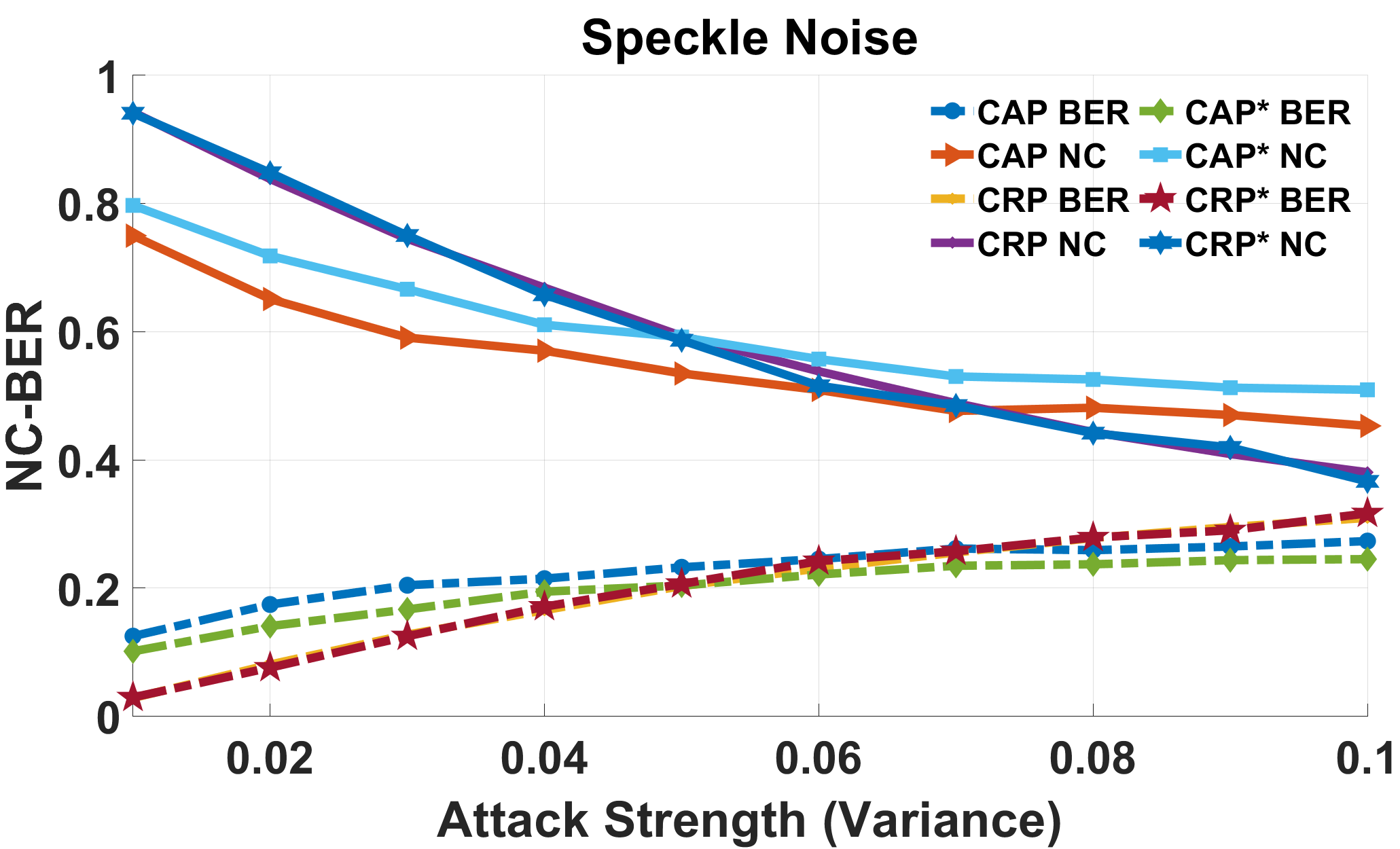} &
			\includegraphics[width=0.30\textwidth]{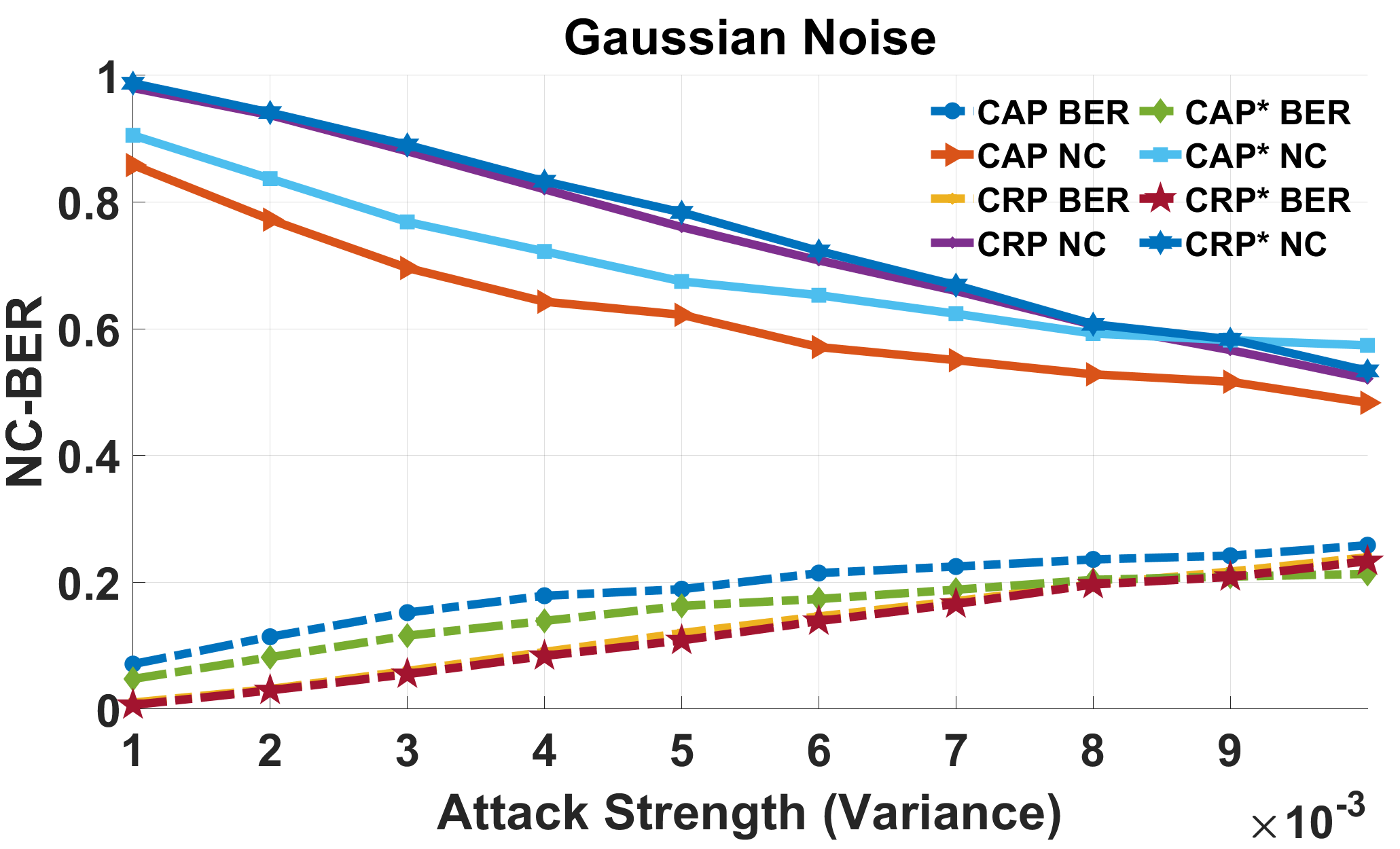} \\
			(c)&(d)&(e)\\
			\includegraphics[width=0.30\textwidth]{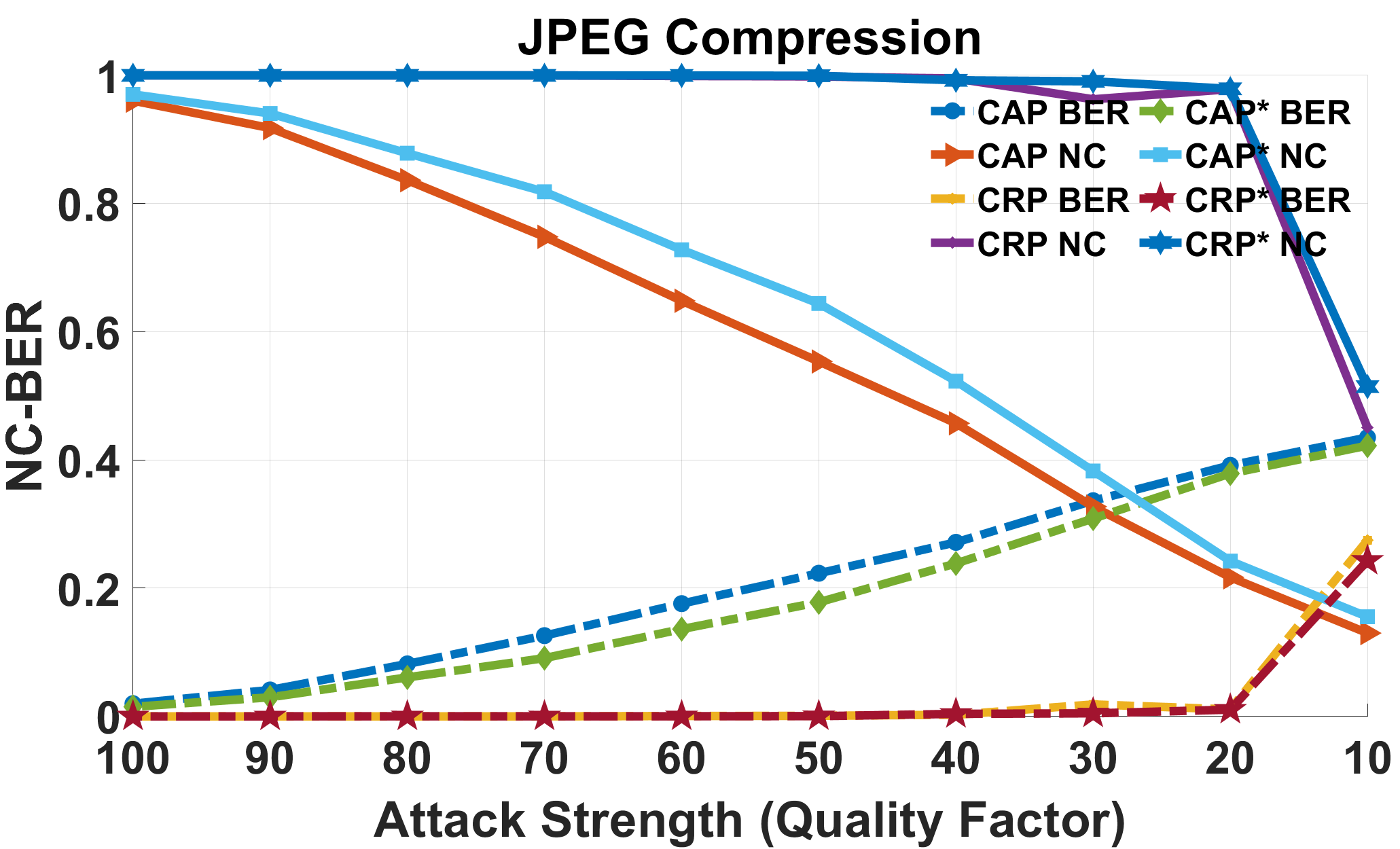} &
			\includegraphics[width=0.30\textwidth]{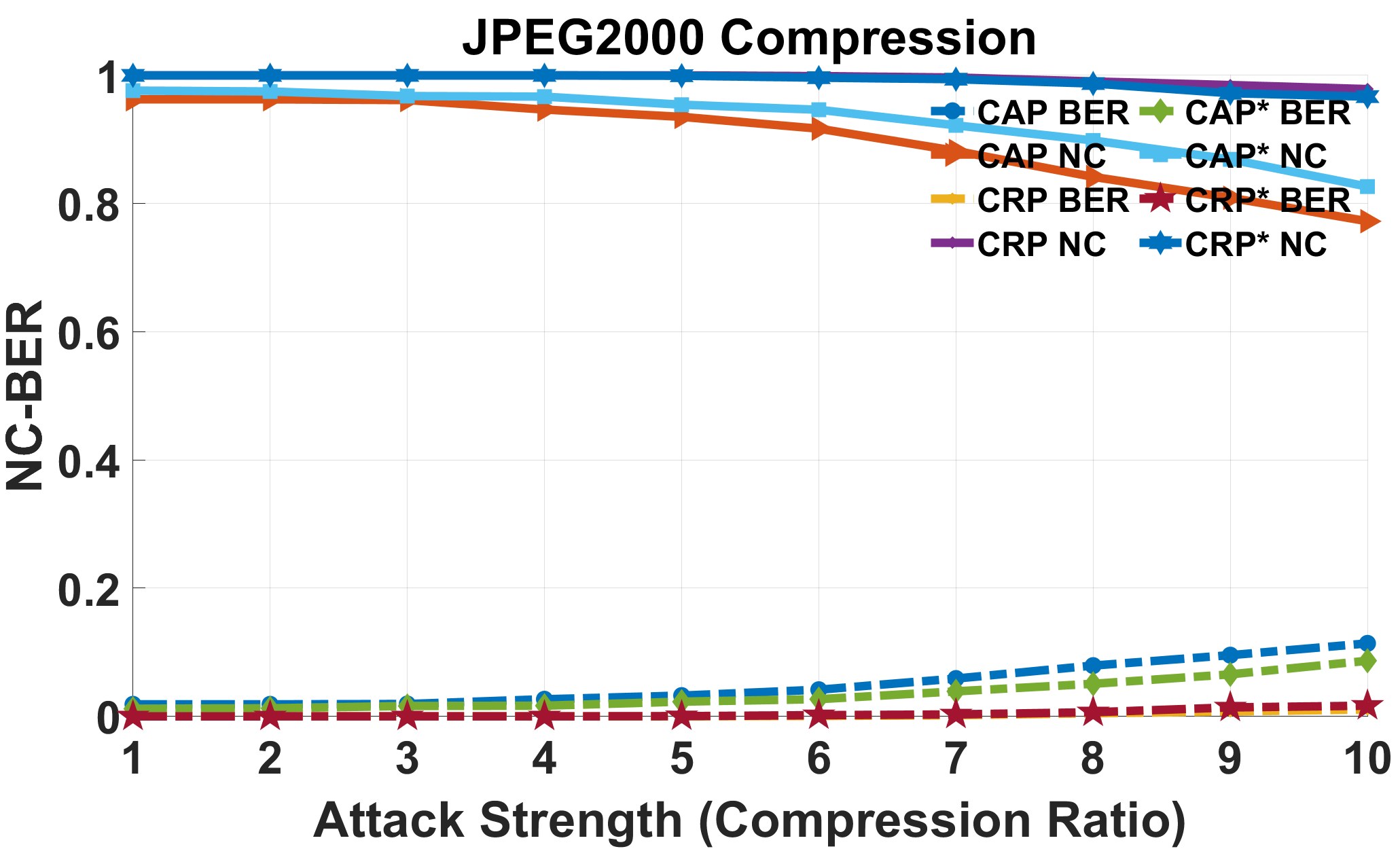} &
			\includegraphics[width=0.30\textwidth]{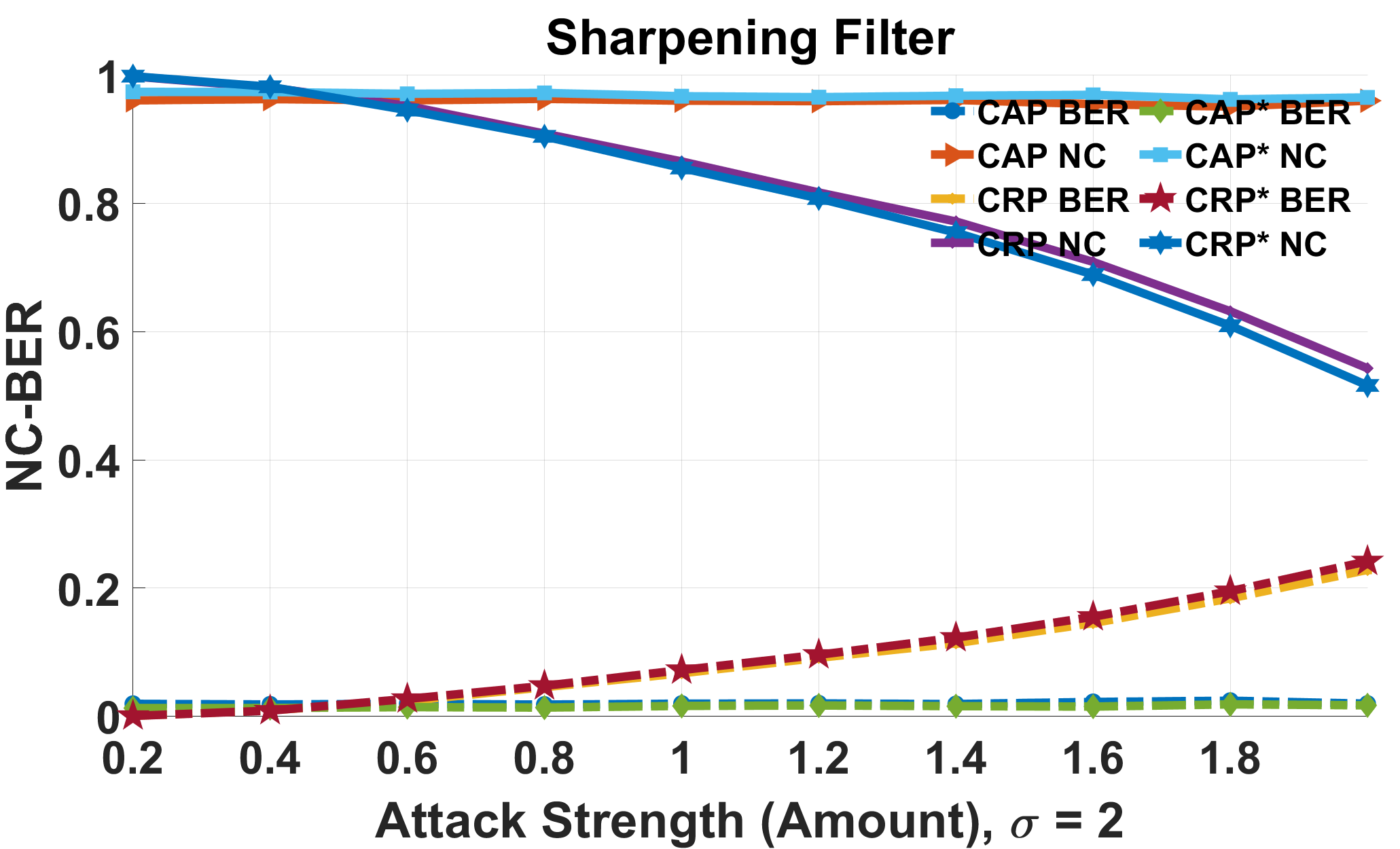} \\
			(f)&(g)&(h)\\
			\includegraphics[width=0.30\textwidth]{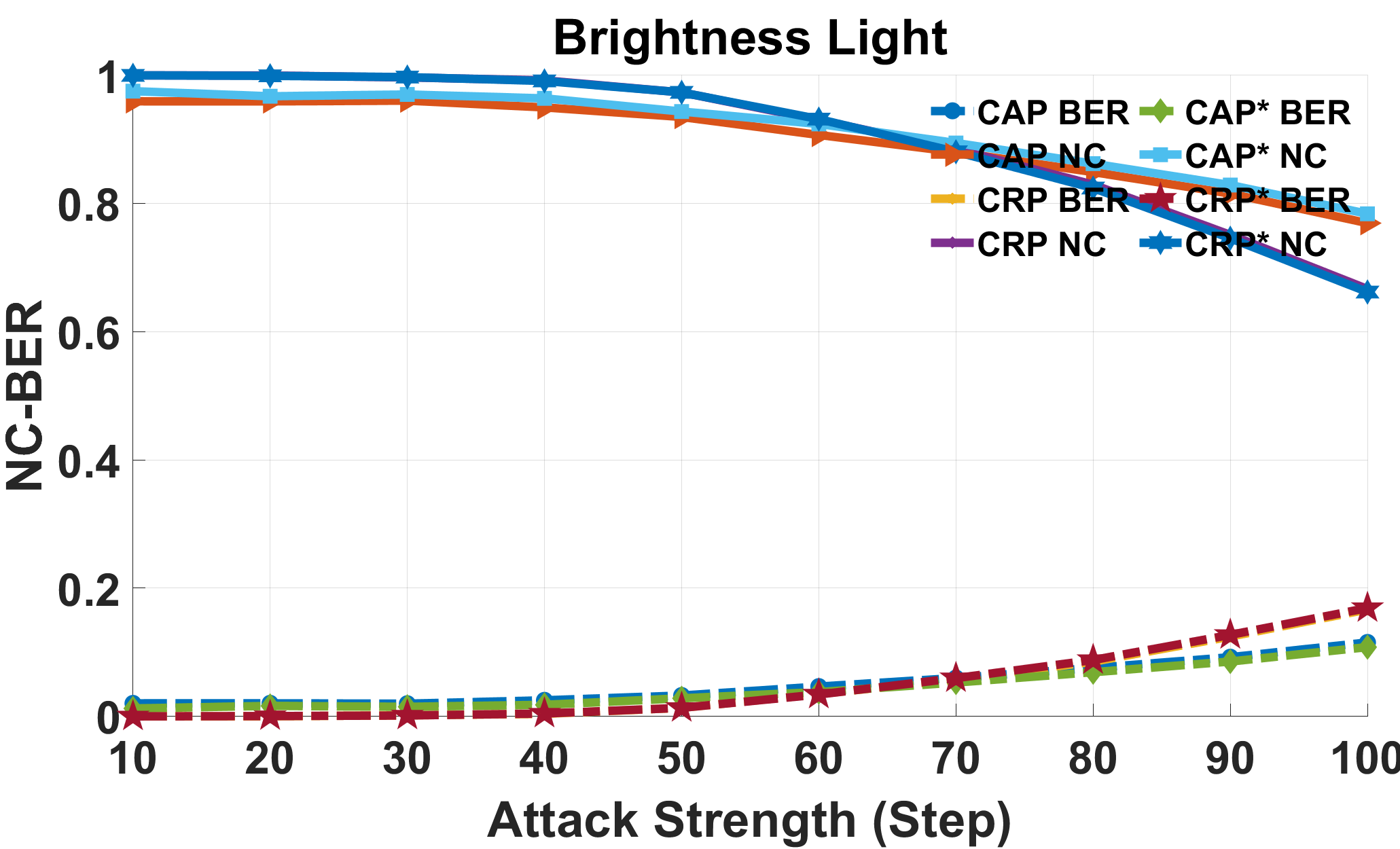} &
			\includegraphics[width=0.30\textwidth]{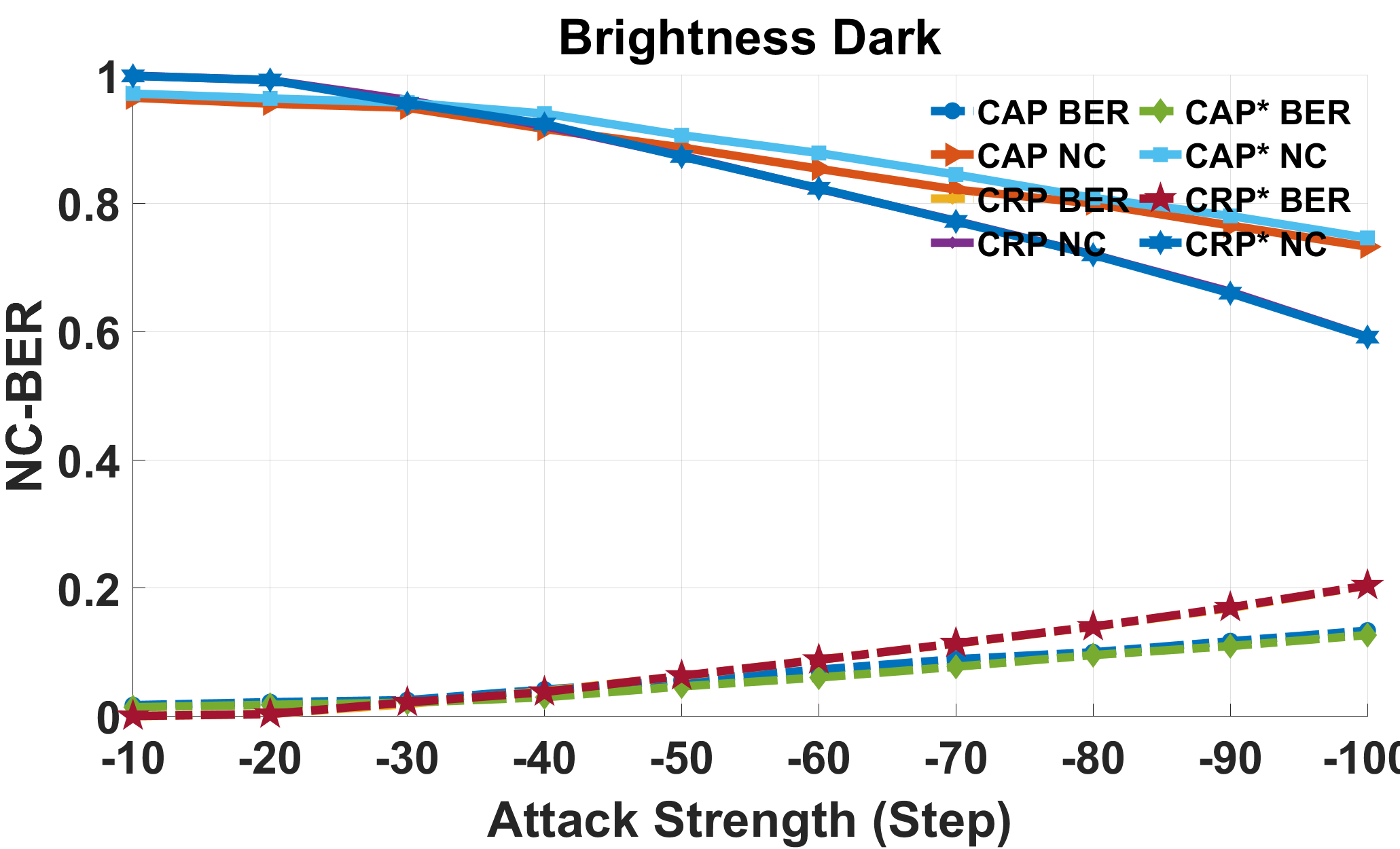} &
			\includegraphics[width=0.30\textwidth]{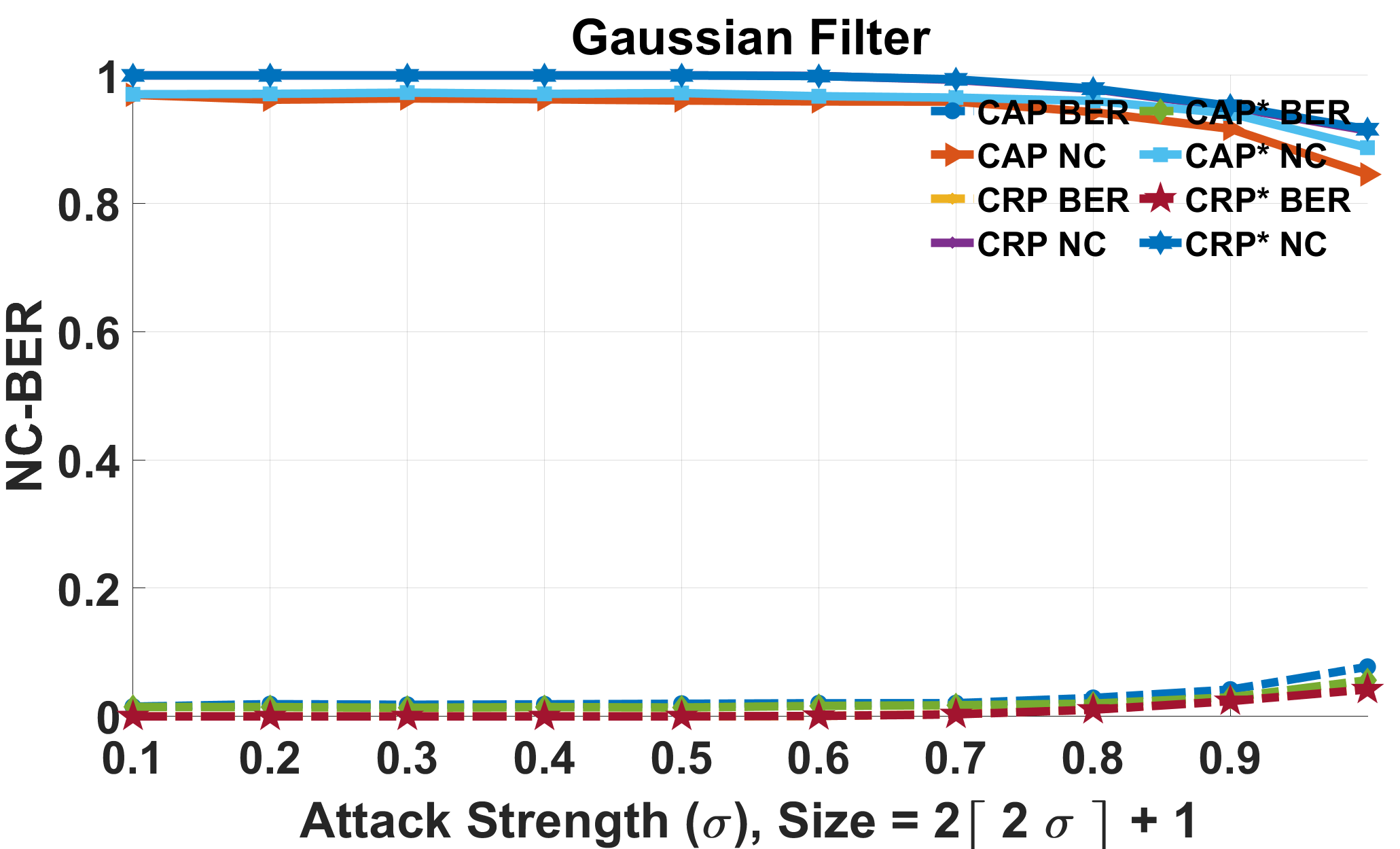} \\
			(i)&(j)&(k)			
		\end{tabular}
		\caption{(a-b) The average quality of watermarked images under attacks, (c-k) the average accuracy of extracted marks under different types of attacks for fifteen candidate images. Note, the symbol $(*)$ represents the accuracy of the extraction mark in dual applications. Also, CAP and CRP are abbreviated of Content Authentication Protection and Copyright Protection, respectively.} 
		\label{fig:robust1}
	\end{figure*}
	\begin{figure*}[t!]
		\begin{tabular}{cc}
			\includegraphics[width=0.45\textwidth]{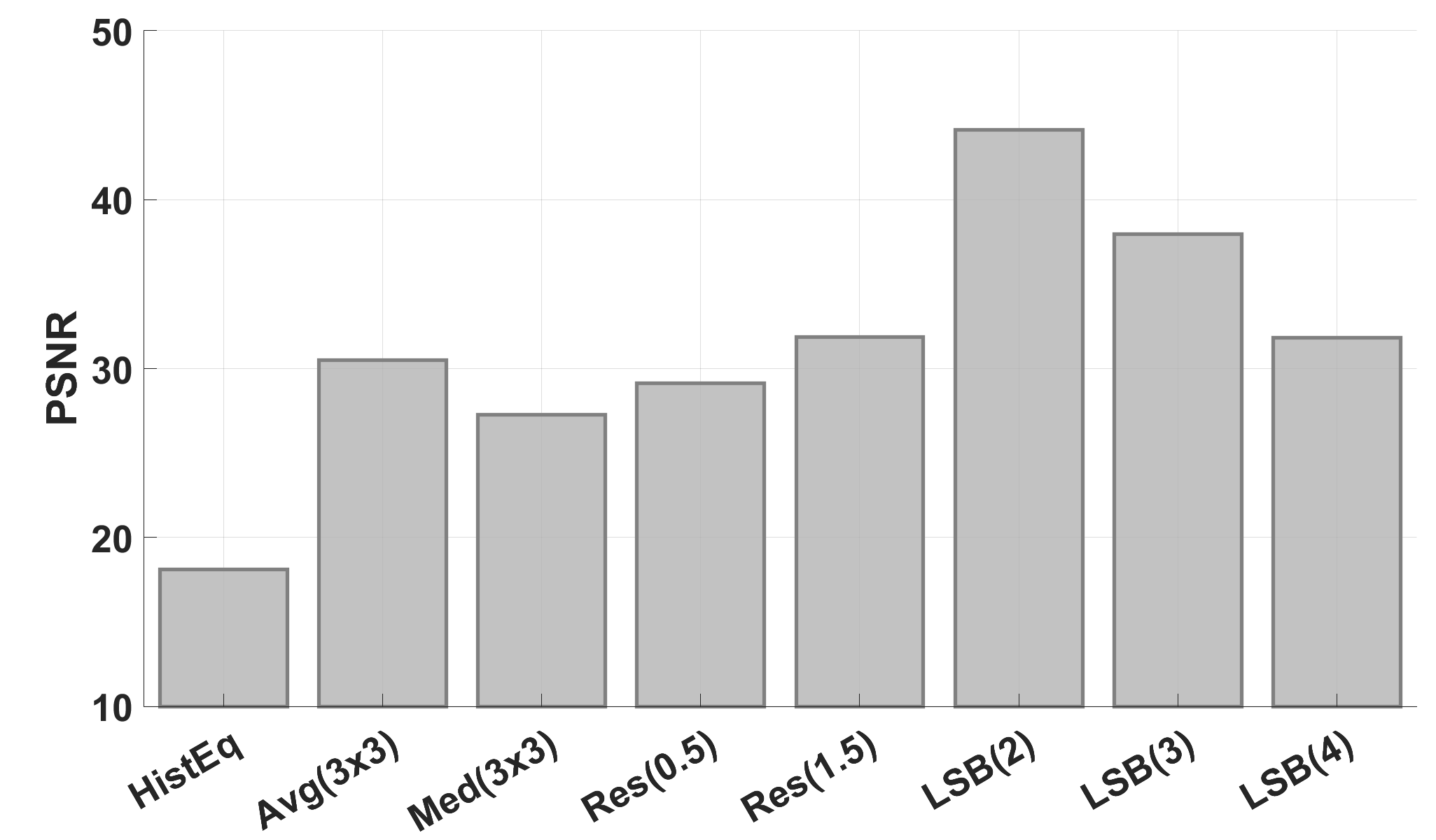} &
			\includegraphics[width=0.45\textwidth]{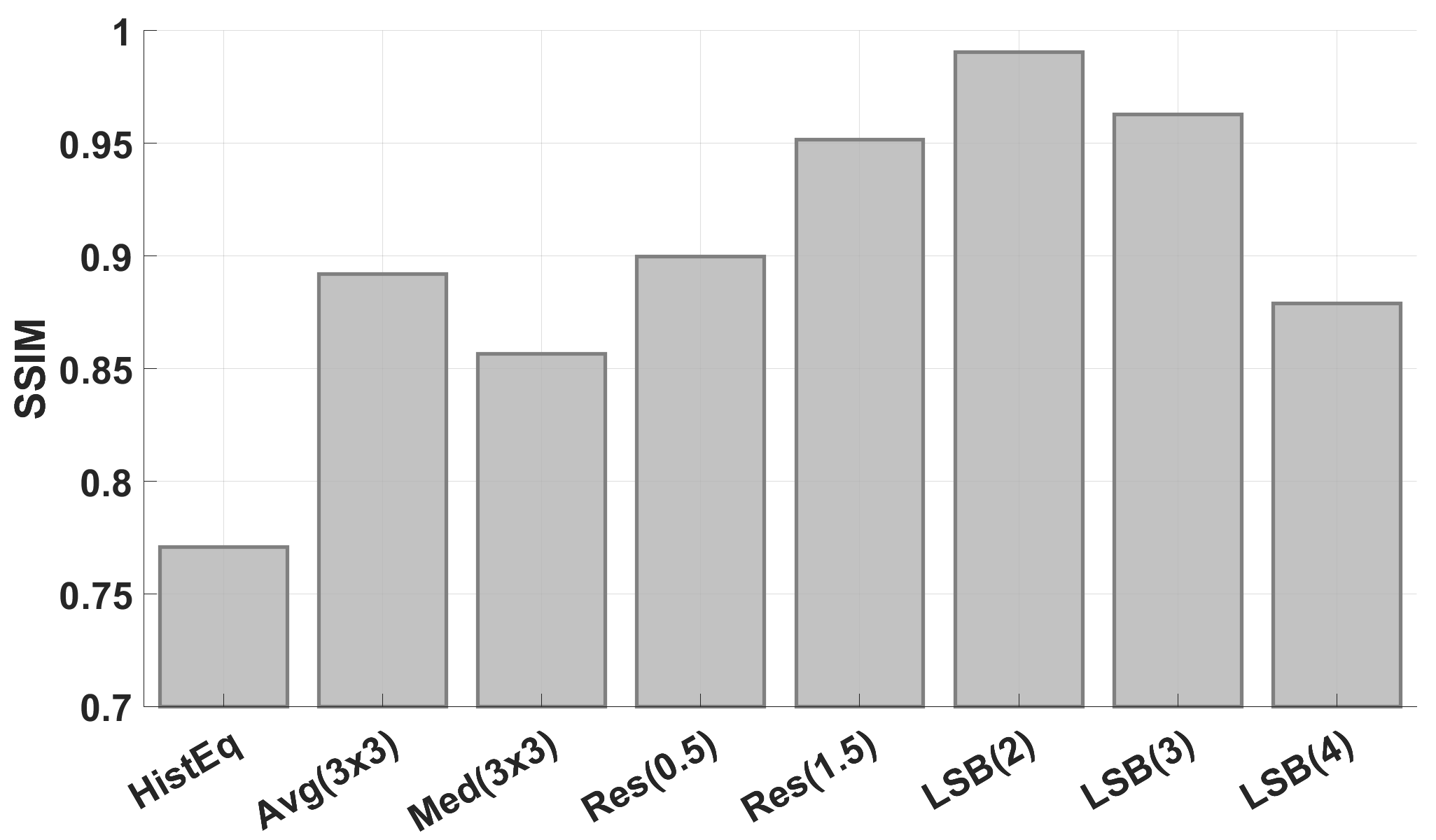} \\
			(a)&(b)\\
			\includegraphics[width=0.45\textwidth]{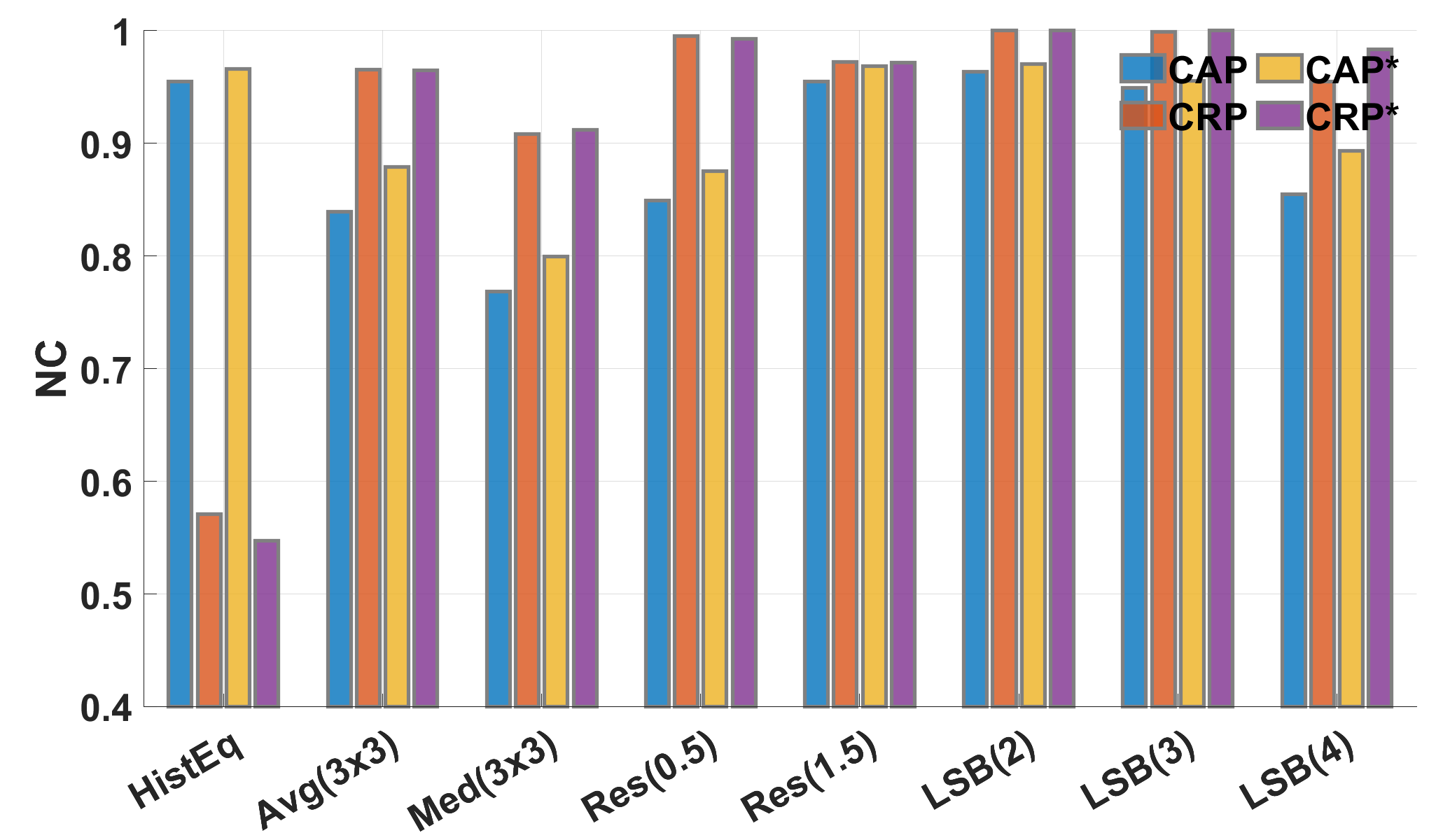} &
			\includegraphics[width=0.45\textwidth]{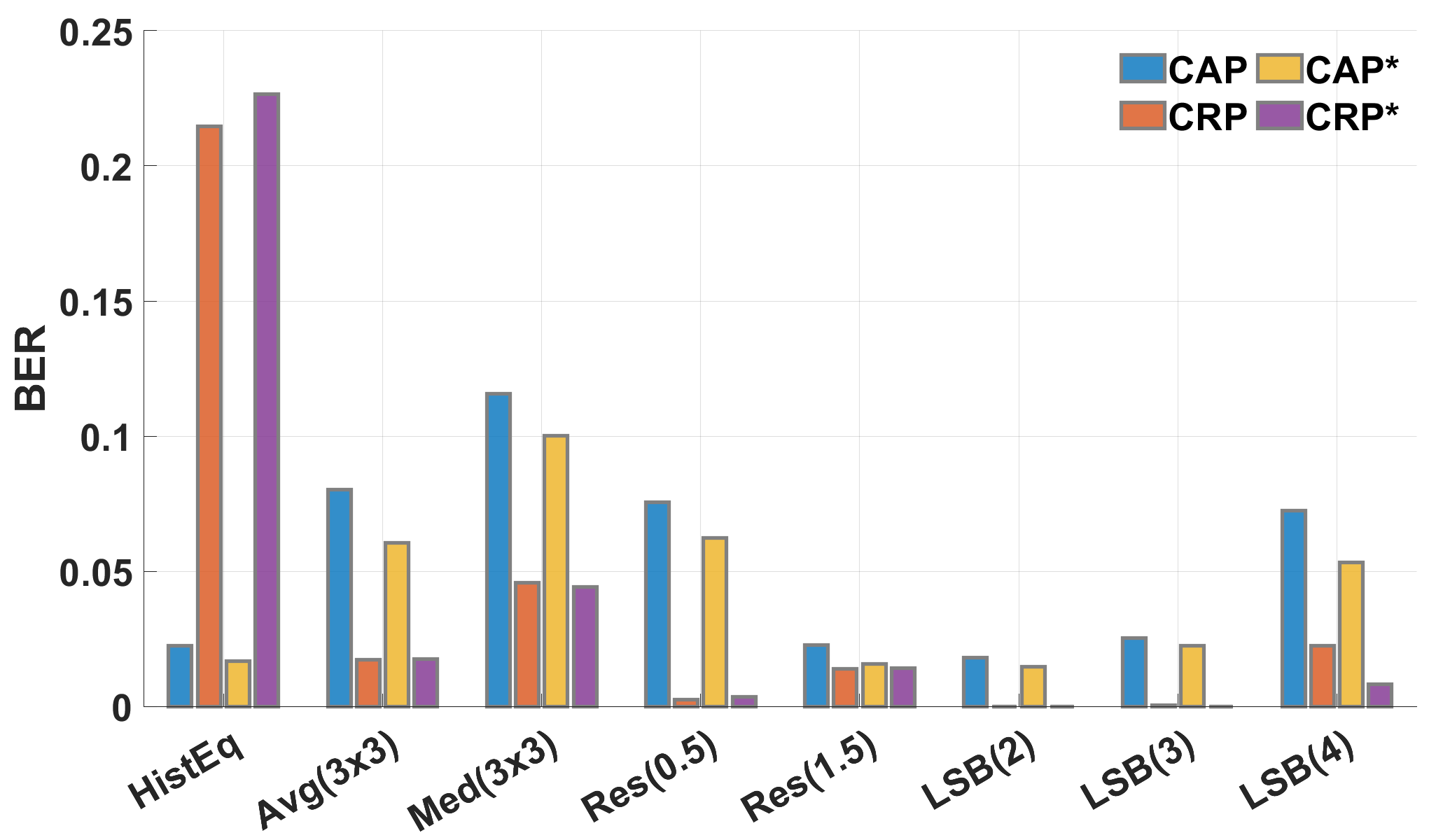} \\
			(c)&(d)		
		\end{tabular}
		\caption{(a-b) The average quality of watermarked images under attacks, (c-d) the average accuracy of extracted marks under different types of attacks for fifteen candidate images. Note, the symbol $(*)$ represents the accuracy of the extraction mark in dual applications. Also, CAP and CRP are abbreviated of Content Authentication Protection and Copyright Protection, respectively.}
		\label{fig:robust2}
	\end{figure*}
	\begin{figure*}[t!]
		\footnotesize
		\renewcommand{\arraystretch}{1.3}
		\centering
		\begin{center}
			\begin{tabular}{ccccc}					
				\includegraphics[width=0.12\textwidth]{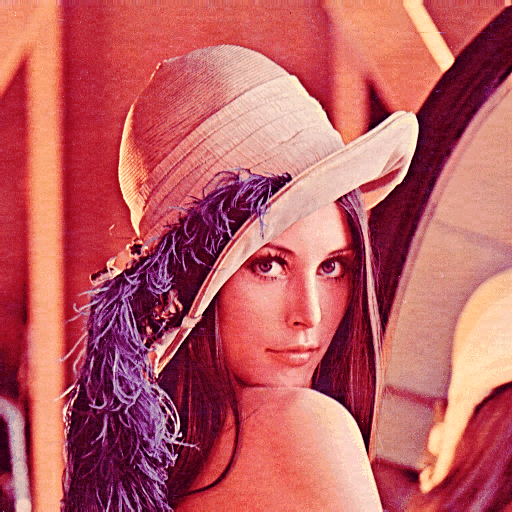}&
				\includegraphics[width=0.12\textwidth]{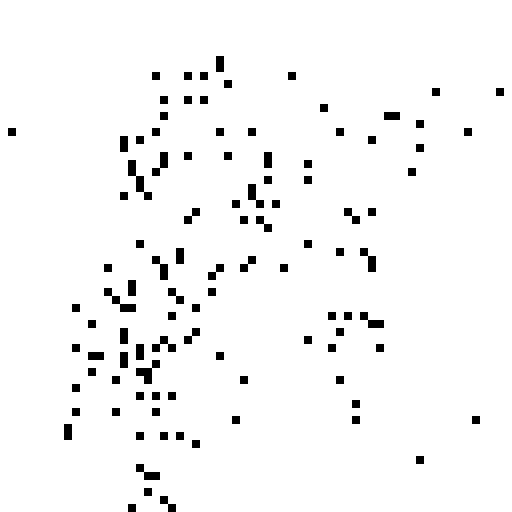}&
				\includegraphics[width=0.12\textwidth]{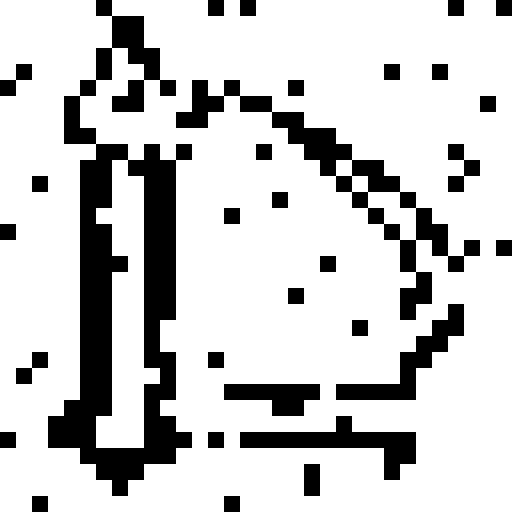}&
				\includegraphics[width=0.12\textwidth]{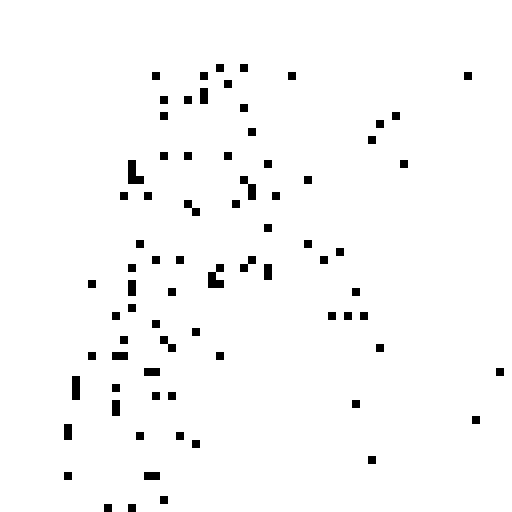}&
				\includegraphics[width=0.12\textwidth]{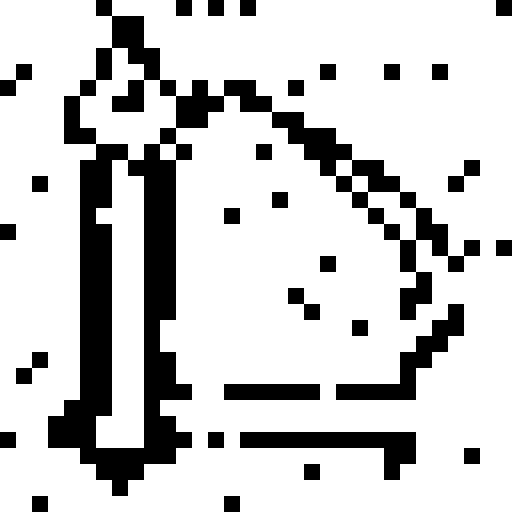}\\
				(22.61, 0.95)&(0.95, 0.03)&(0.88, 0.06)&(0.96, 0.02)&(0.89, 0.05)\\	
				
				\includegraphics[width=0.12\textwidth]{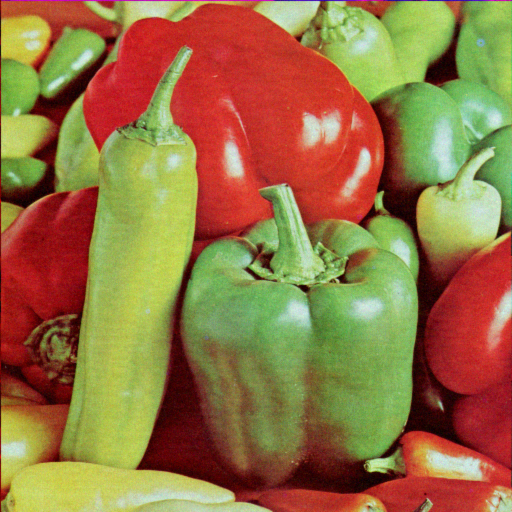}&
				\includegraphics[width=0.12\textwidth]{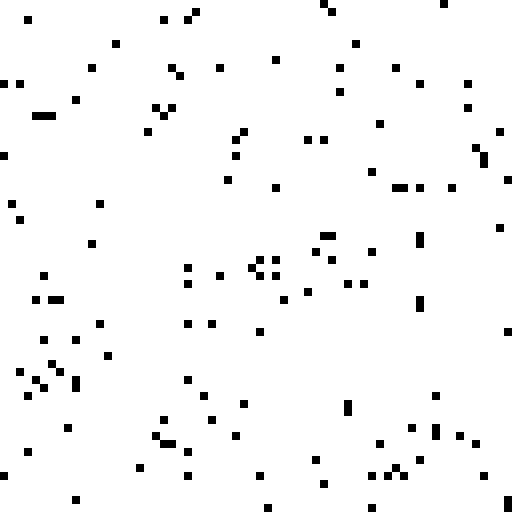}&
				\includegraphics[width=0.12\textwidth]{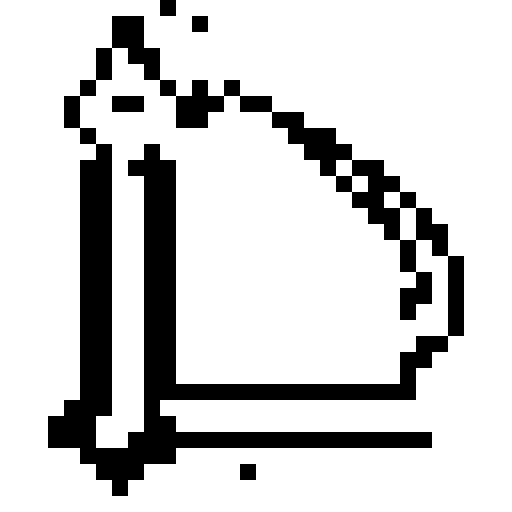}&
				\includegraphics[width=0.12\textwidth]{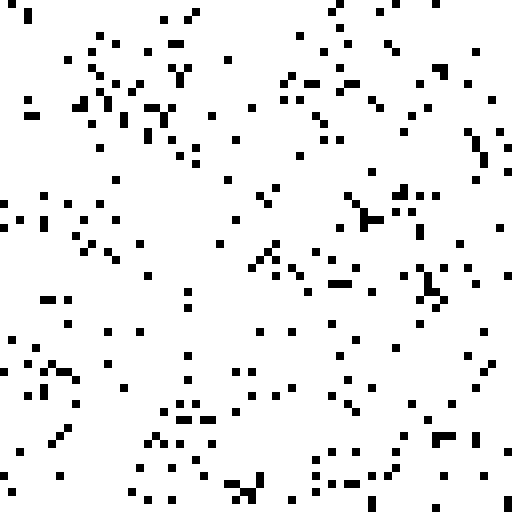}&
				\includegraphics[width=0.12\textwidth]{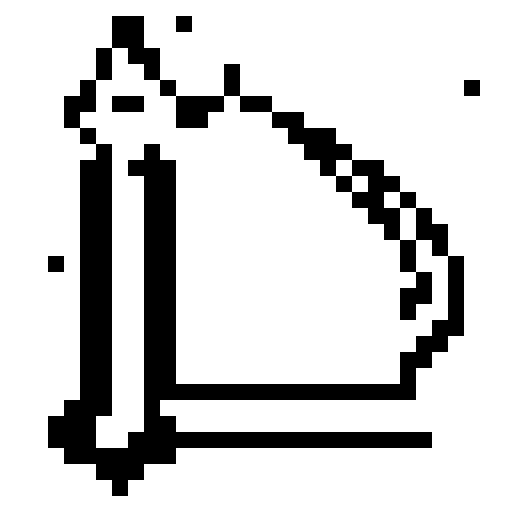}\\					
				(32.58, 0.99)&(0.94, 0.03)&(0.99, 0)&(0.86, 0.07)&(0.99, 0.01)\\

				\includegraphics[width=0.12\textwidth]{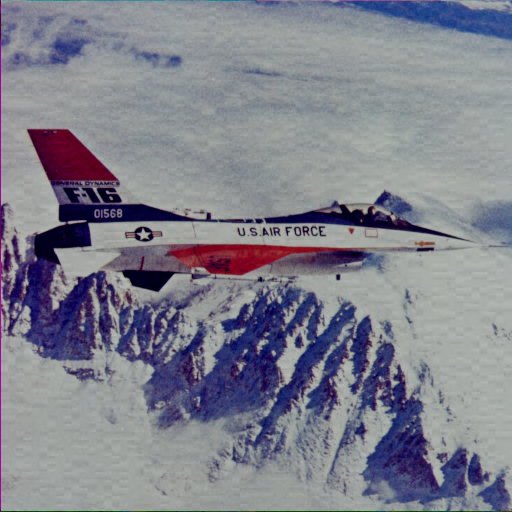}&
				\includegraphics[width=0.12\textwidth]{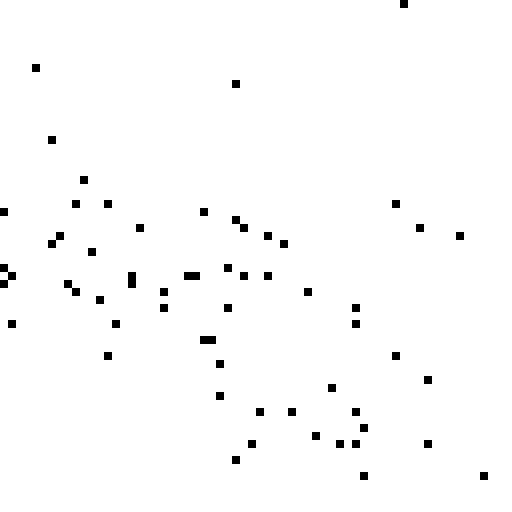}&
				\includegraphics[width=0.12\textwidth]{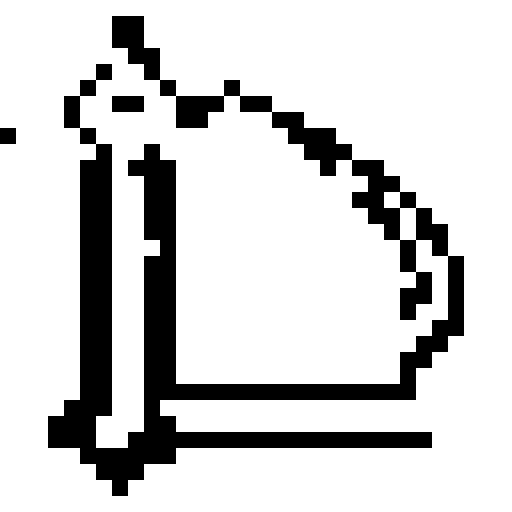}&
				\includegraphics[width=0.12\textwidth]{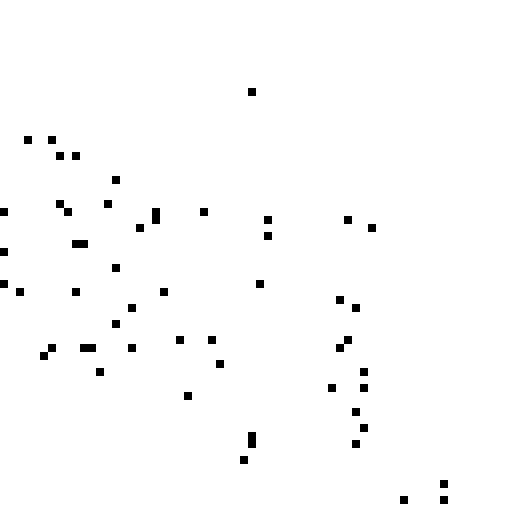}&
				\includegraphics[width=0.12\textwidth]{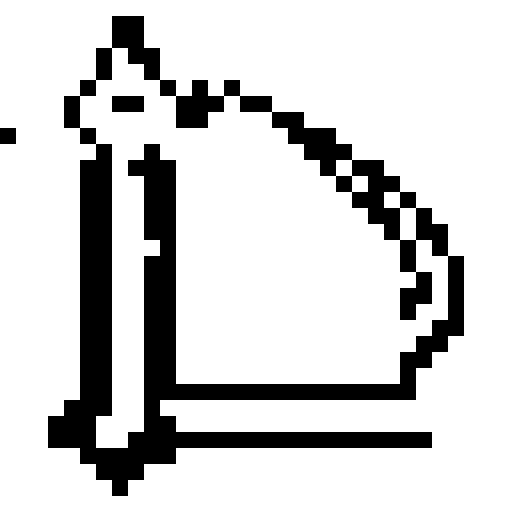}\\
				(14.19, 0.91)&(0.97, 0.01)&(0.99, 0)&(0.97, 0.01)&(0.99, 0)\\
				
				\includegraphics[width=0.12\textwidth]{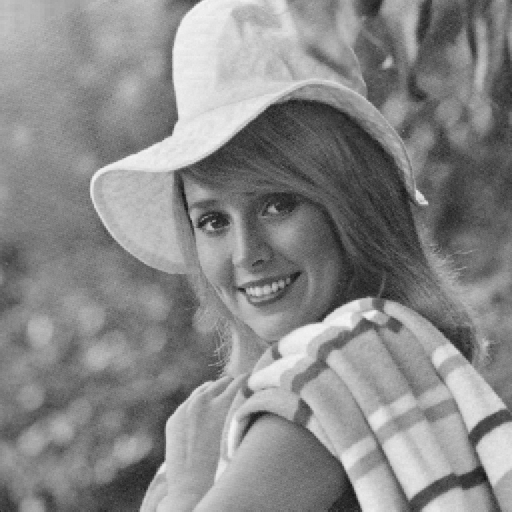}&
				\includegraphics[width=0.12\textwidth]{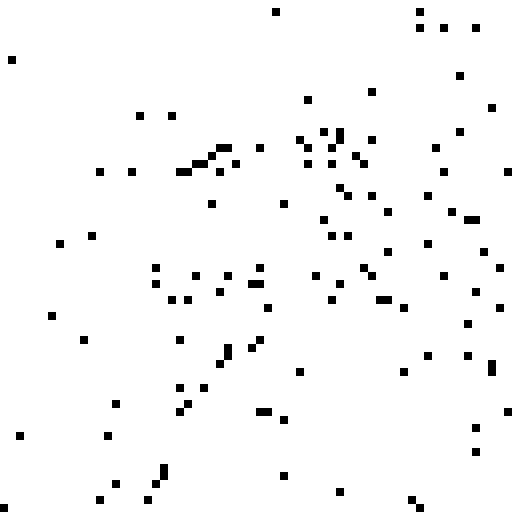}&
				\includegraphics[width=0.12\textwidth]{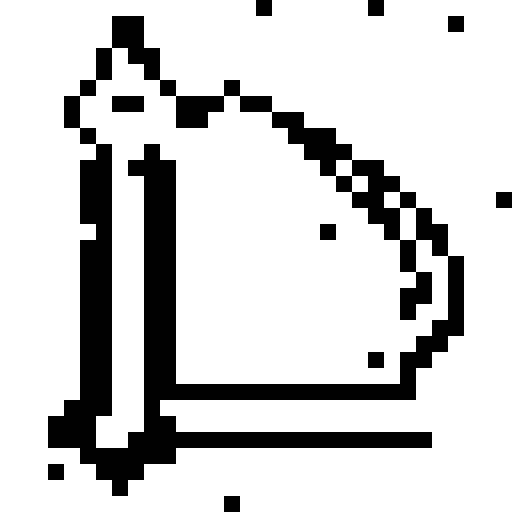}&
				\includegraphics[width=0.12\textwidth]{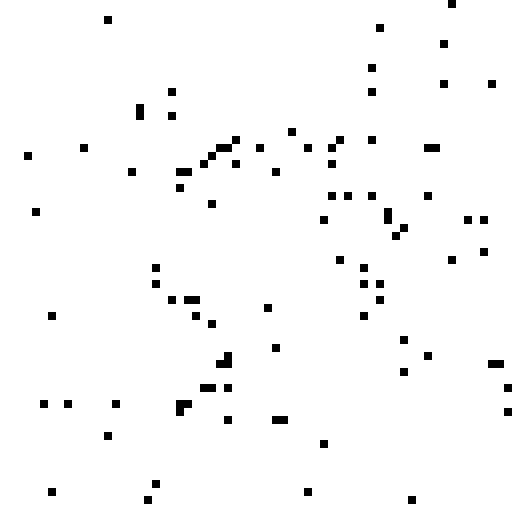}&
				\includegraphics[width=0.12\textwidth]{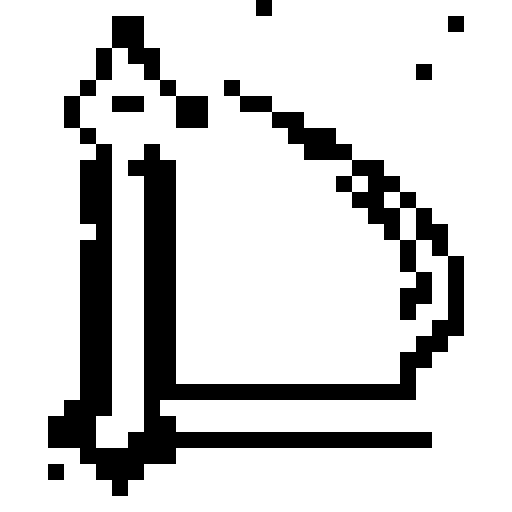}\\
				(33.89, 0.89)&(0.94, 0.03)&(0.98, 0.01)&(0.95, 0.02)&(0.99, 0.01)\\
				
				\includegraphics[width=0.12\textwidth]{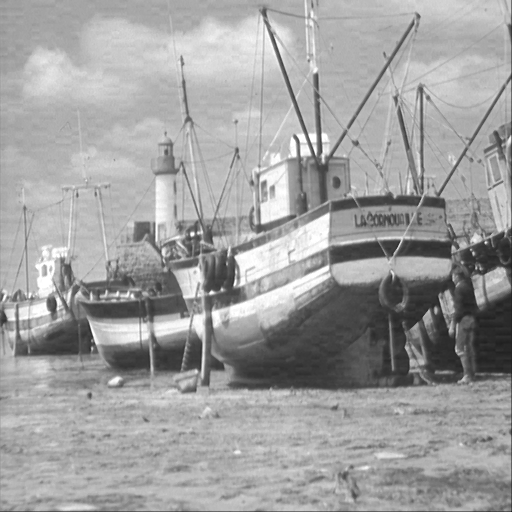}&
				\includegraphics[width=0.12\textwidth]{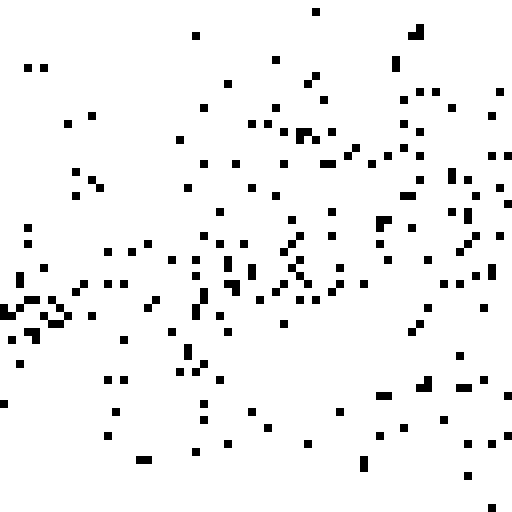}&
				\includegraphics[width=0.12\textwidth]{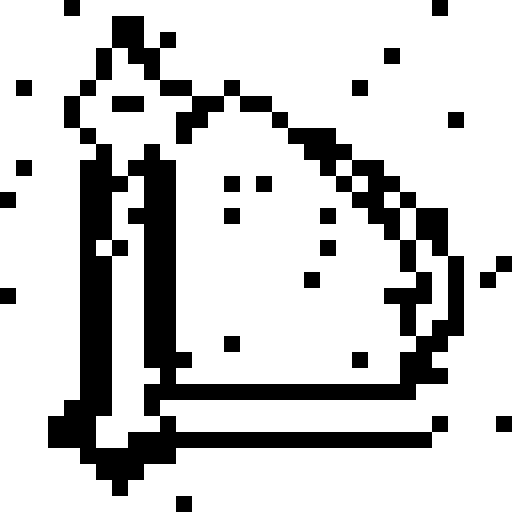}&
				\includegraphics[width=0.12\textwidth]{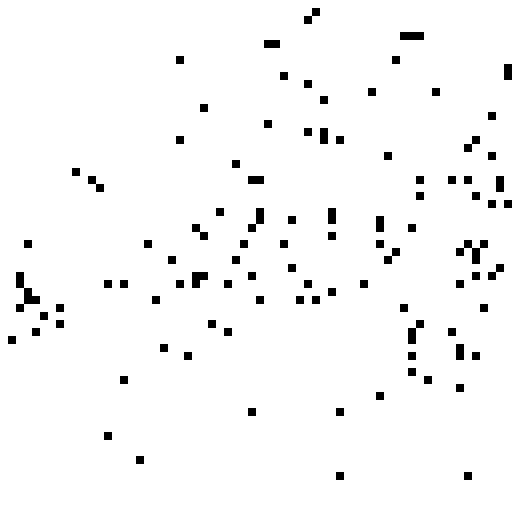}&
				\includegraphics[width=0.12\textwidth]{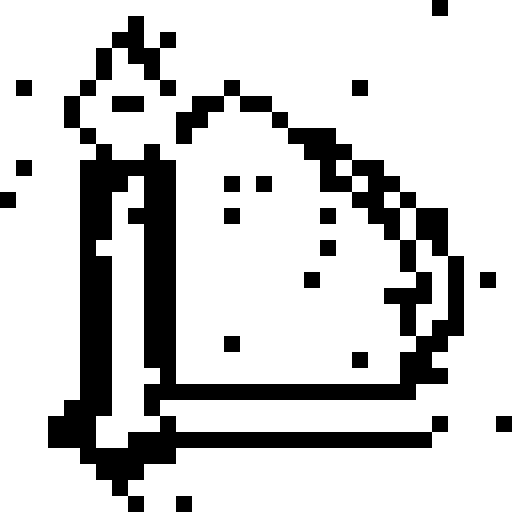}\\
				(18.08, 0.79)&(0.91, 0.05)&(0.92, 0.04)&(0.94, 0.03)&(0.93, 0.03)\\
				(a)&(b)&(c)&(d)&(e)\\
				
			\end{tabular}	
		\end{center}
		\caption{The perceptual quality of the extracted marks under different hybrid attacks for five color and grayscale images. The applied attacks on the samples from top to bottom are: \{Sharpening($\tau$=0.5) and Histogram Equalization\}, \{Gaussian Noise($\sigma$=0.001) and Gaussian Filter($\sigma$=0.5)\}, \{Darken(50) and JPEG2000($CR$=8)\}, \{LSB(3) and Resize(1.5)\}, \{Lighten(30) and Median (3$\times$3)\}. (a) Attacked Image, (b-c) Content Protection and Copyright Protection Marks in single application, respectively. (d-e) Content Protection and Copyright Protection Marks in dual case, respectively. Also, the pair numbers below each figure represents (PSNR, SSIM) and (NC, BER) for bitmap and binary image,  respectively.}
		\label{fig:robust3}	
	\end{figure*}
	\begin{figure*}[t!]
		\begin{tabular}{ccc}
			\includegraphics[width=0.30\textwidth]{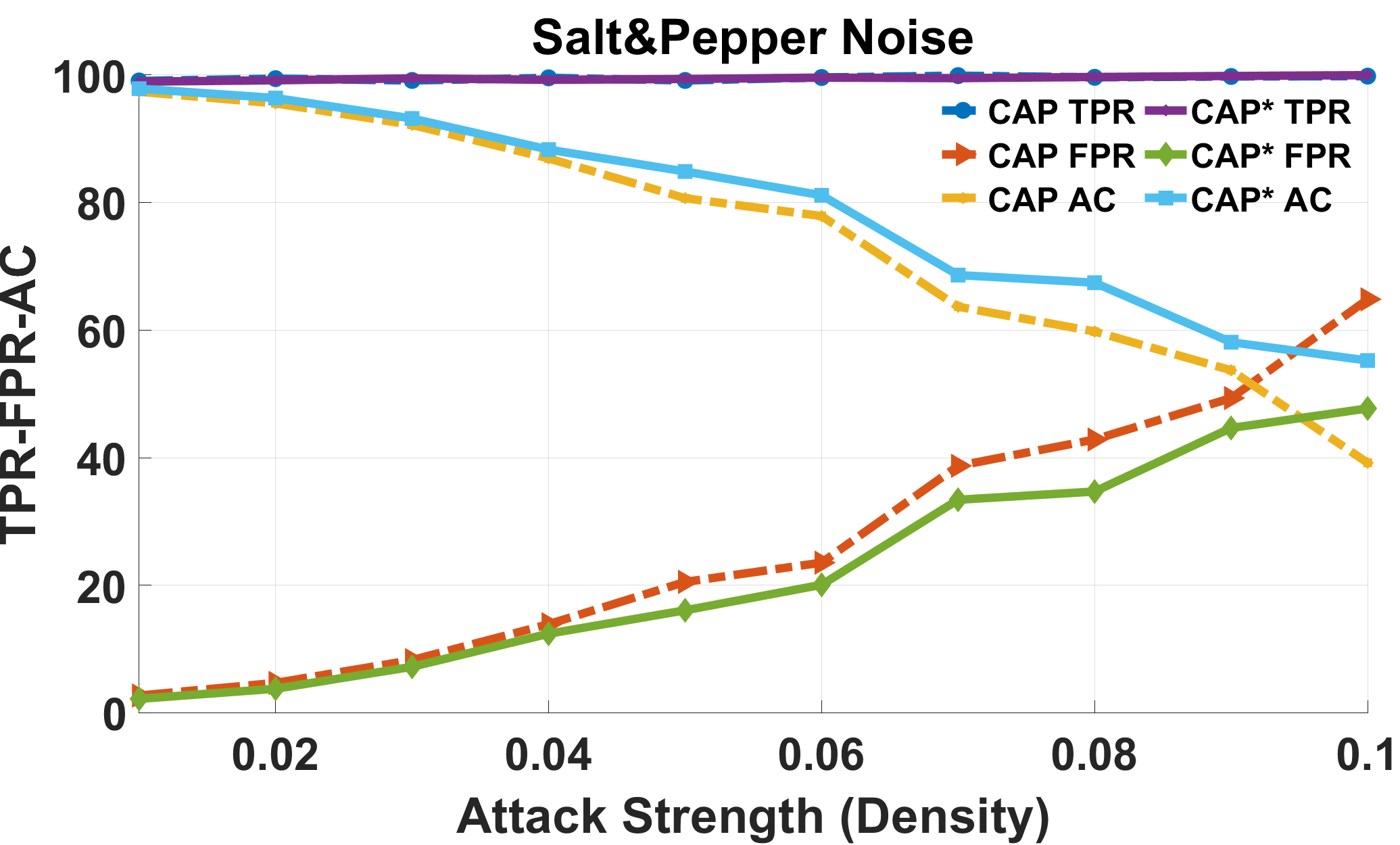} &
			\includegraphics[width=0.30\textwidth]{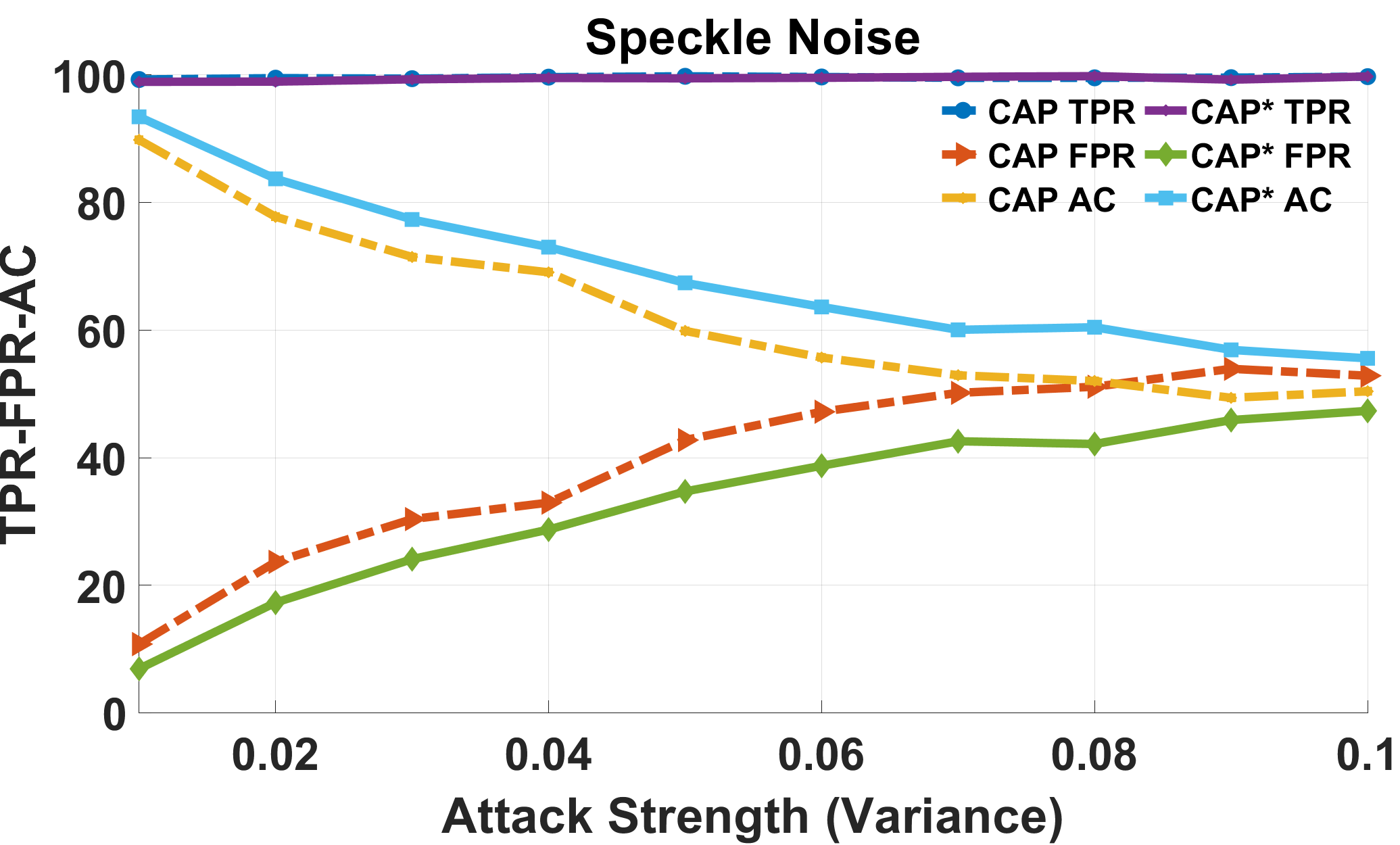} &
			\includegraphics[width=0.30\textwidth]{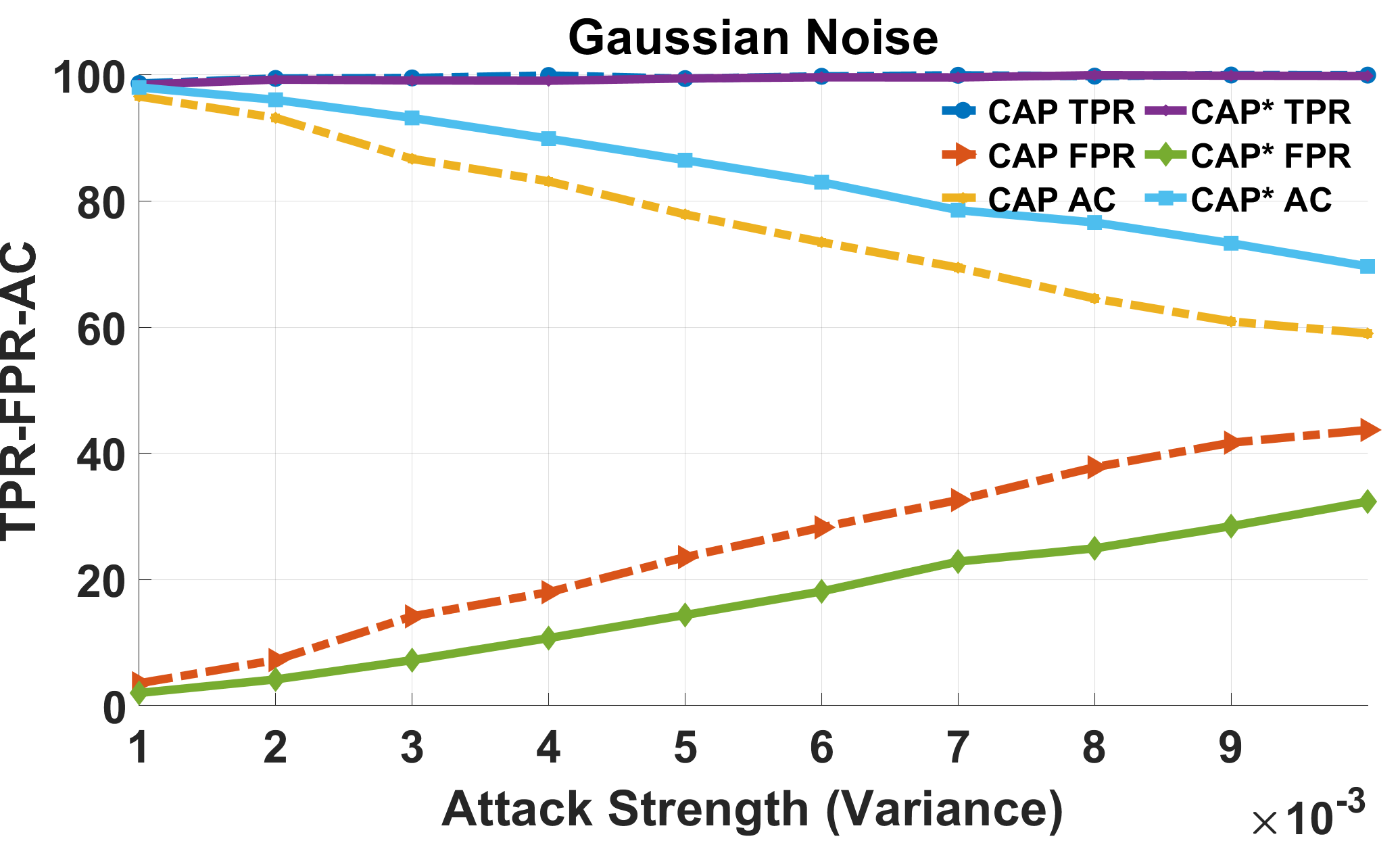} \\
			(a)&(b)&(c)\\
			\includegraphics[width=0.30\textwidth]{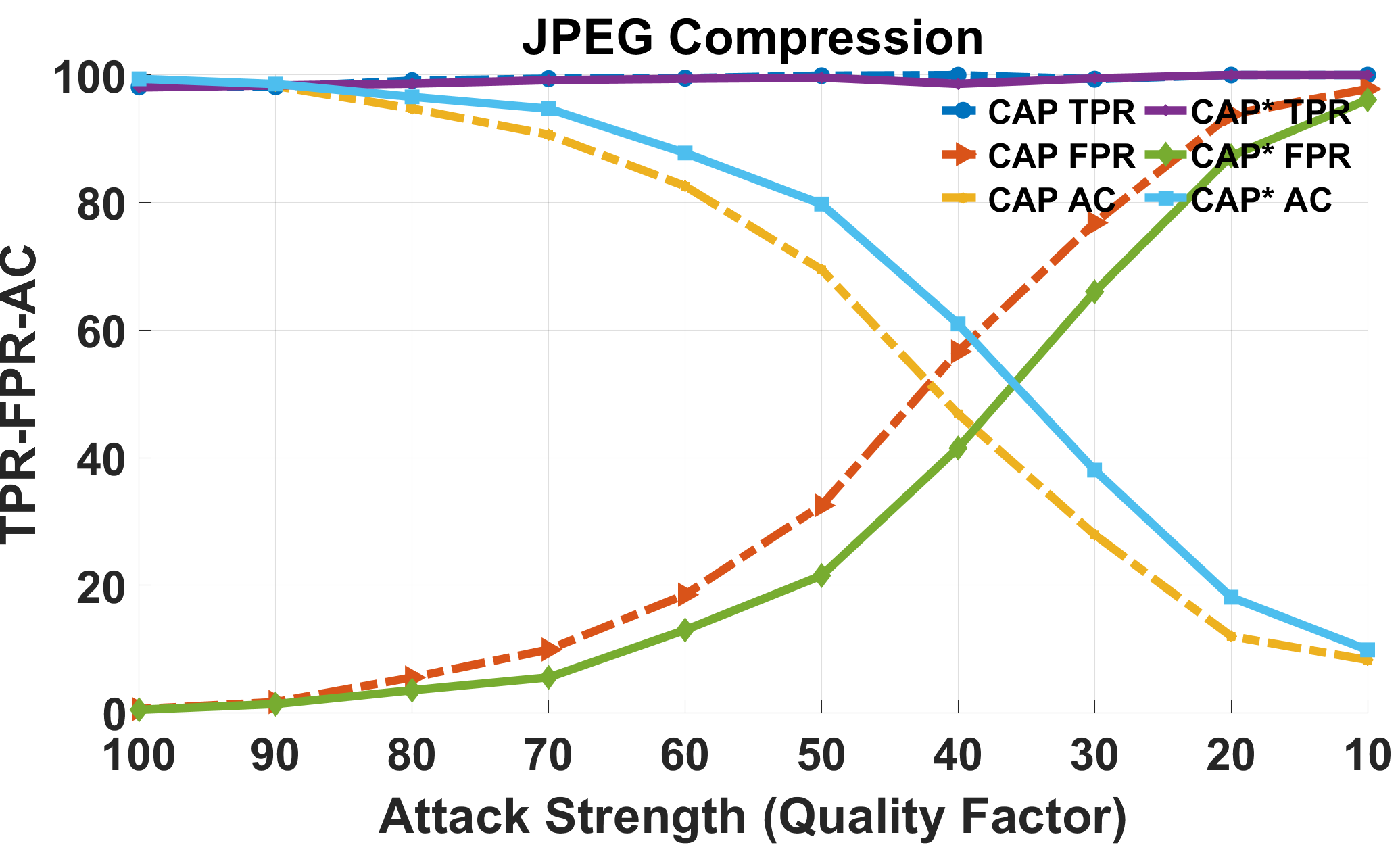} &
			\includegraphics[width=0.30\textwidth]{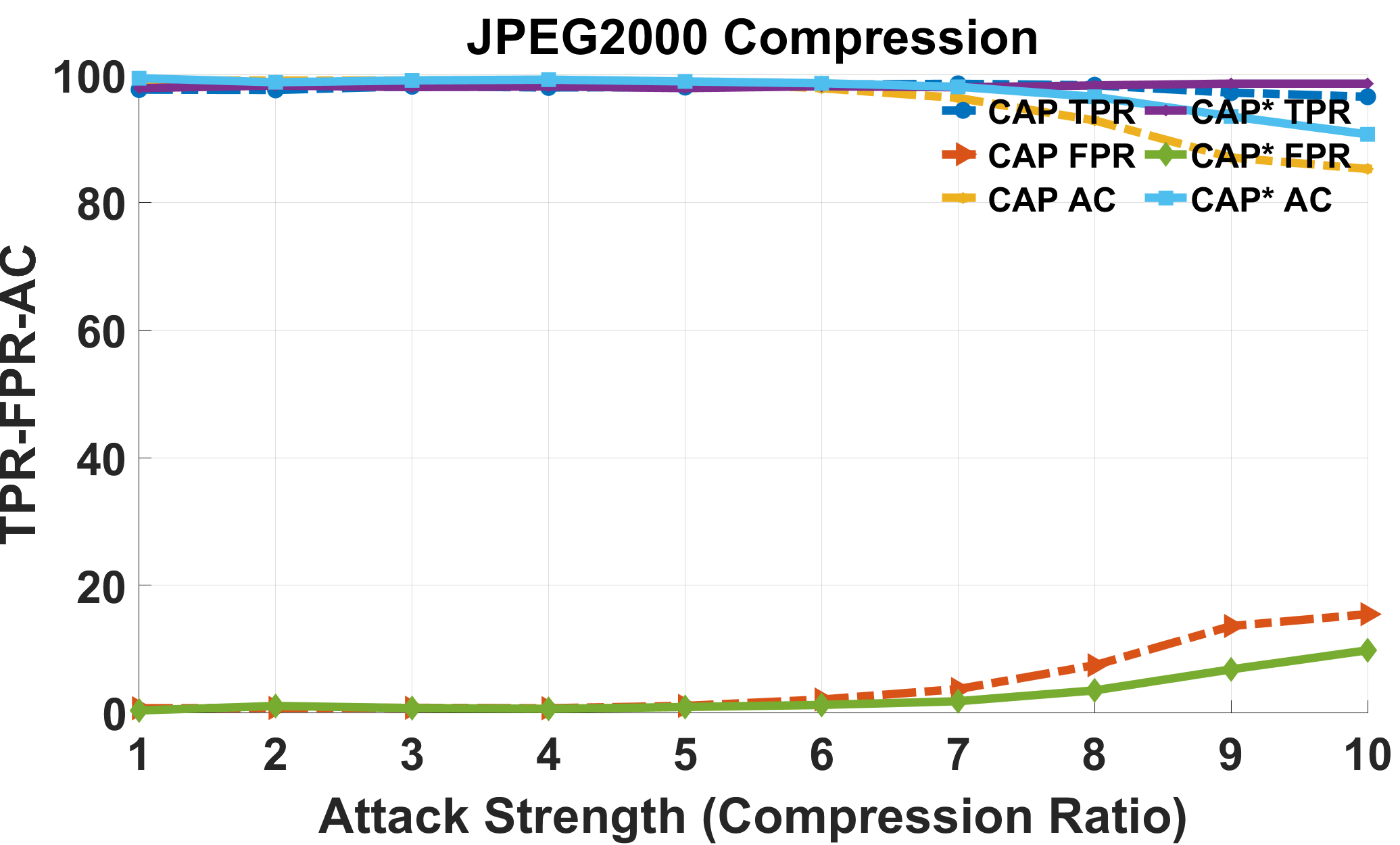} &
			\includegraphics[width=0.30\textwidth]{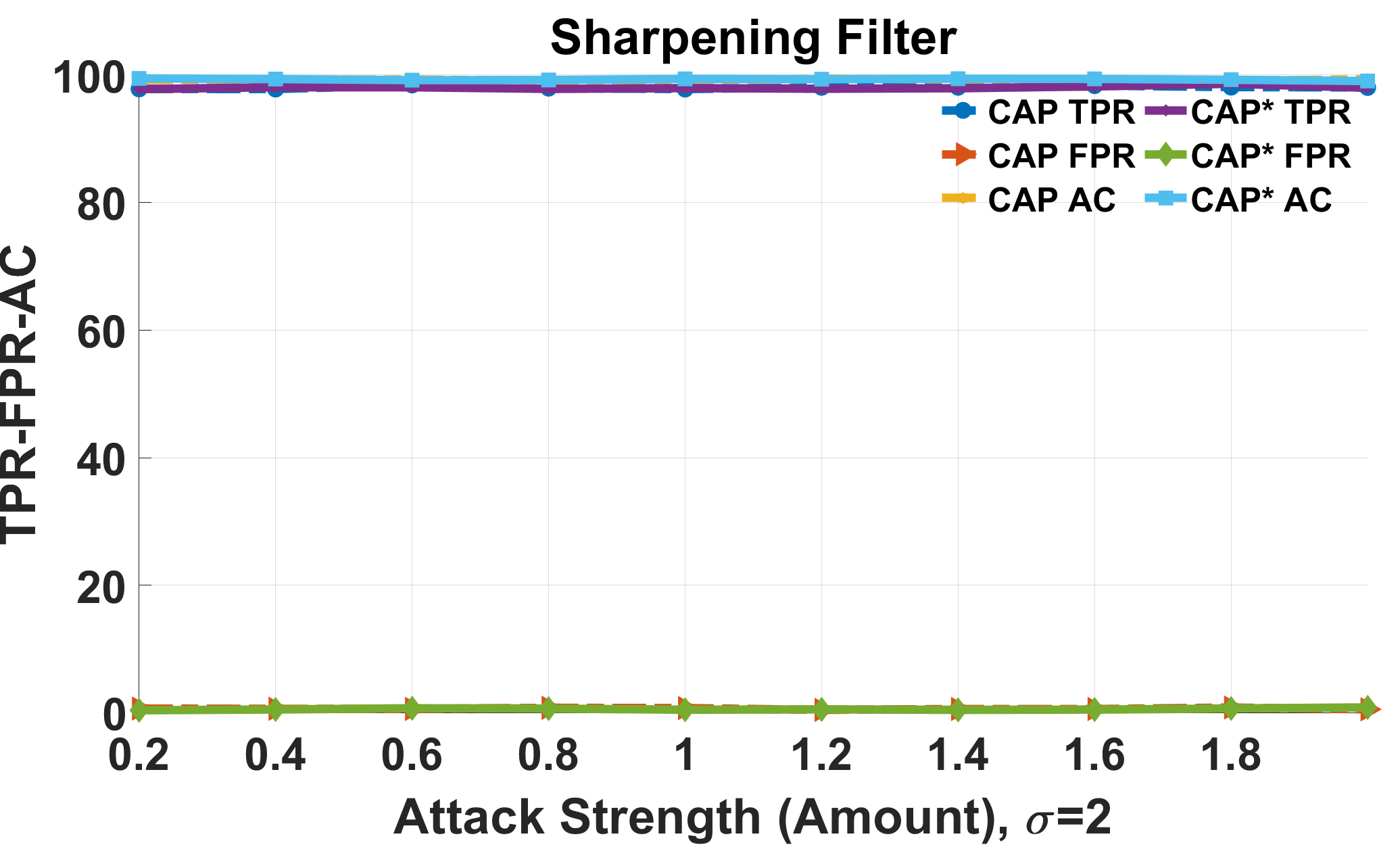} \\
			(d)&(e)&(f)\\
			\includegraphics[width=0.30\textwidth]{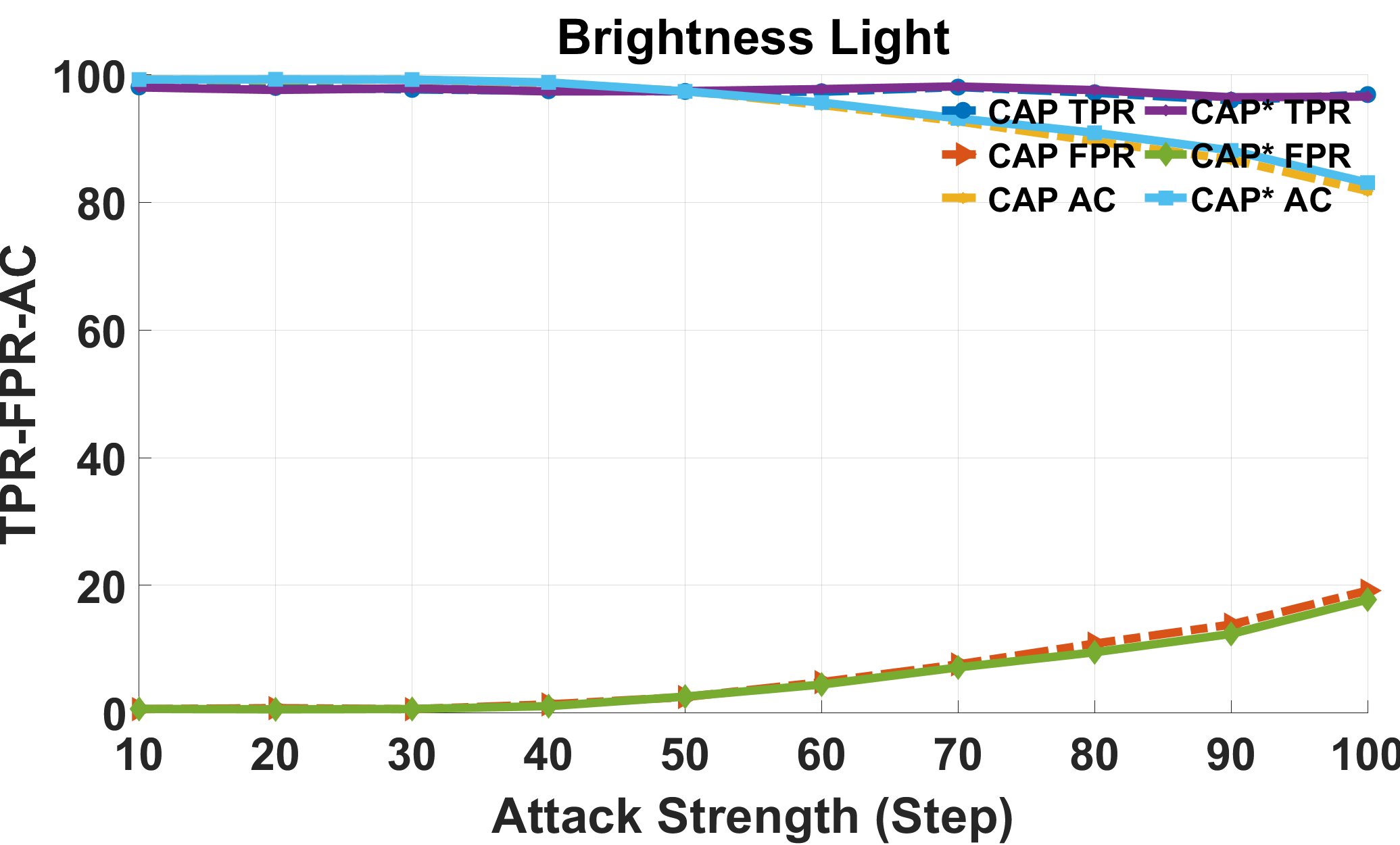} &
			\includegraphics[width=0.30\textwidth]{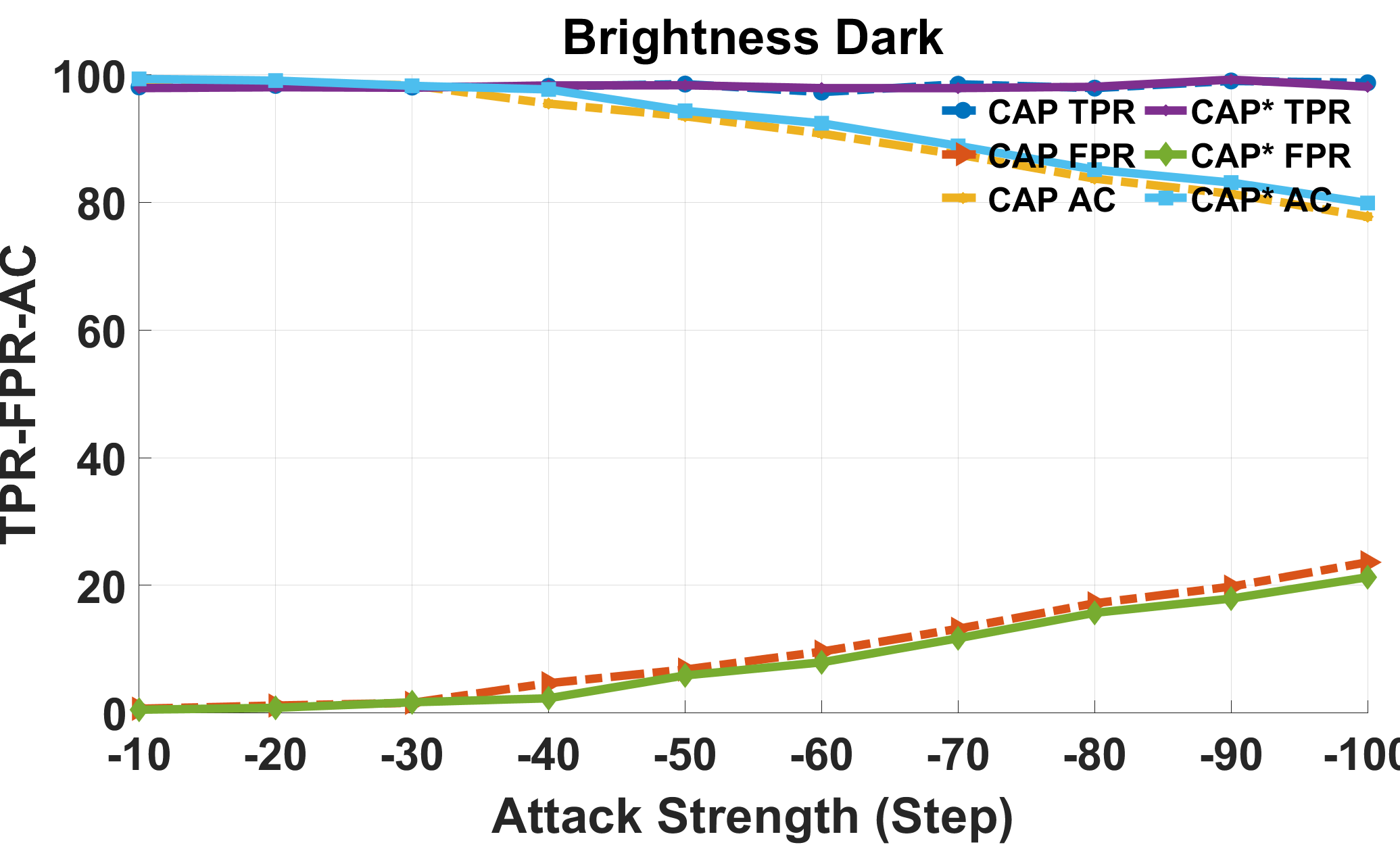} &
			\includegraphics[width=0.30\textwidth]{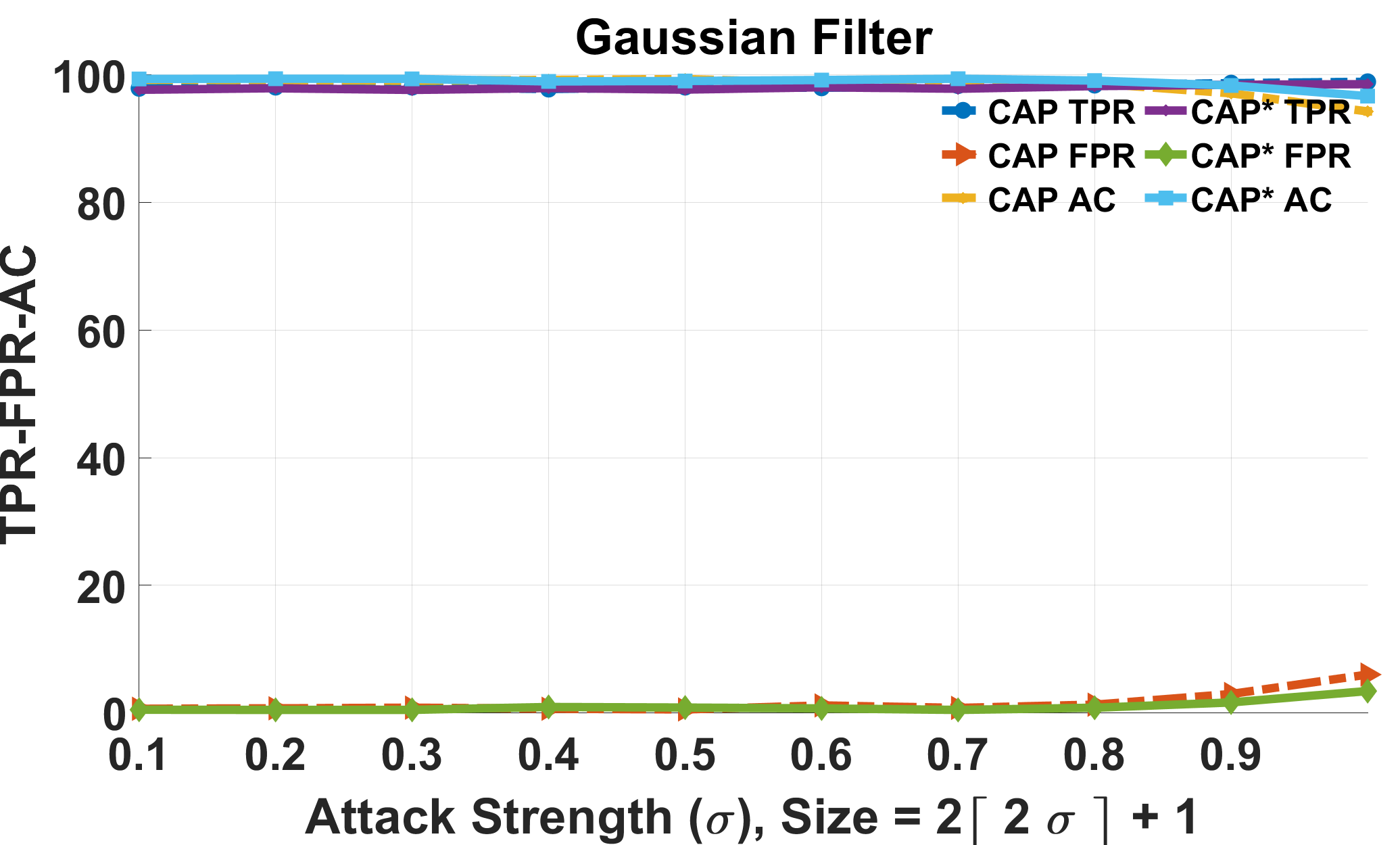} \\
			(g)&(h)&(i)			
		\end{tabular}
		\begin{tabular}{cc}
			\includegraphics[width=0.45\textwidth]{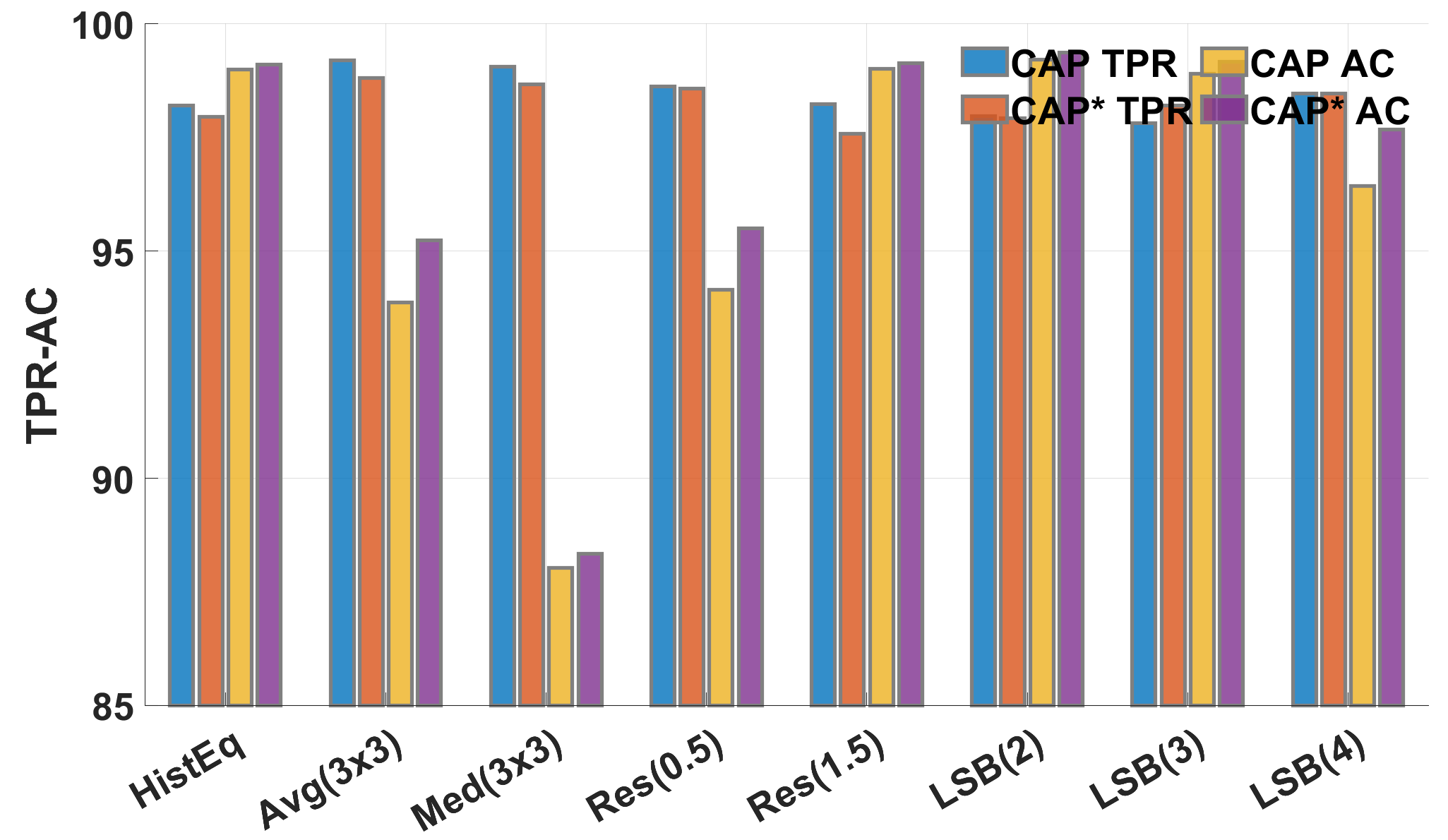}&
			\includegraphics[width=0.45\textwidth]{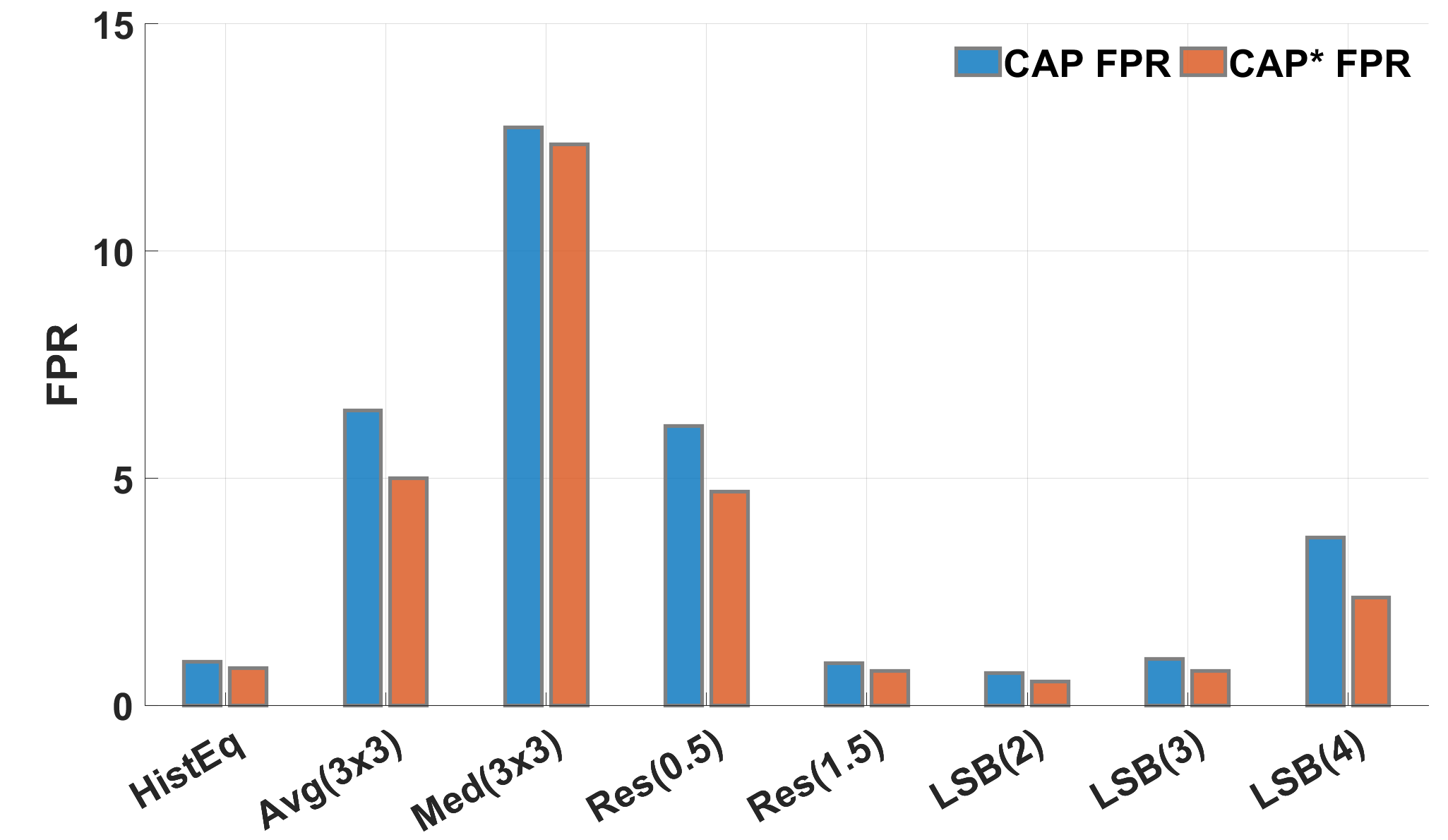}  \\
			(j)&(k)
		\end{tabular}
		\caption{The average of TPR, FPR, and AC of tamper detection phase under different types of attacks for fifteen candidate images. Note, the symbol $(*)$ represents the accuracy of the detection rate in dual application. Also, CAP is abbreviated of Content Authentication Protection.} 
		\label{fig:tamper1}
	\end{figure*}
	\begin{figure*}[t!]
		\center
		\begin{tabular}{cccc}
			\includegraphics[width=0.15\textwidth]{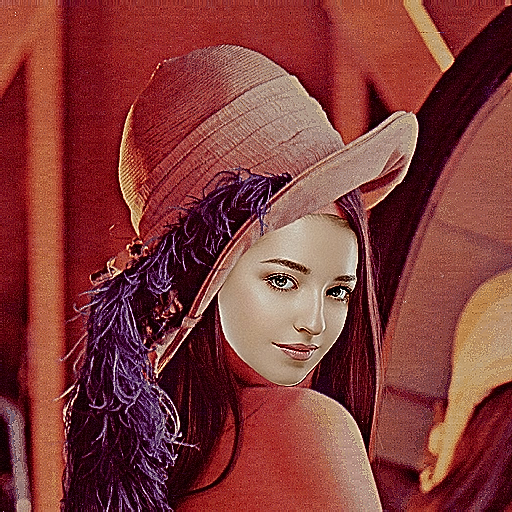} &
			\includegraphics[width=0.15\textwidth]{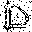} &
			\includegraphics[width=0.15\textwidth]{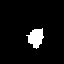} &
			\includegraphics[width=0.15\textwidth]{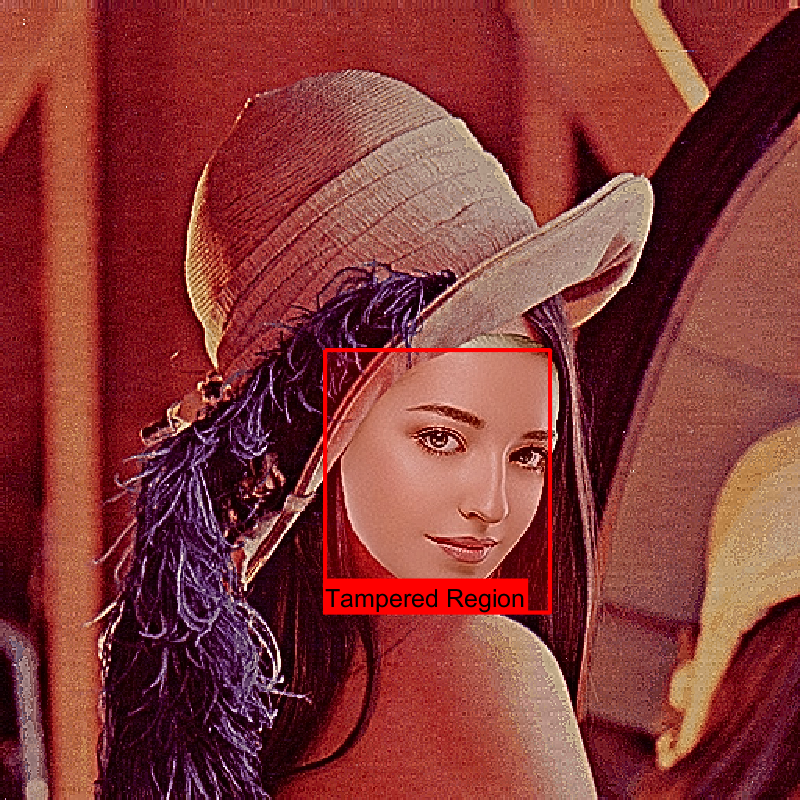} \\
			(13.99, 0.74)&(0.82, 0.09)&(0.91, 0.04)&(81.23, 98.63, 0.1)\\
			\includegraphics[width=0.15\textwidth]{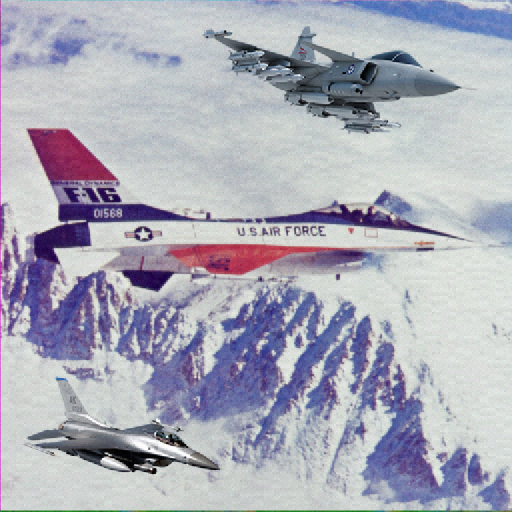} &
			\includegraphics[width=0.15\textwidth]{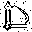} &
			\includegraphics[width=0.15\textwidth]{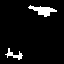} &
			\includegraphics[width=0.15\textwidth]{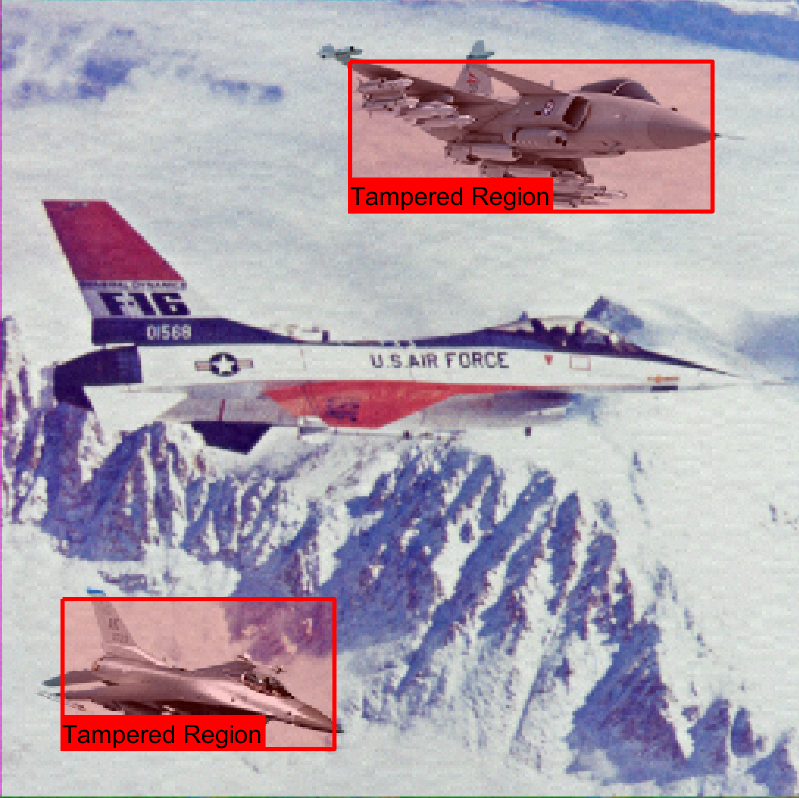} \\				
			(19.09, 0.86)&(0.95, 0.02)&(0.89, 0.05)&(71.09, 97.19, 0.79)\\
			\includegraphics[width=0.15\textwidth]{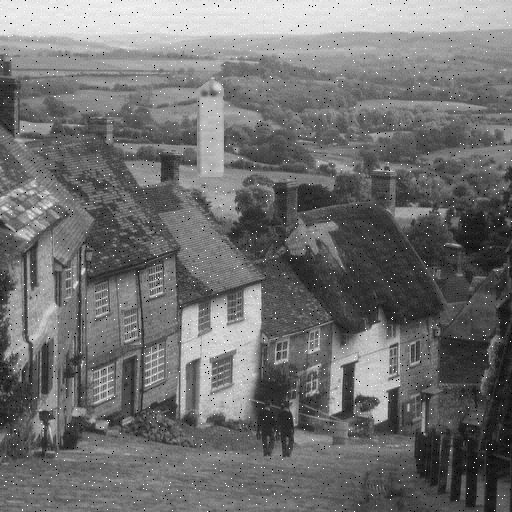} &
			\includegraphics[width=0.15\textwidth]{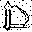} &
			\includegraphics[width=0.15\textwidth]{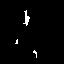} &
			\includegraphics[width=0.15\textwidth]{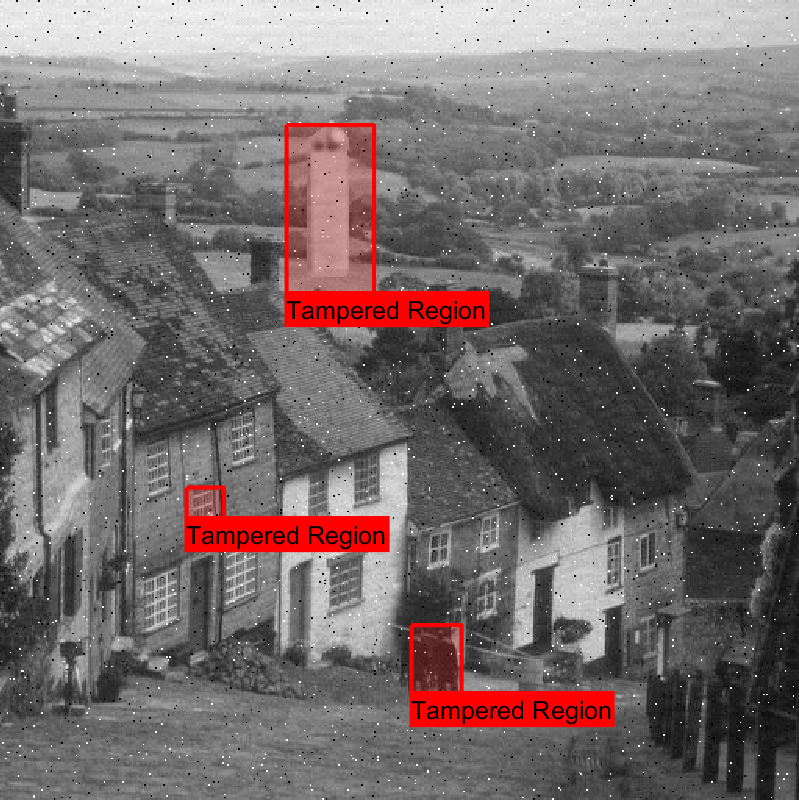} \\		
			(23.91, 0.73)&(0.90, 0.04)&(0.87, 0.06)&(71.43, 98.88, 0.69)\\
			(a)&(b)&(c)&(d)\\
		\end{tabular}
		\caption{The visual show of the tamper detection performance under different tampering with hybrid attacks. (a) Tamper image, (b) Copyright mark, (c) The post-processing of tamper detection phase, and (d) Discovered forged regions. The applied attacks on the tampered image from top to bottom are: \{Darken(50) and Sharpening($\tau=4, \sigma=1$)\}, \{LSB(3) and Resize(1.5)\}, and \{JPEG($CF=70$) and Salt\&Pepper($\rho=0.01$)\}, respectively. The pairs number below each Fig from left to right are: (PSNR, SSIM), (NC, BER), (NC, BER), and (TPR, AC, FPR), respectively.}
		\label{fig:tamper2}
	\end{figure*}
	\subsection{Robustness analysis}
	Over the communication channels, the quality and integrity of transmitted media may be degraded; Moreover, in extreme cases, the profiteers and attackers try to destroy or remove the mark using any kinds of trick. The concept of robustness measures the resistance capability of the algorithm after the slight or even deep modifications that had led these distortions. In general, the robustness refers to the ability of the system to correctly extract the mark from the watermarked image after common intentional and unintentional operations. It judges with renowned metrics such as Bit Error Ratio(BER) and Normalize Cross-Correlation(NC) \cite{ref28}. In this subsection, the performance of WSMN in terms of robustness is investigated under different types of single and hybrid attacks. Brief description of the employed attacks is given as follows:
	\begin{itemize}
		\item Salt\&Pepper Noise, $\rho \in (1, 2, ..., 10)\times10^{-2}$.
		\item Speckle Noise, $\sigma \in (1, 2, ..., 10)\times10^{-2}$.
		\item Gaussian Noise, $\sigma \in (1, 2, ..., 10)\times10^{-3}$.
		\item JPEG Compression, $QF \in (1, 2, ..., 10)\times10$.
		\item JPEG2000 Compression, $CR \in (1, 2, ..., 10)$.
		\item Brightness Lighten, $Step \in (1, 2, ..., 10)\times10$.
		\item Brightness Darken, $Step \in (1, 2, ..., 10)\times10$.
		\item Sharpening Filter, $\tau \in (2, 4, ..., 20) \times10^{-1}$.
		\item Gaussian Filter, $\sigma \in (1, 2, ..., 10)\times10^{-1}$.
		\item Averaging Filter, $(3\times3)$.
		\item Median Filter, $(3\times3)$.
		\item Image Resizing, $(\rfrac{1}{2}, 2)$.
		\item Histogram Equalization.
		\item LSBs, (Least bits are replaced with random numbers).
	\end{itemize}
	After applying these attacks, the mark(s) is extracted and compared with the original one to assess the robustness of WSMN by introduced criteria.
	
	In this way, the quality of the watermarked image under the mentioned attacks are demonstrated in Figs. \ref{fig:robust1} (a-b) and Figs. \ref{fig:robust2} (a-b). As can be seen, in most cases, the applied distortion has lead to a significant reduction in terms of PSNR and SSIM. By looking at details, it can be found from Figs. \ref{fig:robust1} (a-b) and Figs. \ref{fig:robust2} (a-b) the operations of the noise has extreme effects on the quality of signal in comparison to rest attacks. On the contrary, JPEG2000, Gaussian Filter, and LSBs Replacing have resulted insignificant degradation on the integrity of the image. It can be observed, the remained attacks gradually declined in these simulations. In terms of smoothing filters such as Averaging, Median, and even Resizing, the quality of signal reach to 30 dB, approximately. About LSBs Replacing, which effectively destroys the high frequencies, PSNR reduced by step about five from two to four bits ignoring. In these cases, a random number is replaced to the least bits based on the rigidity of attack.
	
	Subsequently, the accuracy of extracted marks under the strict attacks are displayed in Figs. \ref{fig:robust1} (c-k) and Figs. \ref{fig:robust2} (c-d). First of all, the plotted curves prove the fact that the corresponding extracted marks in both single and dual applications have similar behavior. In other words, the difference in the level of accuracy between the extracted marks in each application is so close to each other. It means, the optimization phase well had performed to generate the watermarked image; Although the quality of watermarked image had been negligibly reduced due to the dual embedding, the robustness not only has not decreased but it has been improved. 
	
	Further, it can be gain from Figs, with regard to Salt \& Pepper Noise, Sharpening Filter, and Histogram Equalization attacks the extracted copyright marks have lower NC compared to authentication marks. This is because of the effect of attacks on the low frequencies of the signal. Also, the line graphs demonstrate a considerable margin in the JPEG compression case for both copyright and content authentication protection applications. For other attacks, there is bit competition between both types and slowly fall by increasing the strength of the attack. 
	
	Simulation experiments in Figs. \ref{fig:robust1} (c-k) show that the performance of WSMN under Brightness Darken, Brightness Lighten, Gaussian Filter, JPEG2000, and Sharpening is satisfactory. On the other hand, the system has fragile proficiency in terms of Noise compared to rest attacks when the strength of attacks goes upper than 50\%, approximately. Notice that in the real world, the attack is usually carried out in such a way that HVS is not aroused. Also, the results in Figs. \ref{fig:robust2} (c-d) indicate that despite the meaningful reduction in terms of Histogram Equalization for copyright marks, WSMN has notable performance for other attacks in both applications. All in all, as expected, the extracted watermarks are extremely correlated with the original version and tolerate under the most potent attacks in most cases.
	
	The robustness analysis is further investigated visually under various hybrid attacks in Fig. \ref{fig:robust3}. In this way, the perceptual quality of the marks is presented and the quantitative results are listed. According to the reported results, both marks satisfactorily extracted. Hence, the ownership can clearly prove with the help of an accurate extract copyright mark. Also, by slight morphology operation as post-processing, the forged regions can be highlighted in the suspicious image. To sum up, it is quite evident, the designed intelligent embedding rules lead to acceptable performance under single and hybrid signal processing attacks. In other words, due to emphasizing the texture of the image, utilizing a learning algorithm, and employing multi-objective optimization, it can provide tolerable robustness and admissible quality, simultaneously.
	\subsection{Tamper localization analysis}
	One of the desired aims of designing WSMN was that it is able to detect and localize the tampering parts on the watermarked image. The tamper detection or content authentication algorithm can locate the tamper regions with the help of embedded information. Unfortunately, intruders and profiteers modify some parts of the signal with other images to reach bad intention purposes. Hence, the detector of a watermarking system must have the capability to correctly discover and determine those modification regions where the integrity lost. In general, tampering type classify into three categories as removing (or modify), adding (or drawing), and copy-move (copy-paste) tampers. The copy-move as common tampering refers to copying a region of watermarked image and pasting it on somewhere else into the same image. In extreme cases as vector-quantization, the candidate region comes from another watermarked image which watermarked by different security key(s).

	\begin{table*}[t!]
		\centering
		\footnotesize
		\caption{PSNR values of watermarked images for WSMN and related works. A) Baboon, B) Barbara, C) Lena, D) Pepper, E) Lake, F) F16, G) House, H) Goldhill, and  I) Boat. Note: - and * demonstrate the unavailability of the image in the mentioned works and color results, respectively.\\
			Proof ownership(Copyright protection), Content authentication protection(Integrity protection), and Dual purpose(Multipurpose) applications.}		
		\label{TABLE:comapre1}
		\renewcommand{\arraystretch}{1.3}
		\setlength{\tabcolsep}{4pt}
		\scalebox{1} {
			\begin{tabular*}{\textwidth}{@{\extracolsep{\fill}}c@{}c@{}c@{}c@{}c@{}c@{}c@{}c@{}c@{}c@{}c@{}c@{}c@{}c@{}c@{}c@{}c@{}c@{}c@{}c@{}c@{}}					
				\cline{1-20}	
				\multicolumn{1}{c}{\multirow{3}{*}{{\rotatebox[origin=c]{90}{Image}}}}&\multicolumn{6}{c}{WSMN}&\multicolumn{13}{c}{Related Works}\\
				\cline{2-7}\cline{8-20}
				&\multicolumn{2}{c}{Ownership}&\multicolumn{2}{c}{Integrity}&\multicolumn{2}{c}{Dual}&\multicolumn{7}{c}{Ownership}&\multicolumn{3}{c}{Integrity}&\multicolumn{3}{c}{Dual}\\
				\cline{2-3}\cline{4-5}\cline{6-7}\cline{8-14}\cline{15-17}\cline{18-20}
				&Gray&Color&Gray&Color&Gray&Color&\cite{ref9}*&\cite{ref8}&\cite{ref10}*&\cite{ref12}&\cite{ref14}*&\cite{ref11}&\cite{ref13}&\cite{ref19}&\cite{ref20}&\cite{ref21}&\cite{ref23}&\cite{ref22}&\cite{ref24}*\\
				\cline{1-20}				
				A&38.44&38.38&39.78&39.59&36.14&36.40&39.82&38.45&39.28&36.69&40.19&39.15&-&27.01&41.09&41.3&-&29.2&40.09\\
				B&42.81&42.89&42.51&42.69&38.45&39.66&-&36.95&40.07&38.69&40.32&36.85&42.49&-&-&-&-&28.11&40.20\\
				C&43.62&43.63&43.93&43.92&40.73&40.80&42.31&40.89&40.30&39.74&40.35&41.51&42.58&34.67&41.76&41.04&37.39&29.13&40.45\\
				D&42.90&42.99&43.46&41.32&40.16&40.22&41.37&40.08&40.33&39.90&40.26&38.19&42.67&34.51&41.24&40.51&-&30.11&41.50\\
				E&38.43&42.80&41.63&42.31&36.8&36.98&40.59&-&-&37.71&40.22&-&-&31.54&-&-&-&-&-\\
				F&39.96&38.69&43.45&43.34&38.91&38.37&43.73&38.90&41.03&-&40.63&38.26&-&32.81&41.04&40.35&-&30.01&40.90\\
				G&38.64&38.62&42.25&43.45&35.13&36.89&38.68&-&-&39.08&40.76&-&-&-&-&-&-&-&-\\
				H&39.90&-&41.79&-&36.12&-&-&40.67&-&-&-&39.67&-&-&-&-&-&-&-\\
				I&38.65&-&42.85&-&36.71&-&-&37.98&-&38.13&-&36.59&42.5&31.54&-&-&-&-&40.50\\
				\cline{1-20}			
		\end{tabular*}}
	\end{table*}
	\begin{figure*}[t!]
		\begin{tabular}{ccc}
			\includegraphics[width=0.31\textwidth]{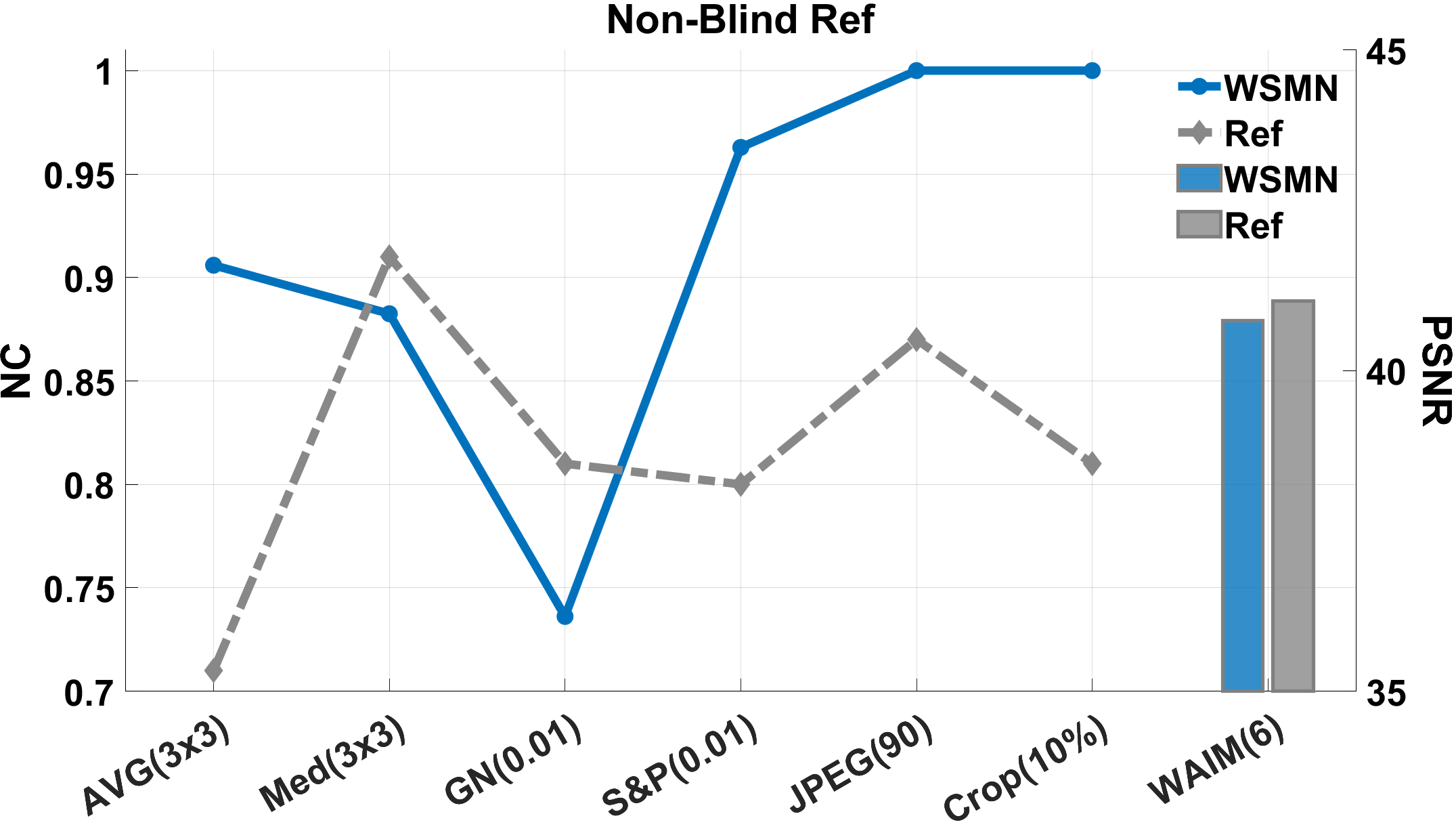} &
			\includegraphics[width=0.31\textwidth]{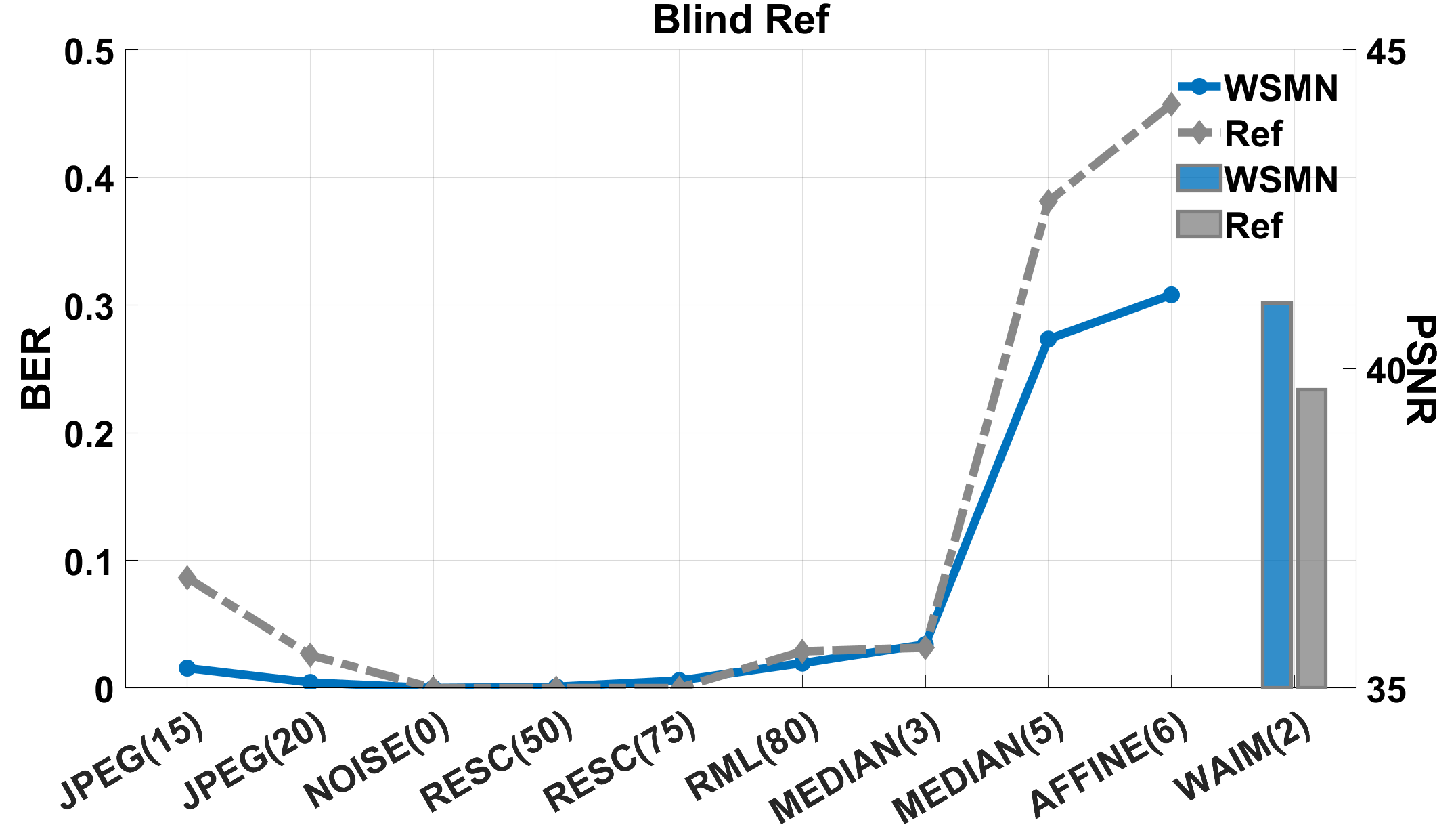} &
			\includegraphics[width=0.31\textwidth]{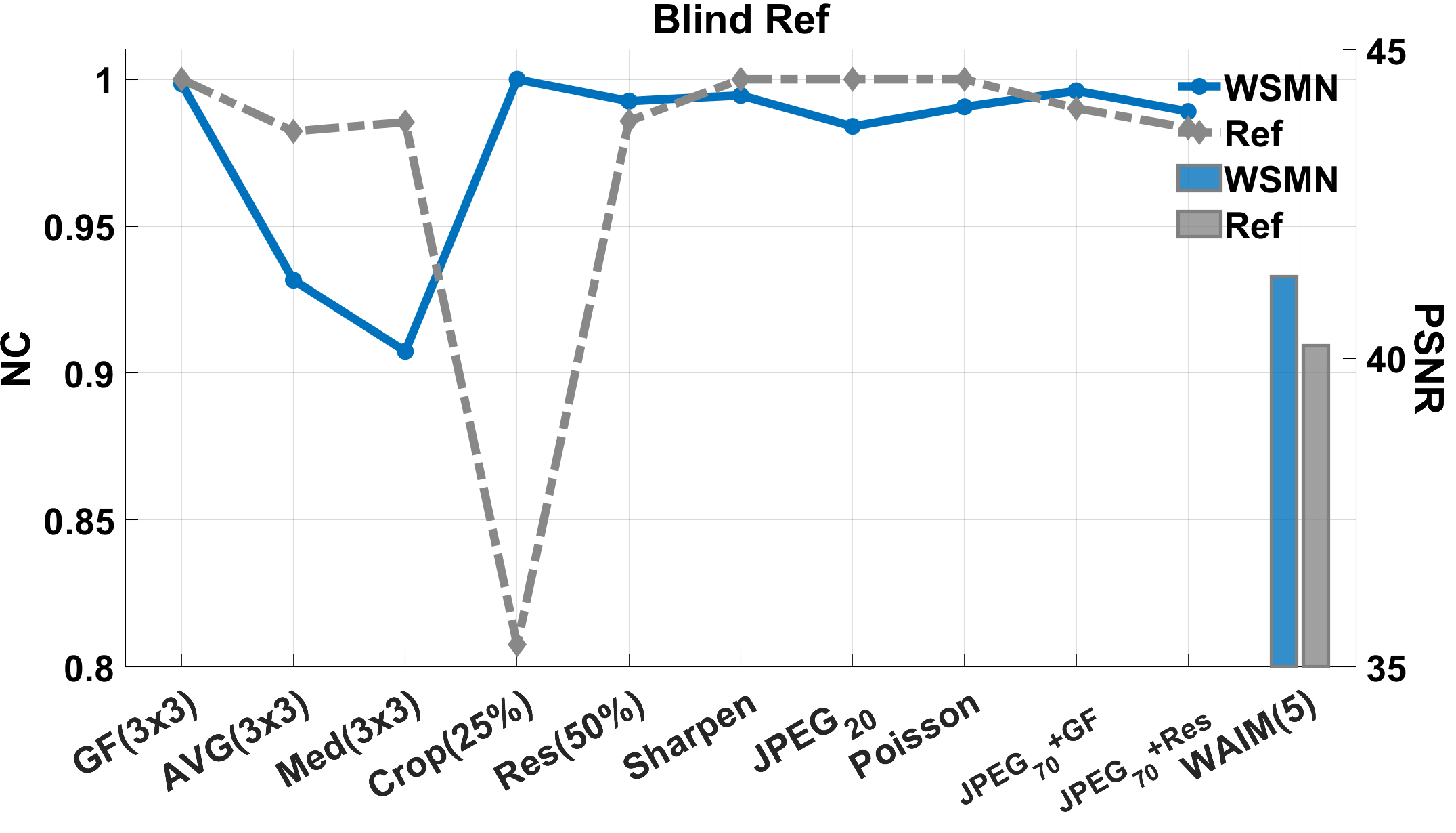} \\
			(a)&(b)&(c)\\\\
			\includegraphics[width=0.31\textwidth]{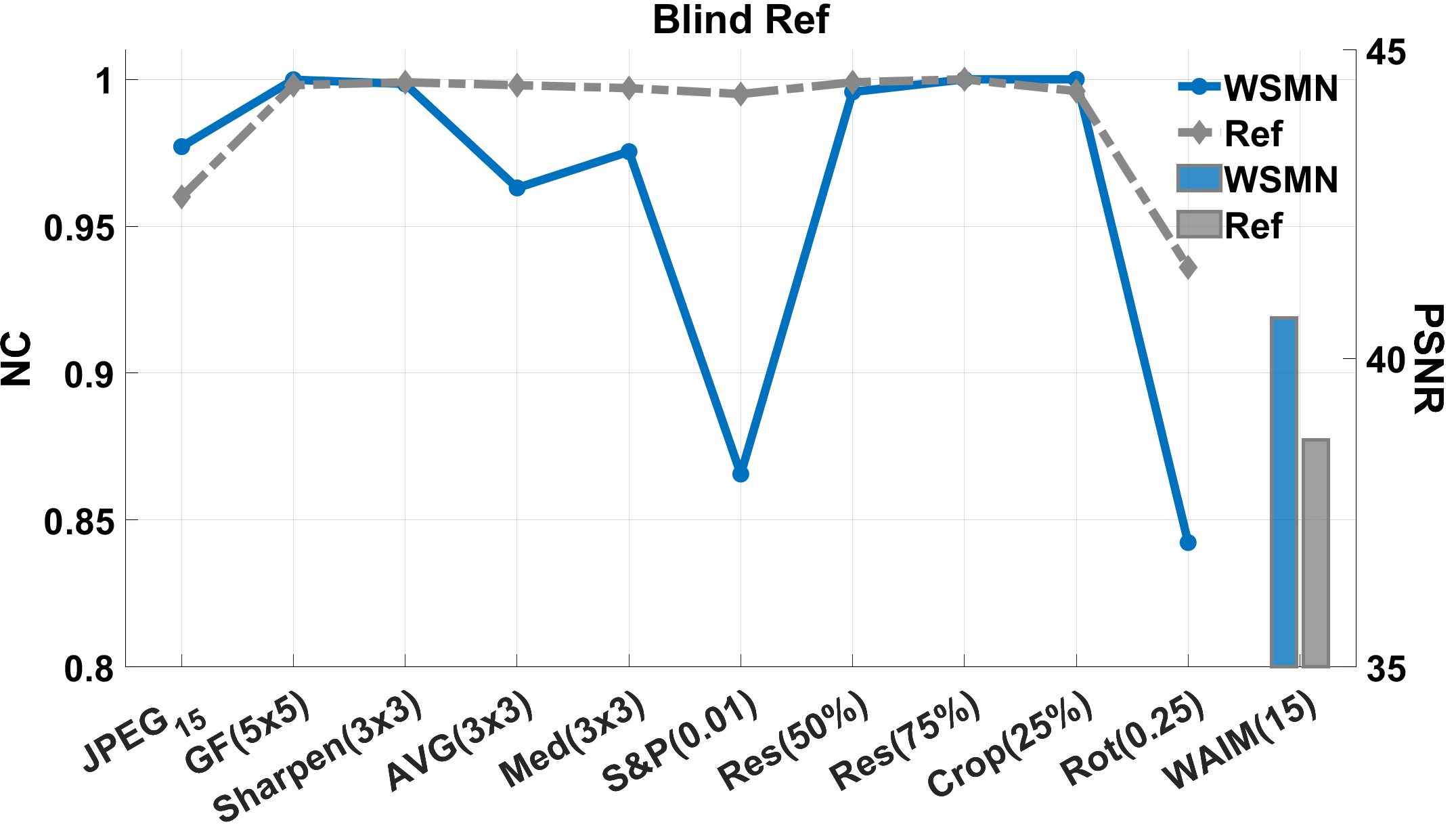} &
			\includegraphics[width=0.31\textwidth]{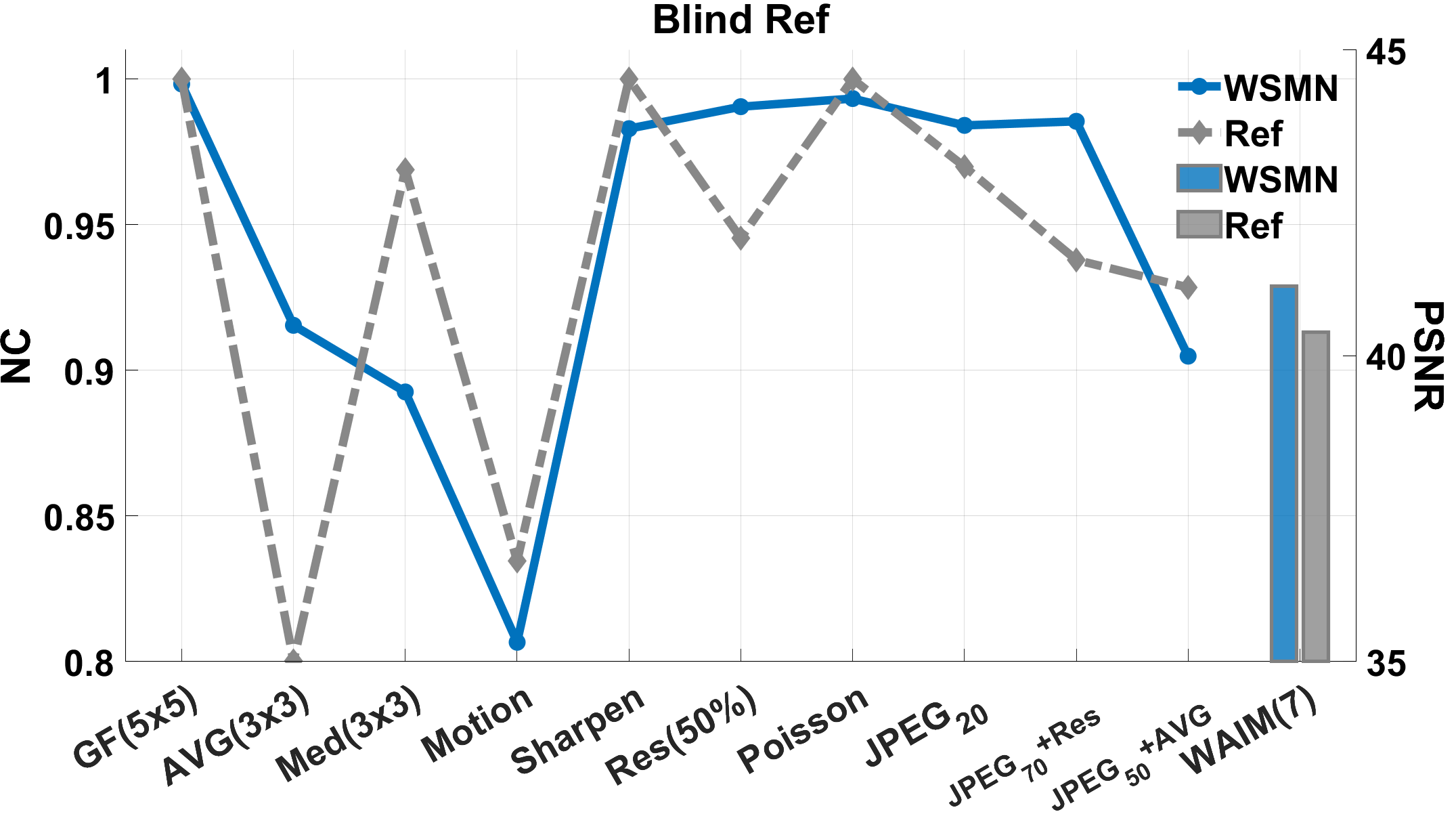} &
			\includegraphics[width=0.31\textwidth]{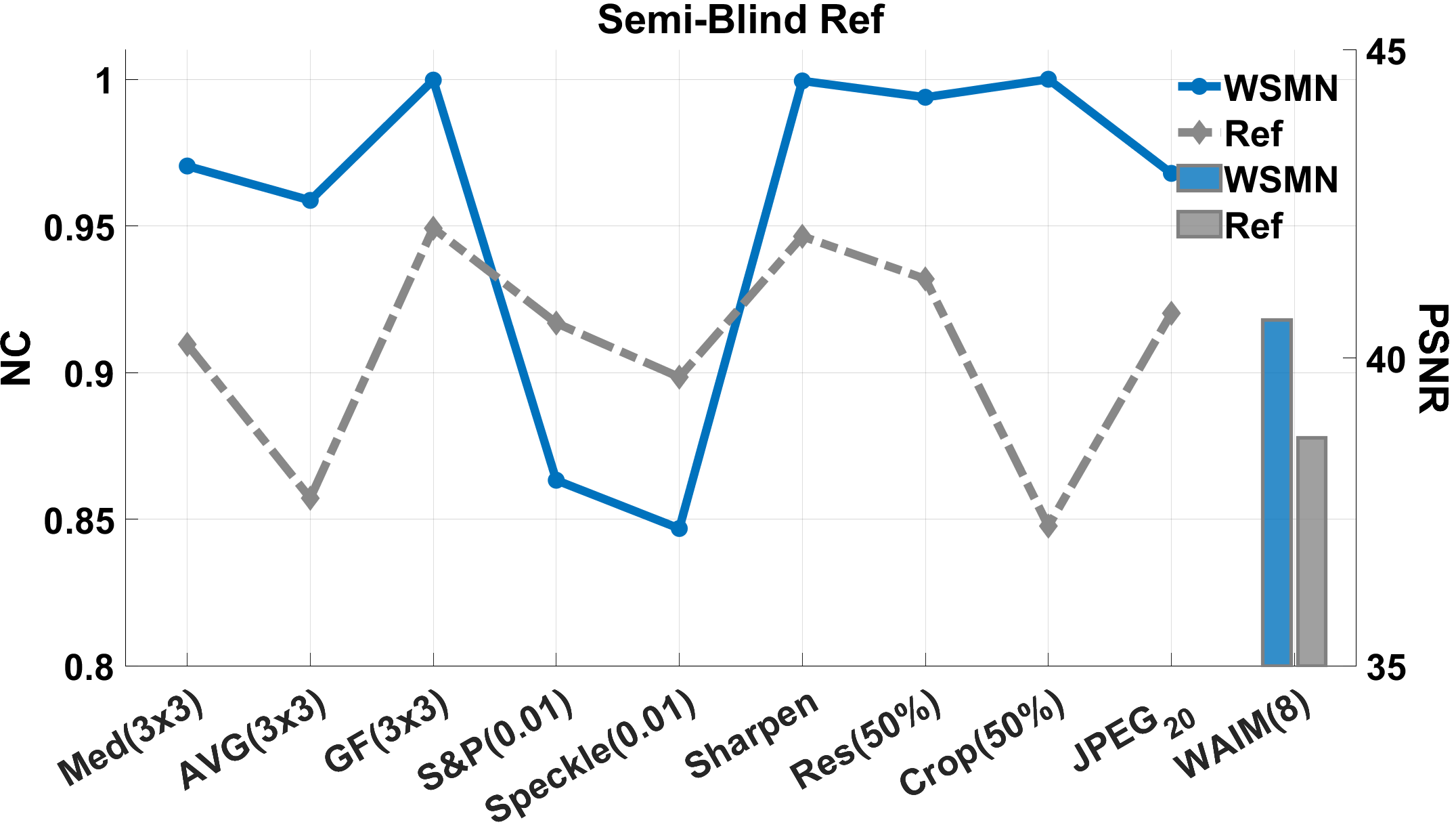} \\
			(d)&(e)&(f)\\		 		
		\end{tabular}
		\center
		\begin{tabular}{cc}
			\includegraphics[width=0.40\textwidth]{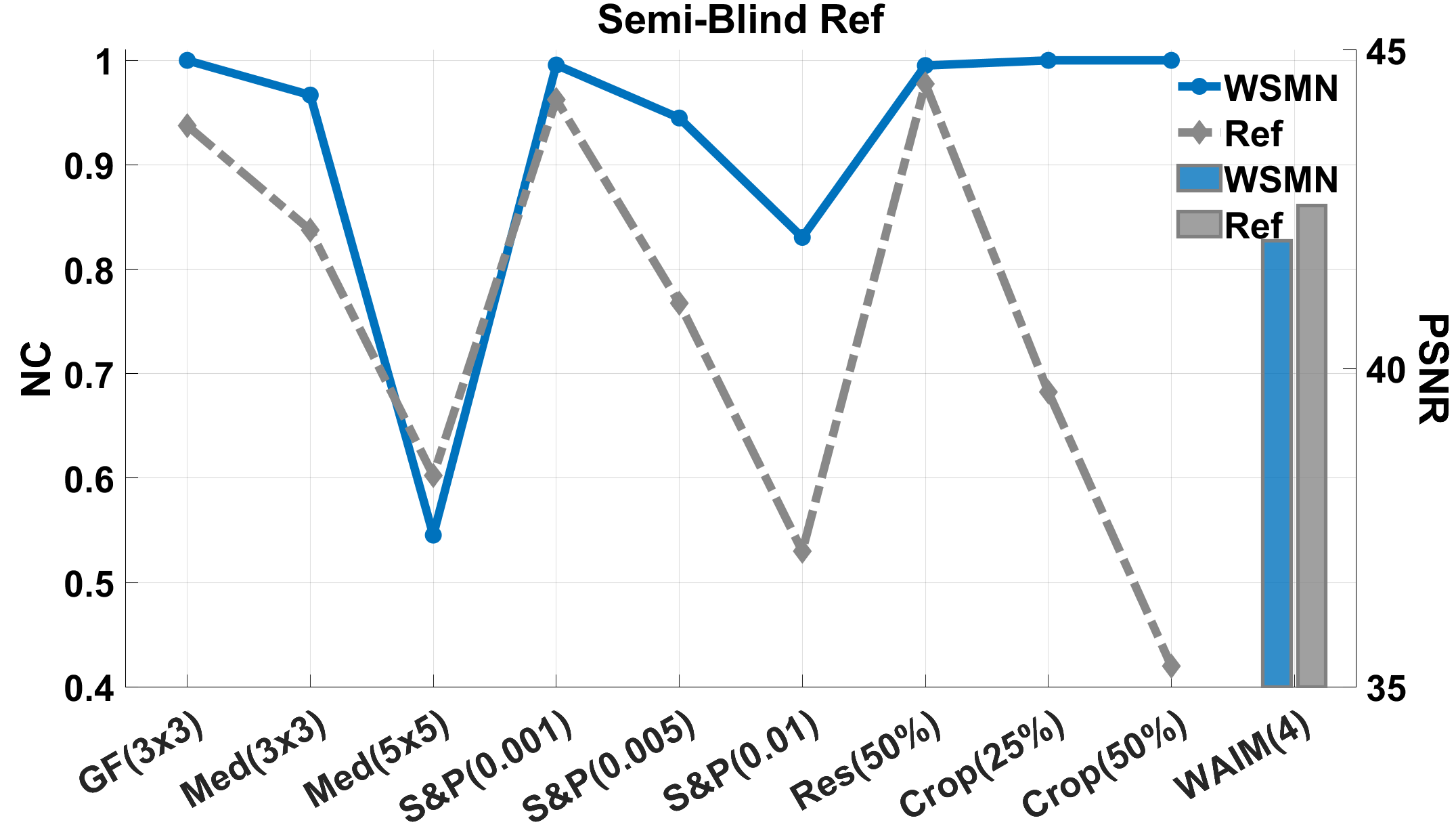} &
			\includegraphics[width=0.40\textwidth]{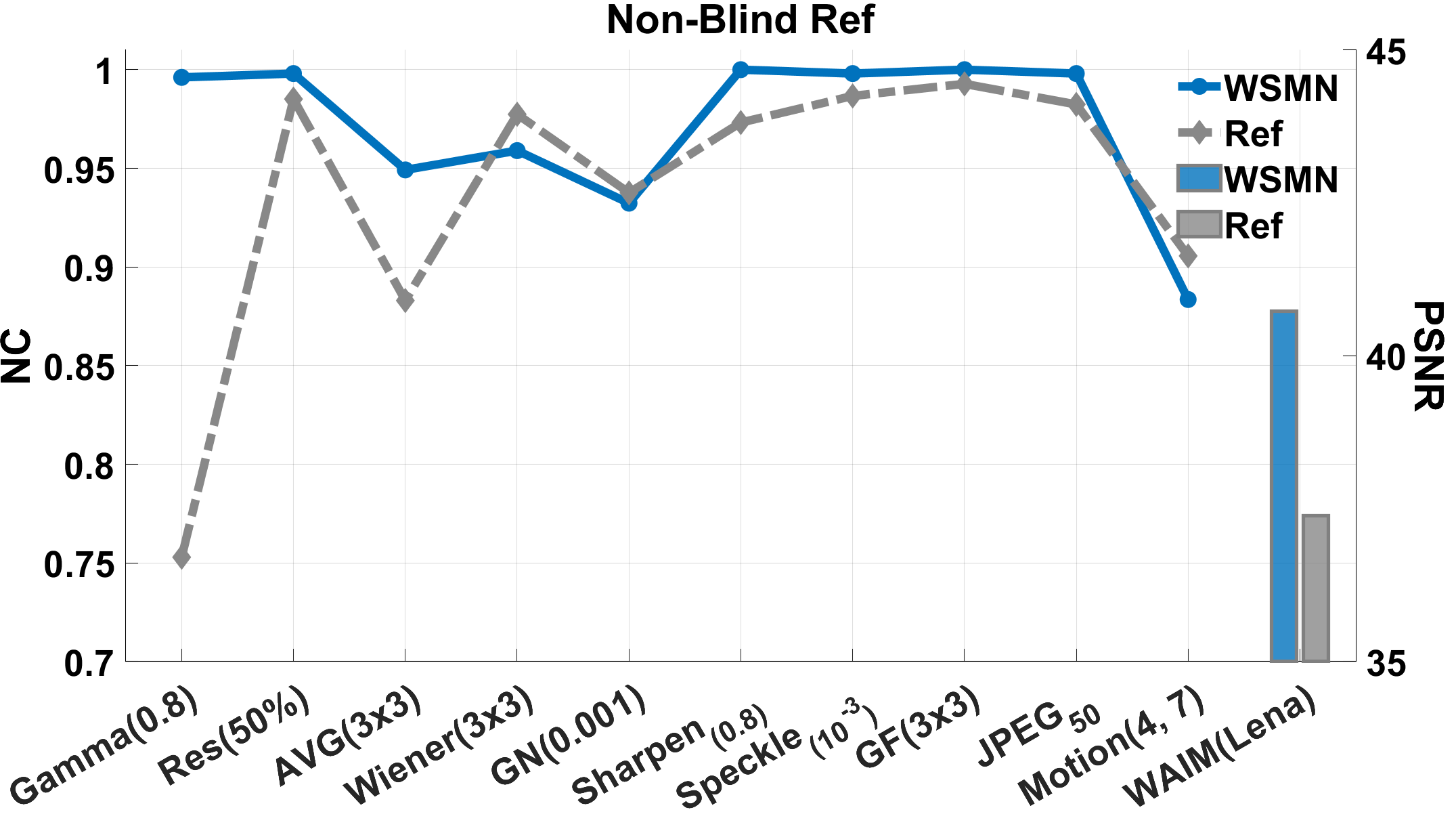} \\
			(g)&(h)\\\\
		\end{tabular}
		
		\begin{tabular}{cc}
			\includegraphics[width=0.40\textwidth]{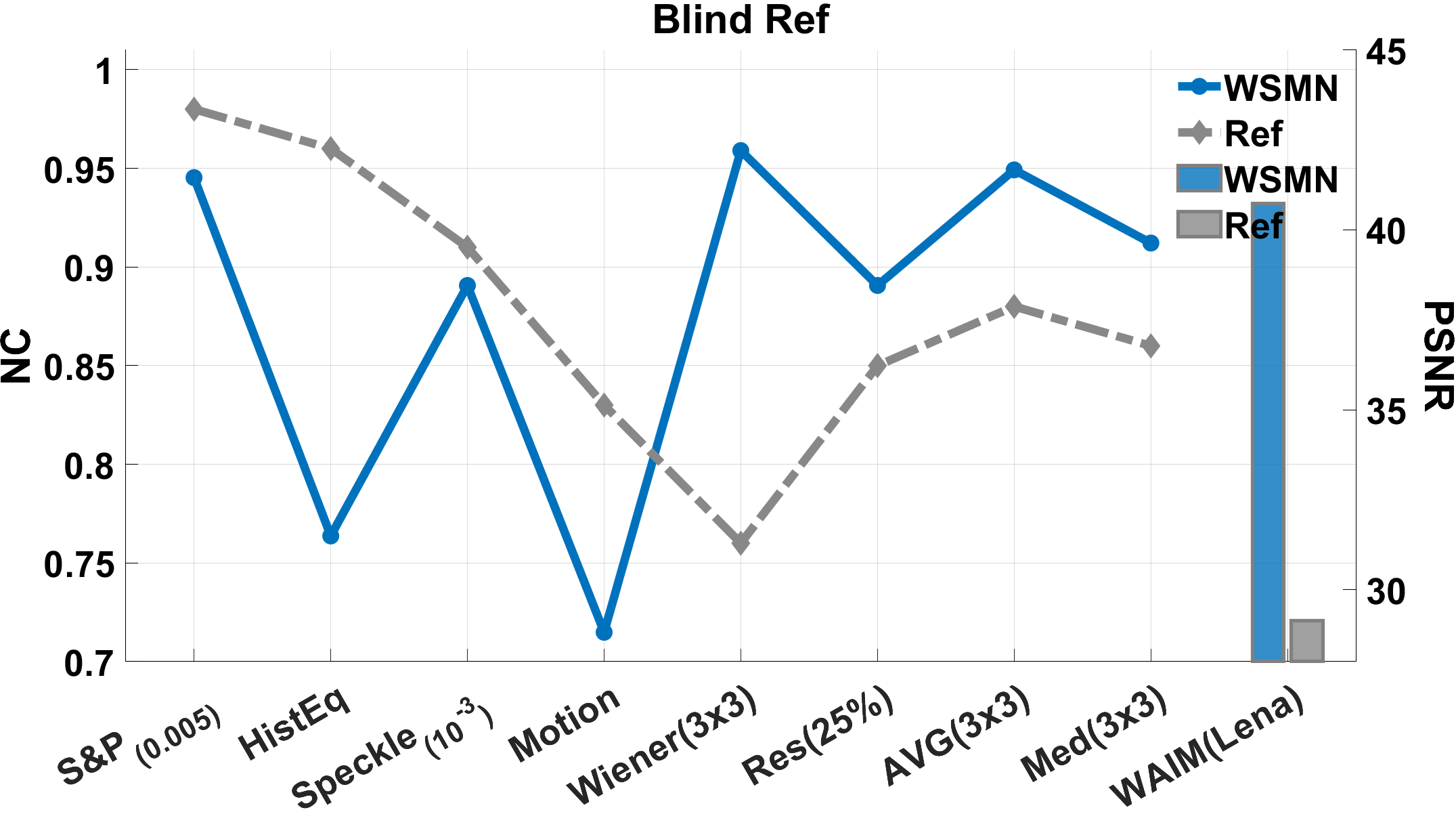} &
			\includegraphics[width=0.40\textwidth]{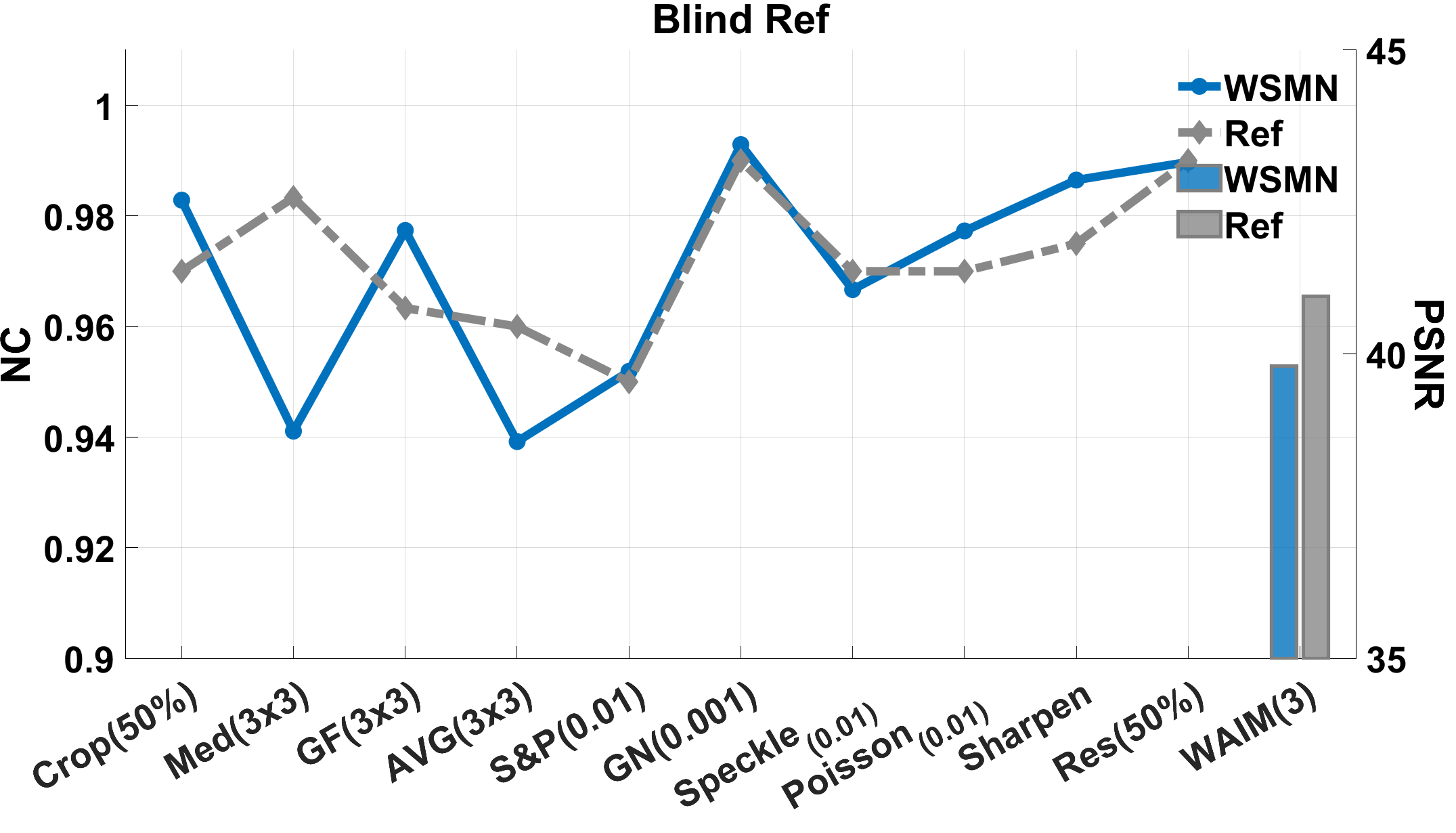} \\
			(i)&(j)
		\end{tabular}
		\caption{The performance of WSMN in terms of robustness (Copyright Protection) under different types of attacks compared to previous works in the same experiments setup. (a) \cite{ref9} Test images: Lena, F16, Lake, Pepper, Baboon, and House, (b) \cite{ref8} Stirmark, Test images: Lena and Baboon \cite{ref35}, (c) \cite{ref10} Test images: Lena, Pepper, Baboon, Barbara, and F16 (d) \cite{ref12} Tested on the whole dataset, (e) \cite{ref14} Test images: Lena, Pepper, Baboon, Barbara, F-16, House, and Lake, (f) \cite{ref11} Test image: Lena, Goldhill, Pepper, Boat, Barbara, House and F16, (g) \cite{ref13} Test images: Barbara, Boat, Lena, and Pepper (h) \cite{ref23} Test image: Lena, (i) \cite{ref22} Test image: Lena, (j) \cite{ref24} Test images: Pepper, Lena, and F16. Note, the bar chart in the right side of each sub-Figs represents PSNR of WAtermarked IMages (n) (WAIM, $n$ shows number of test images) to facilitate the comparison. Also, the list of abbreviation are Average Filter(AVG), Median Filter(Med), Gaussian Filter (GF) Gaussian Noise(GN), Salt\&Pepper Noise(S\&P), Resize (Res), and Rotation (Rot).} 
		\label{fig:compare2}
	\end{figure*}
	\begin{sidewaystable}
		\footnotesize
		\caption{Comparison between WSMN and recent schemes in terms of various features.\\
			Note: R, SF, and F represent Robust, Semi-Fragile, and Frailge, respectively. U/D shows undefined for corresponded schemes. For embedding technique, Q and C  indicate quantization and correlation techniques, respectively. Also, $\times$ denotes the lack of employing intelligent algorithms for various purposes.}
		\label{TABLE:comapre3}
		\renewcommand{\arraystretch}{1.3}
		\scalebox{1} {
			\begin{tabular*}{\textwidth}{@{\extracolsep{\fill} }@{}l@{}c@{}c@{}c@{}c@{}c@{}c@{}c@{}c@{}c@{}c@{}c@{}c@{}c@{}c@{}}
				\cline{1-15}
				Features&WSMN&\cite{ref9}&\cite{ref8}&\cite{ref10}&\cite{ref12}&\cite{ref14}&\cite{ref11}&\cite{ref13}&\cite{ref19}&\cite{ref20}&\cite{ref21}&\cite{ref23}&\cite{ref22}&\cite{ref24} \\
				\cline{1-15}
				Application&Dual&Ownership&Ownership&Ownership&Ownership&Ownership&Ownership&Ownership&Verification&Verification&Verification&Dual&Dual&Dual \\
				Extraction&Blind&Non-Blind&Blind&Blind&Blind&Blind&Semi-Blind&Semi-Blind&Blind&Semi-Blind&Semi-Blind&Non-Blind&Blind&Blind \\
				Robustness&R/SF&R&R&R&R&R&R&R&SF&SF&SF&R/F&R/F&R/F \\
				Copyright&32$\times$32$\times$4&256$\times$256&32$\times$32&32$\times$32&32$\times$32$\times$3&32$\times$32&32$\times$16&32$\times$32&U/D&U/D&U/D&128$\times$128&128$\times$128&64$\times$64 \\
				Verification&64$\times$64&U/D&U/D&U/D&U/D&U/D&U/D&U/D&64$\times$64&64$\times$64&128$\times$128&128$\times$128&256$\times$256&64$\times$64 \\
				Block Size&8$\times$8&1$\times$1$_{LL}$&4$\times$4$_{LL}$&8$\times$8&8$\times$8&8$\times$8&2$\times$2$_{LL_{2}}$&1$\times$1$_{LL_{4}}$&8$\times$8&8$\times$8&4$\times$4&4$\times$4&4$\times$4&8$\times$8 \\
				Technique&Q+C&C&C&C&Q&C&C&Q&C&Q&Q&C+LSBs&LSBs&C+LSBs \\
				Domain&DST+LWT&DWT+SVD&DWT+SVD&DCT&DCT&DCT&LWT&DWT&DWT&DWT+SVD&DWT&Pixel+DWT&Pixel+DCT&Pixel+DCT \\
				Learning Alg&MLP&$\times$&$\times$&$\times$&$\times$&$\times$&SVM&B-ELM&$\times$&$\times$&$\times$&$\times$&$\times$&$\times$ \\
				Texture Alg&K-Means&$\times$&$\times$&Variance&$\times$&Variance&$\times$&$\times$&$\times$&$\times$&$\times$&$\times$&$\times$&$\times$ \\
				Optimization&NSGA-II&$\times$&$\times$&$\times$&ABC&TLBO&$\times$&CVNN&$\times$&$\times$&$\times$&ABC&$\times$&$\times$ \\
				Security&CCS&RSA&Q-Deformed&ACM&$\times$&$\times$&Rand&$\times$&ACM&SVD&Hash+MTA&$\times$&LCM&ACM \\
				Color Image&\checkmark&\checkmark&$\times$&\checkmark&$\times$&\checkmark&$\times$&$\times$&$\times$&$\times$&$\times$&$\times$&$\times$&\checkmark \\
				\cline{1-15}
				
		\end{tabular*}}
	\end{sidewaystable}	
	In this subsection, several simulations are conducted to evaluate the effectiveness of WSMN in terms of detecting tampered parts. For this aim, three distinct measures, including Accuracy (AC), True Positive Rate (TPR), and False Positive Rate (FPR) are adopted. In the first experiment, the performance is analyzed under various attacks. For this aim, the center part of fifteen color and grayscale watermarked images are tampered by region with a size of $100\times100$ pixels. Afterward, the mentioned attacks are performed on the tampered image to simulate the distorting mark and challenge system. The average results of these experiments in terms of AC, TRP, and FPR are indicated in Fig. \ref{fig:tamper1}. As can be seen, in most situations, by increasing the strength of attacks, the tampered regions can still be fully localized with acceptable accuracy. As expected, although WSMN makes proud in the majority of cases, for some Noisy state and JPEG with the strength more than half unable to determine tampered regions, meticulously. In other words, FPR slightly rises during the increasing step of noise attacks and the compression ratio of JPEG. Also, the FPR of the median filter is insignificantly higher than other operations. Something else which should be pinpoint here is that, similar to the accuracy of the extracted copyright mark, the performance of WSMN in terms of localization for dual application is slightly better compared to single-mode. Moreover, by looking at details, it can be found from TPR and FPR in Figs. \ref{fig:tamper1} (a-d), the tampered regions are definitely recognized, but in some cases, due to the strength of attacks, the algorithm mistakenly marked the valid part as tampered regions. 
	
	Further, to prove the performance competency of WSMN, three types of purposeful tampering with imperceptible semantic changes under extreme hybrid attacks have been studied, and the results are visually illustrated in Fig. \ref{fig:tamper2}. In the first experiment, the face of Lena is modified by the face of another girl. Then, Brightness Darken and Sharpening Filter are applied to the tampered image to destroy the authentication mark. In the second analysis, the two warplanes are inserted to the top and bottom of the watermarked F16. In the following, LSBs Replacing and Resizing filter are performed on the forged image. Lastly, to analysis, the capability of the scheme to detect copy-move and vector quantization tampering, a window of the house and the man are duplicated in watermarked Goldhill. Moreover, a building is copied from watermarked Camera Man and pasted in Goldhill. Similarly, the final version of the tampered image is generated under further process by employing JPEG compression and Salt\&Pepper Noise. It is quite evident from Fig. \ref{fig:tamper2}. (b), the extracted copyright marks are still clear after the tampering; Hence, thanks to accurate mark, the ownership can be guaranteed and employed in judicial purposes. Besides, as demonstrated in Fig. \ref{fig:tamper2}. (d), the tampered parts (or multi-tampering) are correctly marked under these extreme hybrid pair attacks. In summary, WSMN able to localize small tampered regions under extreme hybrid attacks regardless of the size and number of forged regions, modification location, intensities, etc.
	
	\subsection{Comparison with the related State-of-the-Art}
	As discussed before, the main properties of a watermarking system are imperceptibly and robustness. Hence, in this subsection, to demonstrate the superiority of WSMN, different comparisons against state-of-the-art methods are performed. The compassion examination is based on three factors, such as quality of watermarked image, the robustness of extracted mark, and the main features of proposed systems, respectively.
	In this way, WSMN is compared with ten approaches under the same experiment's setup, including the application (single and dual), type of image (color or grayscale), the type and determining parameters of applied attacks, etc.
	
	As the first comparison, the imperceptibly of WSMN for both single and dual cases are illustrated in Table \ref{TABLE:comapre1}. For this aim, the same tested images which mostly utilized in related works are candidate; The sign (-) shows the corresponding image is not available in prior works. It is quite evident, WSMN reaches higher PSNR values for both single and dual applications in the majority of cases. For the ownership protection in both single and dual-mode, WSMN has absolute superiority regarding to \cite{ref8,ref12,ref11,ref13,ref23,ref22}. In compression to rest works, the slight decrease is found for Baboon and F16. On the other hand, for content integrity protection, WSMN is approximately proud of whole cases. It should be noted, the authenticate watermark is fragile in the dual application presented schemes \cite{ref23, ref22, ref24}. In other words, these schemes cannot tolerate aggressive attacks and categorized as fragile for this application. Hence, this is one of the main drawbacks of these schemes. Due to the fragile watermarking, there is an insignificant improvement in terms of imperceptible in \cite{ref24} compared to WSMN. As mentioned before, to the best of the authors' knowledge, it is the first time that robust and semi-fragile schemes are employed to prove the ownership and integrity protection, simultaneously.
	
	Furthermore, WSMN is also compared with existing watermarking schemes in terms of robustness in Fig. \ref{fig:compare2}. In this way, various single and hybrid attacks are applied to the same watermarked image; NC and BER of the extracted mark are illustrated as line charts in Figs. \ref{fig:compare2} (a-j). Similarly, the employed test images and the parameters of attack are identically considered for a fair comparison. Moreover, the average PSNR of watermarked images is presented on the right side of the charts to facilitate comparison. By global looking, the results indicate the superior performance of WSMN compare to other single and dual applications regarding the majority of attacks. Moreover, the quality of watermarked images are nearly equal or even extremely higher in some cases. In comparison to \cite{ref8, ref10, ref12, ref14, ref11, ref23, ref22}, WSMN can prepare high imperceptible watermarked image in single and dual modes. Meanwhile, it significantly yields higher accuracy for the majority of attacks such as Gamma, JPEG, Crop, Resize, Sharpen, Smoothing Filters, etc. compared to the mentioned methods in most cases. For the rest method \cite{ref9, ref21, ref24}, which the quality of the watermarked image is slightly upper than WSMN, the plots of Figs. \ref{fig:compare2} (a, g, i) demonstrate the admissible performance compare to reference methods. In contrast, simulation experiments show that, although WSMN achieves high quality of embedded images and watermark robustness under compression, filtering, scaling, and other mentioned attacks, it can be found weakness with noise operations in some cases. It should be noted, thanks to the various chances of copyright marks, WSMN can extract the mark without any mistake under 50\% cropping. In general, despite the fact the most schemes are semi-blind or even non-blind, they cannot reach to acceptable correlation in terms of robustness in the majority of cases.
	
	In the last comparison, the highlighted features of the studied watermarking schemes are summarized as depicted in Table \ref{TABLE:comapre3}. In this way, firstly, the key properties of each watermarking system, including the goal of system, extraction process, and robustness are listed. As mentioned before, non-blind \cite{ref9,ref23} or semi-blind \cite{ref11, ref13, ref20, ref21} watermarking has the least practical compare to blind schemes. More, the fragility of the schemes in \cite{ref23,ref22,ref24} causes the watermark to be destroyed with the slightest attack. In the following, the rest of features such as the size of watermark and blocks, embedding technique, employed domain (transform), texture analysis, utilization of optimization and learning algorithm, the security of system, and supporting color image are demonstrated based on the what has been reported. As seen, the majority of schemes used wavelet and cosine transforms in the embedding phase. Also, the security, optimizing strength parameters, and texture of the image were meaningless in most cases. Unlike proposed techniques, WSMN consistently overcomes existing challenges by considering the texture of the image and performing intelligent algorithms to optimize the embedding watermarking.
	
	In terms of content integrity protection, unfortunately, the majority of watermarking schemes proposed in recent years are fragile and working in the spatial domain by modifying pixels directly. Whereas fragile techniques reach significant localization accurately but do not prove robustness to content preserving operations. Moreover, about-semi fragile schemes \cite{ref19, ref20, ref21}, which presented in the last decade, there is not a similar setup such as the rate and location of tampering to facilitate comparison. To sum up, the accuracy of WSMN in terms of authentication is proved based on imperceptibly and property of the algorithm in Tables \ref{TABLE:comapre1} and \ref{TABLE:comapre3}, respectively.
	
	Finally, based on the experimental results, WSMN outperforms the other schemes in terms of imperceptibility, capacity, and security. Besides, it is quite clear that it leads to the best performance in terms of robustness in the majority of cases. Furthermore, high accuracy results are achieved by WSMN in content integrity applications. Accordingly, it can be concluded that the proposed dual scheme is more notable.
	
	\section{Conclusion and Future Works}
	\label{sec:Conclusion}
	The rapid growth of the Internet, followed by the sharing of multimedia data and access to digital image processing tools, facilitated image spreading and tampering, and consequently, caused copyright infringement and destruction of image integrity. To overcome the challenges and improve the efficiency of the previous methods, this paper presented an optimized multipurpose blind watermarking based on Shearlet transform with the help of smart algorithms, including MLP and NSGA-II. In WSMN, the robust copyright logo and semi-fragile authentication mark are embedded in approximate and details coefficients of Shearlet based on quantization and correlation techniques, respectively. In the embedding phase, rough and smooth blocks are distinguished into different levels based on texture descriptors and K-Means clustering algorithm. Furthermore, the adaptive threshold selection by NSGA-II markedly increases the performance of WSMN. In consequence of these strategies and adaptive adjustment techniques for determining the direction and threshold steps, extremely enhance the quality of watermarked images and system efficiency against different hybrid attacks. WSMN provides four chances for the copyright logo, which yields a high correlation of the extracted mark, specifically in large tampered areas. On the other hand, the authentication mark is extracted with the maximum possible correlation under hybrid attacks by MLP. The experimental results indicate superiority and efficiency of WSMN in comparison with the reviewed methods in the literature concerning the quality of the watermarked image, robustness against different attacks such as basic image processing operations, noises, compression, and tamper detection rate. WSMN is not only robust against the single attack but also is robust against hybrid attacks. Moreover, the high quality of the watermarked image and optimal embedding process are among other advantages of this method. According to the mentioned advantages, WSMN is efficient, secure, and functional for copyright protection and tamper localization of gray and color images.
	
	Despite these advantages, including robustness against compression attacks, noise, and image processing operations, WSMN is too vulnerable against geometric attacks such as rotation and affine. Further studies should focus on improving robustness against these attacks. Additionally, WSMN will further be extended to develop an efficient recovery algorithm for reconstructing tampered regions. To do so, compression algorithms, such as JPEG2000, can be used for generating digest to improve quality while being so compact. Also, to encourage future works, the MATLAB source code of WSMN is available in \href{http://b.bolourian.student.um.ac.ir/}{(Link)}.

\end{document}